\documentclass[twocolumn]{aastex631}
\usepackage{amsmath}
\usepackage{epsf}
\usepackage{color}
\usepackage{graphicx}
\usepackage{color}
\usepackage{enumitem}
\usepackage{mathtools}
\usepackage[mathcal]{eucal}
 \let\mathscr\relax
\usepackage[scr]{rsfso}

\shorttitle{JWST/CEERS disk galaxies at $z>3$}
\shortauthors{Vega-Ferrero et al.}

\usepackage{graphicx}	% Including figure files
\usepackage{hyperref}
\usepackage{amsmath}	% Advanced maths commands
\usepackage{comment}

\begin{document}
% Title of the paper, and the short title which is used in the headers.
% Keep the title short and informative.

\title{On the nature of disks at high redshift seen by JWST/CEERS with contrastive learning and cosmological simulations}

%\suppressAffiliations

\correspondingauthor{Jes\'us Vega-Ferrero}
\email{astrovega@gmail.com}

\author[0000-0003-2338-5567]{Jes\'us Vega-Ferrero}
\affiliation{Instituto de Astrof\'isica de Canarias, 38200, La Laguna, Tenerife, Spain}
\affiliation{Departamento de Astrof\'isica, Universidad de La Laguna, 38205, La Laguna, Tenerife, Spain}
\affiliation{Departamento de F\'{i}sica Te\'{o}rica, At\'{o}mica y \'{O}ptica, Universidad de Valladolid, 47011 Valladolid, Spain}

\author[0000-0002-1416-8483]{Marc Huertas-Company}
\affiliation{Instituto de Astrof\'isica de Canarias, 38200, La Laguna, Tenerife, Spain}
\affiliation{Departamento de Astrof\'isica, Universidad de La Laguna, 38205, La Laguna, Tenerife, Spain}
\affiliation{LERMA, Observatoire de Paris, CNRS, PSL, Universit\'e de Paris, France}

\author[0000-0001-6820-0015]{Luca Costantin}
\affiliation{Centro de Astrobiolog\'{\i}a (CAB), CSIC-INTA, Ctra. de Ajalvir km 4, Torrej\'on de Ardoz, E-28850, Madrid, Spain}

\author[0000-0003-4528-5639]{Pablo G.~P\'erez-Gonz\'alez}
\affiliation{Centro de Astrobiolog\'{\i}a (CAB), CSIC-INTA, Ctra. de Ajalvir km 4, Torrej\'on de Ardoz, E-28850, Madrid, Spain}

\author[0000-0002-3803-6952]{Regina Sarmiento}
\affiliation{Instituto de Astrof\'isica de Canarias, 38200, La Laguna, Tenerife, Spain}
\affiliation{Departamento de Astrof\'isica, Universidad de La Laguna, 38205, La Laguna, Tenerife, Spain}

\author[0000-0001-9187-3605]{Jeyhan S. Kartaltepe}
\affiliation{Laboratory for Multiwavelength Astrophysics, School of Physics and Astronomy, Rochester Institute of Technology, 84 Lomb Memorial Drive, Rochester, NY 14623, USA}

\author[0000-0003-1065-9274]{Annalisa Pillepich}
\affiliation{Max-Planck-Institut f{\"u}r Astronomie, K{\"o}nigstuhl 17, 69117 Heidelberg, Germany}

\author[0000-0002-9921-9218]{Micaela B. Bagley}
\affiliation{Department of Astronomy, The University of Texas at Austin, Austin, TX, USA}

\author[0000-0001-8519-1130]{Steven L. Finkelstein}
\affiliation{Department of Astronomy, The University of Texas at Austin, Austin, TX, USA}

\author[0000-0001-8688-2443]{Elizabeth J.\ McGrath}
\affiliation{Department of Physics and Astronomy, Colby College, Waterville, ME 04901, USA}

\author[0000-0003-1643-0024]{Johan H. Knapen}
\affiliation{Instituto de Astrof\'isica de Canarias, 38200, La Laguna, Tenerife, Spain}
\affiliation{Departamento de Astrof\'isica, Universidad de La Laguna, 38205, La Laguna, Tenerife, Spain}

%In alphabetic order
%%%%%%%%%%%%%%%%%%%%%%%%%%%%%%%%%%%%%%%%%%%%%%%%%%%%%%%%%%%%%%

\author[0000-0002-7959-8783]{Pablo Arrabal Haro}
\affiliation{NSF's National Optical-Infrared Astronomy Research Laboratory, 950 N. Cherry Ave., Tucson, AZ 85719, USA}

\author[0000-0002-5564-9873]{Eric F.\ Bell}
\affiliation{Department of Astronomy, University of Michigan, 1085 S. University Ave, Ann Arbor, MI 48109-1107, USA}

\author[0000-0002-2861-9812]{Fernando Buitrago}
\affiliation{Departamento de F\'{i}sica Te\'{o}rica, At\'{o}mica y \'{O}ptica, Universidad de Valladolid, 47011 Valladolid, Spain}
\affiliation{Instituto de Astrof\'{i}sica e Ci\^{e}ncias do Espa\c{c}o, Universidade de Lisboa, OAL, Tapada da Ajuda, PT1349-018 Lisbon, Portugal}

\author[0000-0003-2536-1614]{Antonello Calabr{\`o}}
\affiliation{Osservatorio Astronomico di Roma, via Frascati 33, Monte Porzio Catone, Italy}

\author[0000-0003-4174-0374]{Avishai Dekel}
\affiliation{Centre for Astrophysics and Planetary Science, Racah Institute of Physics, The Hebrew University, Jerusalem, 91904, Israel}
\affiliation{Santa Cruz Institute for Particle Physics, University of California, Santa Cruz, CA 95064, USA}

\author[0000-0001-5414-5131]{Mark Dickinson}
\affiliation{NSF's National Optical-Infrared Astronomy Research Laboratory, 950 N. Cherry Ave., Tucson, AZ 85719, USA}

\author[0000-0002-9013-1316]{Helena Dom\'{i}nguez S\'{a}nchez}
\affiliation{Centro de Estudios de F\'{i}sica del Cosmos de Arag\'{o}n (CEFCA), Plaza de San Juan, 1, E-44001 Teruel, Spain}

\author[0000-0002-7631-647X]{David Elbaz}
\affiliation{Laboratoire AIM-Paris-Saclay, CEA/DRF/Irfu - CNRS - Universit\'e Paris Cit\'e, CEA-Saclay, pt courrier 131, F-91191 Gif-surYvette, France}

\author[0000-0001-7113-2738]{Henry C. Ferguson}
\affiliation{Space Telescope Science Institute, 3700 San Martin Drive, Baltimore, MD 21218, USA}

\author[0000-0002-7831-8751]{Mauro Giavalisco}
\affiliation{University of Massachusetts Amherst, 710 North Pleasant Street, Amherst, MA 01003-9305, USA}

\author[0000-0002-4884-6756]{Benne W. Holwerda}
\affiliation{Physics \& Astronomy Department, University of Louisville, 40292 KY, Louisville, USA}

\author[0000-0002-8360-3880]{Dale D. Kocesvski}
\affiliation{Department of Physics and Astronomy, Colby College, Waterville, ME 04901, USA}

\author[0000-0002-6610-2048]{Anton M. Koekemoer}
\affiliation{Space Telescope Science Institute, 3700 San Martin Drive, Baltimore, MD 21218, USA}

\author[0000-0002-2499-9205]{Viraj Pandya}
\altaffiliation{Hubble Fellow}
\affiliation{Columbia Astrophysics Laboratory, Columbia University, 550 West 120th Street, New York, NY 10027, USA}

\author[0000-0001-7503-8482]{Casey Papovich}
\affiliation{Department of Physics and Astronomy, Texas A\&M University, College Station, TX, 77843-4242 USA}
\affiliation{George P.\ and Cynthia Woods Mitchell Institute for Fundamental Physics and Astronomy, Texas A\&M University, College Station, TX, 77843-4242 USA}

\author[0000-0003-3382-5941]{Nor Pirzkal}
\affiliation{ESA/AURA Space Telescope Science Institute}

\author[0000-0001-5091-5098]{Joel Primack}
\affiliation{Santa Cruz Institute for Particle Physics, University of California, Santa Cruz, CA 95064, USA}

\author[0000-0003-3466-035X]{{L. Y. Aaron} {Yung}}
\affiliation{Astrophysics Science Division, NASA Goddard Space Flight Center, 8800 Greenbelt Rd, Greenbelt, MD 20771, USA}

\begin{abstract}

Visual inspections of the first optical rest-frame images from JWST have indicated a surprisingly high fraction of disk galaxies at high redshifts. Here, we alternatively apply self-supervised machine learning to explore the morphological diversity at $z \geq 3$. Our proposed data-driven representation scheme of galaxy morphologies, calibrated on mock images from the TNG50 simulation, is shown to be robust to noise and to correlate well with the physical properties of the simulated galaxies, including their 3D structure. We apply the method simultaneously to F200W and F356W galaxy images of a mass-complete sample ($M_*/M_\odot>10^9$) at $ 3 \leq z \leq 6$ from the first JWST/NIRCam CEERS data release. We find that the simulated and observed galaxies do not exactly populate the same manifold in the representation space from contrastive learning. We also find that half the galaxies classified as disks ---either CNN-based or visually--- populate a similar region of the representation space as TNG50 galaxies with low stellar specific angular momentum and non-oblate structure. Although our data-driven study does not allow us to firmly conclude on the true nature of these galaxies, it suggests that the disk fraction at $z \geq 3$ remains uncertain and possibly overestimated by traditional supervised classifications. Deeper imaging and spectroscopic follow-ups as well as comparisons with other simulations will help to unambiguously determine the true nature of these galaxies, and establish more robust constraints on the emergence of disks at very high redshift.

\end{abstract}


%%%%%%%%%%%%%%%%%%%%%%%%%%%%%%%%%%%%%%%%%%%%%%%%%%
%%%%%%%%%%%%%%%%% BODY OF PAPER %%%%%%%%%%%%%%%%%%
%%%%%%%%%%%%%%%%%%%%%%%%%%%%%%%%%%%%%%%%%%%%%%%%%%

\section{Introduction}

Understanding how galaxy diversity emerges across cosmic time is one of the main goals of galaxy formation. How and when do stellar disks form? What are the main drivers of bulge growth? How and when did galaxy morphology and star formation get connected? Despite significant progress in the past years, thanks in particular to deep surveys undertaken with the Hubble Space Telescope (HST, e.g., \citealp{ 2007ApJS..172....1S, 2011ApJS..197...35G, Koekemoer2011}), these questions remain largely unanswered. The general picture is that massive star-forming galaxies in the past were more irregular in their stellar structure (e.g.,  \citealp{ 1996ApJS..107....1A, 2003ApJS..147....1C}) than today's disks even if observed in the optical rest-frame \citep{Buitrago2013,2015ApJ...809...95H}. Galaxies above $z\sim1$ also show the presence of giant star-forming clumps (e.g, \citealp{2015ApJ...800...39G,2018ApJ...853..108G,2020MNRAS.499..814H,2021MNRAS.501..730G}) which might indicate a turbulent and unstable Interstellar Medium (ISM, e.g., \citealp{2010MNRAS.404.2151C,2014ApJ...780...57B}). Although the gas shows signatures of rotation at $z \sim 2$ (e.g.,\citealp{2015ApJ...799..209W}), the settling of disks seems to be a process happening at least from $z\sim2$ (e.g., \citealp{2012ApJ...758..106K,Buitrago2014,2017ApJ...843...46S,Costantin2022b}) coincident with the decrease of gas fractions in massive galaxies (e.g., \citealp{2019A&A...622A.105F,2010MNRAS.407.2091G}). Another important result of the past years is that the presence of bulges in galaxies is strongly anti-correlated with the star formation activity at all redshifts probed (e.g., \citealp{2014ApJ...788...28V,2017ApJ...840...47B,Costantin2020,Costantin2021,2022MNRAS.513..256D}). This suggests that bulge formation and quenching are tightly connected physical processes (e.g., \citealp{2020ApJ...897..102C}).   

With its unprecedented sensitivity, spatial resolution, and infrared coverage, the James Webb Space Telescope (JWST) is offering a new window to the stellar structure of galaxies in the first epochs of cosmic history \citep{Gardner2006}. For the first time, we are able to explore the stellar morphologies of the first galaxies formed in the universe, which should enable new constraints on the physical processes governing galaxy assembly at early times and hopefully a better understanding of the physical processes leading to the formation of the first stellar disks and bulges. Some very recent works have already started this exploration by performing visual classifications \citep{2022ApJ...938L...2F,Ferreira2023,Kartaltepe2022}, by applying supervised Machine Learning trained on HST images \citep{2023ApJ...942L..42R} of galaxies observed in the first JWST deep fields or by using Convolutional Neural Networks (CNNs) trained on HST/WFC3 labeled images and domain-adapted to JWST/NIRCam \citep{Huertas-Company2023}. One of the main results of these early works is that JWST seems to be detecting star-forming disks even at $z>3$, which would push the time of disk formation to very early epochs. Two questions naturally arise from these first works:

\begin{itemize}
    \item Are the galaxies seen by JWST true disks, i.e. flat, rotating systems? The aforementioned results are based primarily on qualitative morphological classifications, with quantitative tracers of morphology (e.g., S\`ersic fits) incorporated to further inform differences between the visually defined classes. However, galaxies might look morphologically disky but have significantly different stellar kinematics than local disk galaxies. Distinguishing edge-on flat disks from more prolate systems is also a very challenging task that could bias the results (e.g. \citealp{vanderWel2014,Zhang2019}). 

    \item Do modern cosmological simulations reproduce the observed galaxy diversity at $z>3$? Although some preliminary comparisons exist, a fair comparison in the observational plane is required to fully address this question (e.g., \citealp{Rodriguez-Gomez2019,2019MNRAS.489.1859H,2021MNRAS.501.4359Z}).
    
\end{itemize}

In this work, we attempt to provide new insights into these two main questions. To that purpose, we apply a novel data-driven approach based on contrastive learning \citep{2021ApJ...911L..33H,Sarmiento2021} to a mass-complete sample of JWST galaxies at $z\geq3$ observed within the Cosmic Evolution Early Release Science \citep[CEERS;][]{Finkelstein2017,Finkelstein2022,Finkelstein2023} survey. By calibrating the method with mock galaxies \citep{Costantin2022a} from the TNG50 cosmological simulation \citep{Nelson2019a,Pillepich2019,Nelson2019b} and by choosing the proper augmentations (i.e., transformations applied to the images such as rotations, flux normalizations, noise, etc.), we are able to build a morphological description which is more robust to noise and galaxy orientation than more traditional approaches. Our morphological representation can then be correlated with the physical properties of galaxies from the simulation to provide new insights about the physical nature of disk-like galaxies, and to explore the agreements and disagreements between observations and simulations.  

The paper proceeds as follows: in \autoref{sec:data} we describe the galaxy datasets used in this work; \autoref{sec:self-supervised} describes the contrastive learning setting used to derive unsupervised representations of galaxy morphologies; \autoref{sec:observations} explores the properties of the obtained representations on observed JWST/CEERS galaxies; a comparison of the self-supervised representations for simulated and observed galaxies is presented in \autoref{sec:comparison_representations}; the results and implications are discussed in~\autoref{sec:true_disks}; finally, a summary and the final conclusions are presented in~\autoref{sec:clonlusions}.

%%%%%%%%%%%%%%%%%%%%%%%%%%%%%%%%%%%%%%%%%%%%%%%%%%

\section{Data}
\label{sec:data}

\subsection{CEERS}
\label{sec:CEERS_data}

We use JWST imaging data from NIRCam obtained within CEERS \citep{Finkelstein2017,Finkelstein2022,Finkelstein2023}.  This consists of short and long-wavelength images in both NIRCam A and B modules, taken over ten pointings. Each pointing was observed with seven filters: F115W, F150W, and F200W on the short-wavelength side, and F277W, F356W, F410M, and F444W on the long-wavelength side. Here we only use the F200W and F356W filters. A full description of this public data release\footnote{https://ceers.github.io/} and the data reduction can be found in \citet{bagley23} and \citet{Finkelstein2022}.

In addition to the galaxy images, we use two different catalogs with physical properties of galaxies: 

\begin{itemize}

\item CEERS catalog (CEERS): a photometric catalog \citep{bagley23,Finkelstein2022} with derived stellar masses and photometric redshifts ($z_\mathrm{phot}$) obtained through Spectral Energy Distribution (SED) fitting of the latest data reduction photometry (Pablo G. P\'erez-Gonz\'alez private communication). For a fair comparison with the simulated TNG50 dataset (see~\autoref{sec:sims}),  we select 1\,664 galaxies with $3 \leq z \leq 6$, stellar masses $M_* \geq 10^9 \mathrm{M_{\odot}}$ and $F200W[AB] < 27~\mathrm{mag}$. The magnitude cut ensures a large enough Signal-to-Noise Ratio (SNR) to enable reliable morphological classification (see \citealt{Kartaltepe2022}). Obvious stars are removed using the same procedure as in \citet{Huertas-Company2023}. Also in \citet{Huertas-Company2023}, the completeness limit is estimated at roughly $M_* \sim 10^{8.5} \mathrm{M_{\odot}}$ over the $0 < z < 6$ redshift range. The cut in mass imposed of $M_* \geq 10^9 \mathrm{M_{\odot}}$ is well above this completeness limit and, therefore, the main conclusions presented hereafter in this study should not be affected by incompleteness. For the galaxies in this dataset, we also use the CNN-based morphological classifications that split them into four classes: spheroids (\textit{Sph}), \textit{Bulge + Disk}, \textit{Disk} and disturbed (\textit{Irr}). See \citealt{Huertas-Company2023}, for more details. 

\item Visual classification catalog (VISUAL): a redshift-selected $z \geq 3$ morphological catalog presented in \citet{Kartaltepe2022} containing 850 galaxies in common between CANDELS \citep{Grogin2011,Koekemoer2011} and CEERS observations. This is intended to directly compare our morphological description to the visual classification of  \citet{Kartaltepe2022}. Redshifts and stellar masses are extracted from CANDELS v2 for the HST F160W-selected galaxies in the EGS field \citep[see][for full details on the photometric redshift measurements and resulting catalogs]{Kodra2023}. The visual classifications presented in \citet{Kartaltepe2022} of each galaxy are performed by three people. A given classification is assigned if two out of three people select that option. Galaxies classified in this way are broken down into the following morphological groups: Disk only, Disk+Spheroid, Disk+Irregular, Disk+Spheroid+Irregular, Spheroid only, Spheroid+Irregular and Irregular only. See \cite{Kartaltepe2022} for more details on the different classification tasks and morphological groups. After selecting those galaxies with $M_* \geq 10^9 \mathrm{M_{\odot}}$, $3 < z < 6$, and reliable visual classifications we end up with a dataset of 545 galaxies.

\end{itemize}

Both catalogs also include morphological measurements of S\`ersic index ($n_e$), semi-major axis ($a$) and axis-ratio ($b/a$) derived with \texttt{galfit} \citep{Peng2010}. More information about the fits can also be found in \cite{Kartaltepe2022}.

The distributions of stellar masses of the galaxies in the CEERS and the VISUAL datasets are shown in \autoref{fig:stellar_masses}. The number of galaxies at the different redshifts analyzed is shown in \autoref{tab:input_data}.

\subsection{Mock JWST images of TNG50 galaxies}
\label{sec:sims}

\begin{figure}[t!]
\centering
	\includegraphics[width=\columnwidth]{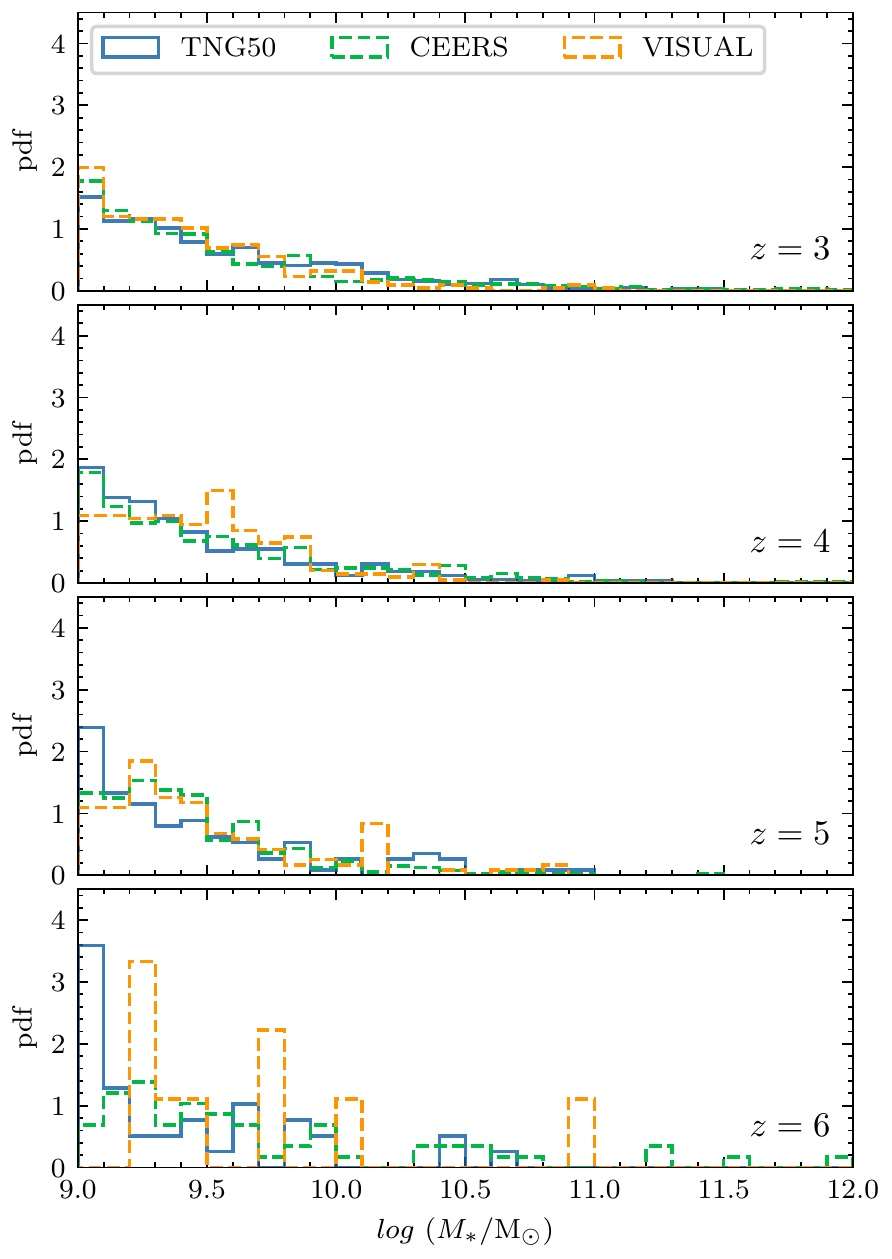}

    \caption{Probability density function of the logarithm of the stellar mass of the simulated galaxies in the TNG50 dataset (blue histogram), and the observed galaxies in the CEERS and VISUAL datasets (green and orange dashed histogram, respectively). Different panels correspond to the redshifts analyzed, $z=(3,4,5,6)$.}
    \label{fig:stellar_masses}
\end{figure}

We use the TNG50-1\footnote{\url{https://www.tng-project.org/}} suite of simulation (hereafter TNG50; \citealt{Nelson2019a,Pillepich2019,Nelson2019b}) and their mock NIRCam observations at $z>3$ galaxies following the observational strategy of CEERS. The mock images\footnote{Data publicly released at \url{https://www.tng-project.org/costantin22}} were produced by modeling the gas cells and star particles in the simulation as presented in \citet{Costantin2022a}. We consider four snapshots of the TNG50 simulation corresponding to $z = (3,4,5,6)$ and galaxies with stellar masses $M_* \geq 10^9 \mathrm{M_{\odot}}$. In total, the original dataset consists of 1\,326 galaxies (see \autoref{tab:input_data}). Each selected galaxy is then observed along 20 different line-of-sight orientations to increase the statistics which produces a dataset of 26\,520 galaxy images that we consider as independent objects for the purpose of this work. As described in \citet{Costantin2022a}, parametric and non-parametric morphological parameters for this dataset are derived using the standard configuration of \texttt{statmorph}\footnote{\texttt{statmorph} is available at \url{https://statmorph.readthedocs.io}.}\citep{Rodriguez-Gomez2019}.  

For this study, we use the noiseless images in the F200W and F356W bands from the \citet{Costantin2022a} dataset with a pixel scale of 0.031 and 0.063 arcsec pix$^{-1}$, respectively. We decided to use these two filters simultaneously since they probe the UV, optical, and near-IR rest-frame at $z>3$ (see Figure 2 in \citealt{2022ApJ...938L...2F}), allowing to probe simultaneously the distribution of young and old stars and offering a complete view of galaxy morphology for our data-driven approach. Another reason for including the F200W is its spatial resolution. The F200W filter is the filter with the longest wavelength at the 0.031 arcsec pix$^{-1}$ resolution of the NIRCam short wavelength channels (F115W and F150W). Although re-sampled into the 0.031 arcsec pix$^{-1}$ resolution after drizzling, the NIRCam long wavelength channels (F277W, F356W, F410W, and F444W) a worse spatial resolution (originally 0.063 arcsec pix$^{-1}$ resolution) than the NIRCam short wavelength channels.

The original field of view of each image is twice the total half-mass radius (i.e., dark matter, gas, and star particles included) of the corresponding galaxy. This roughly corresponds to a field of view 10 times larger than the stellar half-mass radius (i.e., only star particles included). However, our image classification scheme requires a fixed image size. Therefore, we select galaxy images with a field-of-view larger than $64 \times 64$ and $32 \times 32$ pixels in the F200W and F356W bands, respectively, and generate cutouts of those sizes. Then, to match both observations of the same galaxy, images in the F356W band are re-sampled to the same pixel scale as the F200W images (see \autoref{sec:augmentation}, for more details). Given the original field of view of each galaxy image, after cutting them to the input fixed size of our network ($64 \times 64$ pixels), the cutouts will certainly include all the luminous matter in the stamps.

\begin{figure}[t!]
	\includegraphics[width=\columnwidth]{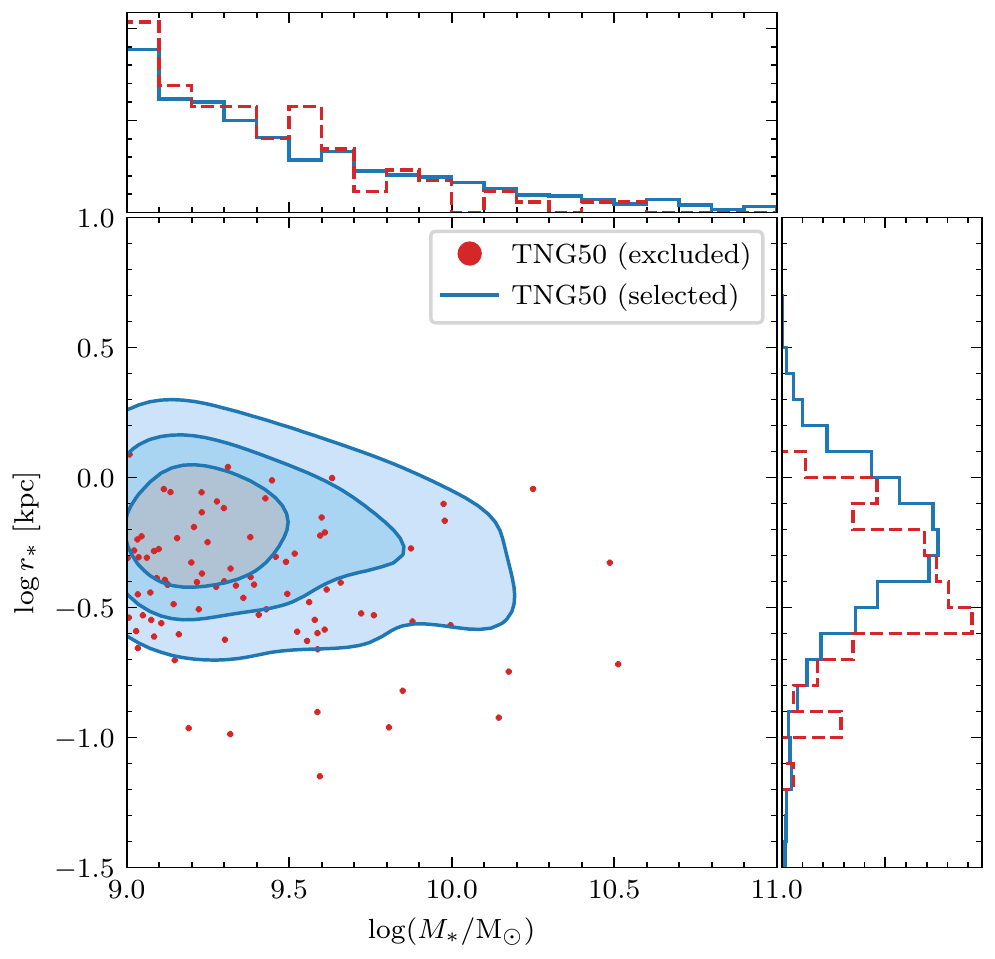}
    \caption{Logarithm of the physical size (stellar half-mass radius, $r_*$, in kpc) versus the stellar mass ($M_*$, in $\mathrm{M_\odot}$) of the TNG50 galaxies. Blue-filled contours show the $25\%$, $50\%$, and $75\%$ probabilities of the TNG50 selected galaxies. Red data points correspond to the excluded galaxies in terms of the size of the field-of-view. See \autoref{tab:input_data}.}
    \label{fig:excluded_sample}
\end{figure}

According to these criteria, $\lesssim 7\%$ of the galaxies (most of them at $z = 3$) are dropped out from our initial sample. The total number of galaxies considered is finally 1\,238 distributed within $z = 3-6$ (see \autoref{tab:input_data}), which translates into 24\,760 projections. Although the number of objects we remove is small, we check in \autoref{fig:excluded_sample} if a specific population is systematically excluded. The figure shows the size-mass relation of the selected TNG50 dataset along with the excluded galaxies based on the size of the field of view. The excluded galaxies are not necessarily the most compact and/or less massive galaxies in the dataset. However, a fraction of them with lower-than-average stellar extent is indeed removed based on our selection and would reach otherwise sizes of a few hundred parsecs.

For comparison, we show in \autoref{fig:stellar_masses} the distribution of the stellar masses in the simulated TNG50 dataset, and the observed CEERS and VISUAL datasets. Note the good agreement between the TNG50 and the CEERS samples, even if we are comparing here the stellar masses directly extracted from the TNG50 simulations and those obtained through SED fitting of the latest JWST data. Also remarkable is the agreement, despite the selection effects, between the CEERS, VISUAL, and TNG50 datasets.
\begin{table}
    {\centering
    \caption{Summary of the sample of TNG50 simulated galaxies and CEERS observed galaxies with $M_* \ge 10^9 \mathrm{M_\odot}$. The first column indicates the redshift ($z$); the second column shows the total number of galaxies in the simulated TNG50 dataset; the third column refers to the number of selected galaxies according to image size limitations (i.e., $64 \times 64$ and $32 \times 32$ pixels in the F200W and F356W bands, respectively); the fourth column shows the number of galaxies in the CEERS dataset; the fifth column indicates the number of galaxies in the VISUAL dataset. Note that for the CEERS and VISUAL datasets galaxies are split into the following redshift bins: $z = 3$ for $3.0 \le z < 3.5$, $z = 4$ for $3.5 \le z < 4.5$, $z = 5$ for $4.5 \le z = 5.5$ and $z = 6$ for $5.5 \le z < 6.0$.}
    
    \label{tab:input_data}
    \begin{tabular}{ccccc} % number of columns, alignment for each
	\hline
	$z$ & TNG50 & TNG50 & CEERS & VISUAL\\
        & (All) & (Selected) &&\\
	\hline
	3-6 & 1326 & 1238 & 1664 & 545 \\
	3 & 829 & 760 & 741 & 216\\
	4 & 343 & 326 & 463 & 201\\
	5 & 115 & 113 & 398 & 119\\
	6 & 39 & 39 & 62 & 9\\
	\hline
    \end{tabular}}
\end{table}

%%%%%%%%%%%%%%%%%%%%%%%%%%%%%%%%%%%%%%%%%%%%%%%%%%
\section{Self-supervised learning representation of mock JWST images of simulated TNG50 galaxies}
\label{sec:self-supervised}

In this section, we describe the main methodology we developed to obtain a data-driven morphological description of galaxies that is robust to noise and other nuisance parameters. 

\subsection{Contrastive learning framework}
\label{sec:contrastive}

Our approach is based on an adaptation of the Simple framework for Contrastive Learning of visual Representations \citep[SimCLR;][]{Chen2020a}. Very briefly, the idea behind the SimCLR framework is to obtain robust representations of images without labels by applying random augmentations as explained below. See \citet{Huertas-Company2023b}, for a recent review of this technique applied to Astrophysics.

Given an image, random transformations are applied to it to generate a pair of two augmented images, $(x_i, x_j)$. Each image in the pair is passed through a Convolutional Neural Network (CNN) to compress the images into a set of vectors, $(h_i, h_j)$. Then, a non-linear fully connected layer (i.e., projection head) is placed to get the representations $(z_i, zj)$. The representations are learned iteratively by maximizing agreement between the augmented views of the same image example $(z_i, zj)$ and minimizing agreement between all other pairs considered as negatives. This is achieved via a so-called contrastive loss in the latent space:

\begin{equation}
    l_{i,j} = -\log \frac{\exp(\langle \boldsymbol{z}_i, \boldsymbol{z}_j\rangle/h)} {\Sigma_{k=1, k \neq i}^{2N}\,\,\exp(\langle \boldsymbol{z}_i, \boldsymbol{z}_k\rangle/h)},
    \label{eq:contrastive_loss}
\end{equation}
where $\langle \boldsymbol{u},\boldsymbol{v} \rangle$ denotes the dot product between $L^2$-normalized $\boldsymbol{u}$ and $\boldsymbol{v}$, and $h$ denotes the temperature parameter that regulates the distribution of the output representations \citep[see][for more details]{Hinton2015,Wu2018}. The final loss is computed in batches of size $N$ across all positive pairs, both $(i, j)$ and $(j, i)$, while the rest of the augmented examples are treated as negative examples, which are denoted by $k$. 

For this study, we follow the implementation from \cite{Sarmiento2021}, which was successfully applied to astronomical data.  The CNN encoder consists of four convolutional layers with kernel sizes 5, 3, 3, 3, and 128, 256, 512, and 1\,024 filters per layer, respectively. Max-pooling layers and Exponential Linear Unit (ELU) activation functions are placed after each convolutional layer. Therefore, the representations before the projection head ---$(h_i, h_j)$--- for each galaxy image are encoded into 1\,024 features. Subsequently, the projection head (composed of three fully connected layers of 512, 128, and 64 neurons per layer) transforms the galaxy representations to a latent space ---$(z_i, z_j)$--- where the contrastive loss is computed. 

\subsection{Data augmentation and network training}
\label{sec:augmentation}

The choice of data augmentations is a key element in contrastive learning training \citep{Chen2020a} as it allows us to turn the representations independent of some nuisance effects. In the context of this study, our goal is to obtain a morphological representation that is robust to signal-to-noise, rotation, and size, and does not depend on color. To reach this objective, we calibrate our algorithm on the mock TNG50 dataset (\autoref{sec:sims}) since it allows us to access noiseless versions of the images and, therefore, marginalize the noise.

For each simulated TNG50 galaxy image, we produce two augmented images ($x_i, x_j$) from the noiseless version. One of the images from the pair ---the one used to produce the noise-added image--- is then convolved by the corresponding PSF in each filter (extracted from the observations \citealt{Finkelstein2022} in each band), while the other image ---the one used to produce the noiseless image--- is not convolved by the PSF to keep as much as spatial information as possible. Both images are re-scaled in flux (as described below) independently. Then, we add source Poisson noise only to the image used for the noise-added version. To match the same pixel scale in the two filters, we then re-bin the images in the F356W filter to the same pixel scale as the images in the F200W filter. Finally, for the noise-added image, we include realistic noise by adding random patches of the sky extracted from the 10 CEERS pointings. Below we described more precisely the augmentations applied to the images:

\begin{itemize}

    \item \textit{Rotation}: we first apply a random flipping (horizontal or vertical, but not both) and a random rotation with 100\% chance independently to the TNG50 noiseless image and the patch of the sky extracted from the CEERS pointings. Also the noiseless and the noise-added version are rotated and flipped differently. This augmentation is intended to ensure the model is invariant to the galaxy orientation;

    \item \textit{Flux}: we randomly apply flux scaling to the noise-added version of the images after the PSF convolution, but before noise is added. The flux factor applied to the noise-added images in the two filters is randomly sampled from the flux distribution of the TNG50 dataset in the F356W band. The same flux factor is applied to both the F200W and F356W filters. This augmentation is intended to stress the robustness to S/N. It may also help ---as we will show in the following--- to make the representations independent of galaxy size since the regions above the noise level will vary with the flux variations. %The TNG50 F346W flux is computed as the sum of the fluxes in each pixel of the $64 \times 64$ noiseless images. Note that this flux distribution has not been derived (as typically) within a given aperture, but it serves as an initial distribution to randomly vary the flux values of each galaxy according to the overall flux distribution of the TNG50 dataset. 

    \item \textit{Noise}: as described in \autoref{sec:data}, for each galaxy image in the F200W and F356W bands we have a noiseless version that does not include any instrumental effects or noise. Using the available CEERS data, we construct mock CEERS galaxy images as a combination of the TNG50 noiseless images and random patches of the ten observed CEERS pointings. For the noise-added version, we first add source Poisson noise and, then, we add real-time realistic noise (that may also include other sources/interlopers) to each of the $64 \times 64$ noiseless galaxy images by summing up one randomly chosen patch (different in each augmentation) from the CEERS pointing of the same size. From the contrastive learning point of view, these images with a real background are considered as an augmented copy of their noiseless analogs during the training process. These augmentations should enforce the representations to be robust to signal-to-noise (S/N) as well as to background and foreground companions.
    
    \item \textit{Color}: finally, in order to prevent the network from learning color information and/or the intrinsic brightness of the galaxies directly from the images, we apply two additional augmentations to both the noiseless and the noise-added images. First, each band is normalized individually after the augmentations are applied. Second, the noiseless and noise-added images are normalized independently and individually for each galaxy. Consequently, the maximum pixel value in every galaxy image (both noiseless and noise-added) is equal to one for each band. 
    
\end{itemize}

We note that we decided not to apply direct size augmentation, (i.e. such as zoom-in or zoom-out), as that would force us to up-sample or down-sample the images with less or more than $64 \times 64$ pixels size, respectively, which creates some artifacts that the network is able to learn. The choice of the cutouts' size when producing the galaxy images is always complicated. One might prefer to make the stamps proportional to the size of the galaxy (see \citealt{Vega-Ferrero2021}, for an example). However, that requires reliable measurements of the galaxy sizes and also the re-sampling of the cutouts to the same size in pixels. Contrarily, it is possible to produce all the cutouts with the same size in pixels, independently of the galaxy size. By doing so, it is not needed to re-sample the cutouts since they already have the same dimensions. We decide to use a fixed size for the cutout to avoid: first, artifacts and noise correlations originating from the re-sampling phase that could mislead the contrastive learning representations; and second, losing spatial resolution of galaxies with large sizes after the re-sampling phase.

\begin{figure*}[t!]
\centering
	\includegraphics[width=2\columnwidth]{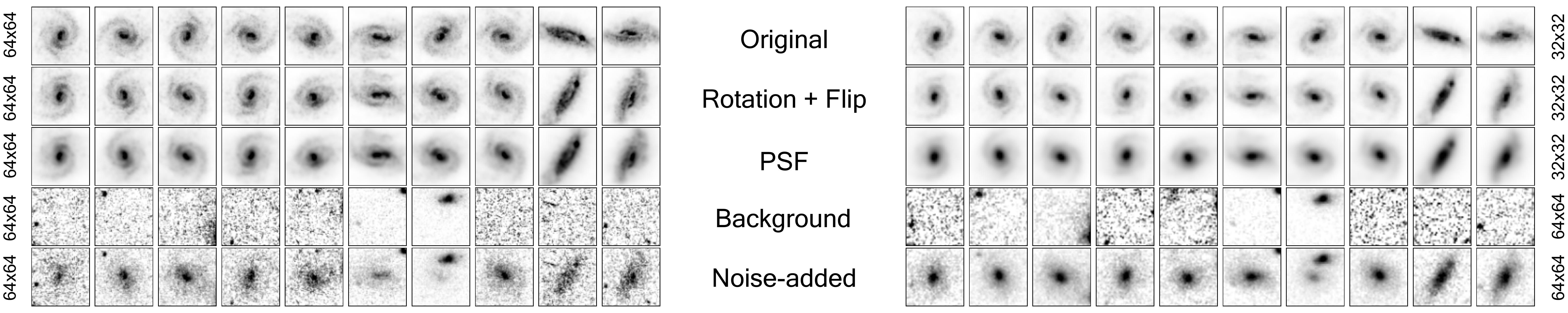}

    \caption{Images and augmentations of a TNG50 galaxy at $z = 3$ with $M_* \approx 6 \times 10^9 \mathrm{M_{\odot}}$ in the F200W (left-hand panel) and F356W (right-hand panel) filters. The different rows (from top to bottom) correspond to: the original/noiseless image; the original image after rotation and flipping (vertical or horizontal); the image after PSF convolution; the background CEERS patch of the sky; the noise-added image, i.e., the sum of the PSF convolved image (flux variation and source Poisson noise applied) and the background patch of the sky. All the panels have a $64 \times 64$ pixels size, except the first three rows for the F356W filter that have the original size of $32 \times 32$ pixels before re-binning to $64 \times 64$ pixels. The pixel values have been $asinh$ transformed with a 0.5\% clipping.}
    \label{fig:z3_ID3}
\end{figure*}

In \autoref{fig:z3_ID3}, we show several projections of a galaxy at $z = 3$ with $M_* \approx 6 \times 10^9 \mathrm{M_{\odot}}$ extracted from the TNG50 simulation. In some of the augmented versions (along with PSF convolution, realistic noise, random rotations, flux variation, etc.), it is possible to distinguish several companions within the stamps. This is the result of adding randomly chosen patches of the CEERS pointings to the TNG50 noiseless images. These examples come from an extended bright galaxy that appears significantly dimmer in some of the augmented images due to the flux variations applied in the augmentations. In summary, the contrastive model should be able to extract a meaningful representation $(h_i, h_j)$ that minimizes the distance between the representations $(z_i, z_j)$ of the two images (original/noiseless and noise-added) of the same galaxy, even if they appear as different as shown in some panels in \autoref{fig:z3_ID3}.

Our contrastive SimCLR model is trained and tested with the mock JWST images for 24\,760 different projections of 1\,238 galaxies within $3 \leq z \leq 6$ and with stellar masses $M_* \geq 10^9 \mathrm{M_{\odot}}$ in the two observed bands (F200W and F356W) with a temperature parameter $h=0.5$ (that controls the strength of penalties on hard negative pairs). We randomly split our dataset into a training and a test sample consisting of 1\,100, and 138 galaxies, respectively. This translates into a training and a test dataset of 22\,000 and 2\,760 galaxy images, respectively. None of the projections of the galaxies in the test set has ever passed through the network during training. Additionally, we ensure that only one projection of the same galaxy enters each batch during training and that all the galaxy images are passed through the network at every epoch. We do so to avoid the algorithm learning the orientation of the same galaxy as seen from different line-of-sight projections since some of the projections are just a simple rotation of the galaxy in the sky.

The input tensors in our contrastive model have, therefore, a dimension of $(N,64,64,2)$, with N being the batch size, 64 and 64 being the dimensions of the input images, and 2 the number of channels or filters (i.e., the F200W and the F356W images). We train our algorithm with a batch size of $N=550$ (i.e., half the number of galaxies in the training set) for 1\,500 epochs in a GPU NVIDIA T4 Tensor Core with 16 GB of RAM. Random data augmentation is applied every 50 epochs and, therefore, we produce 30 ($1500/50$) different augmentations of the whole dataset to increase the variability during the training process. To reduce the dynamic range and to be sensitive to both the center and outskirts of the galaxy, before training, we apply a $asinh$ (inverse hyperbolic \textit{sin}) transformation and a minimum-maximum normalization to each galaxy pair in each band. 

In order to reduce the impact of possible discrepancies between the galaxies in the simulated TNG50 and observed CEERS datasets, when applying our model to data, we fine-tuned the model trained previously with a mixed dataset of simulated TNG50 and observed CEERS galaxies. By doing so, we expect the model to learn important features from the observed CEERS galaxy images that were not present in the training set consisting only of simulated TNG50 galaxy images and, therefore, mitigate possible domain drift-related effects. We fine-tune the model with a training set consisting of 1\,500 images randomly extracted from the previous training set (of noiseless and noise-added images) and 1\,500 images randomly selected from the CEERS dataset of 1\,664 galaxy images. Note that for the observed dataset we do not have noiseless images so, we fed the model with pairs of CEERS galaxy images to which different augmentations (only flip and rotation) are applied. We fine-tune the model up to 600 epochs (enough to converge) with a batch size of $N=500$ galaxy images. In each epoch, random augmentations are applied to the observed CEERS pairs of images to increase the variability. For the simulated TNG50 images, the set of 1\,500 also varies from epoch to epoch by selecting different galaxies and augmentations for each galaxy in the training TNG50 dataset, but not from the reserved test set which is always kept apart from the training.

%%%%%%%%%%%%%%%%%%%%%%%%%%%%%%%%%%%%%%%%%%%%%%%%%%

\subsection{Visualization of the representation space}

\begin{figure*}[t!]
\centering
	\includegraphics[width=\columnwidth]{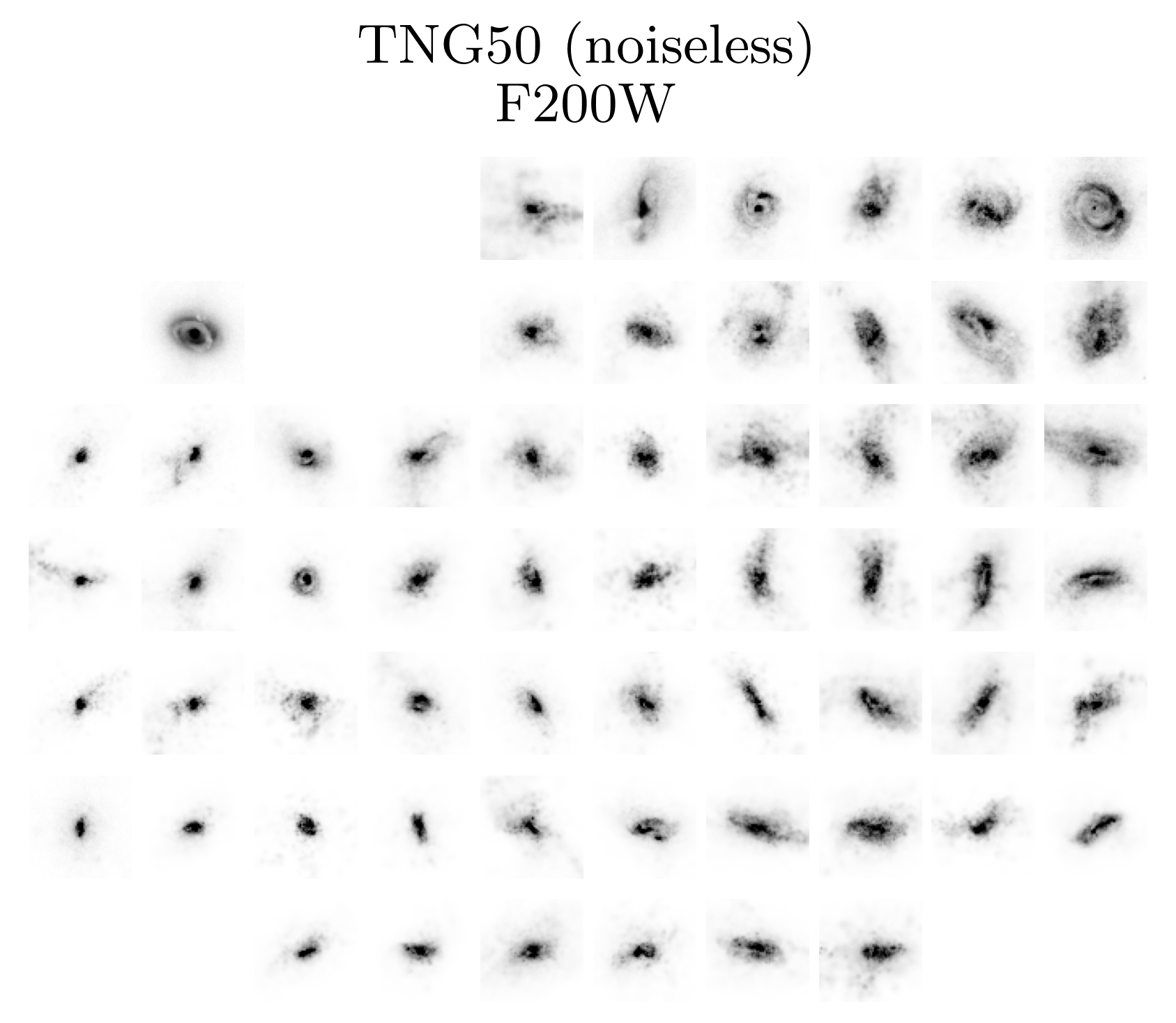}
	\includegraphics[width=\columnwidth]{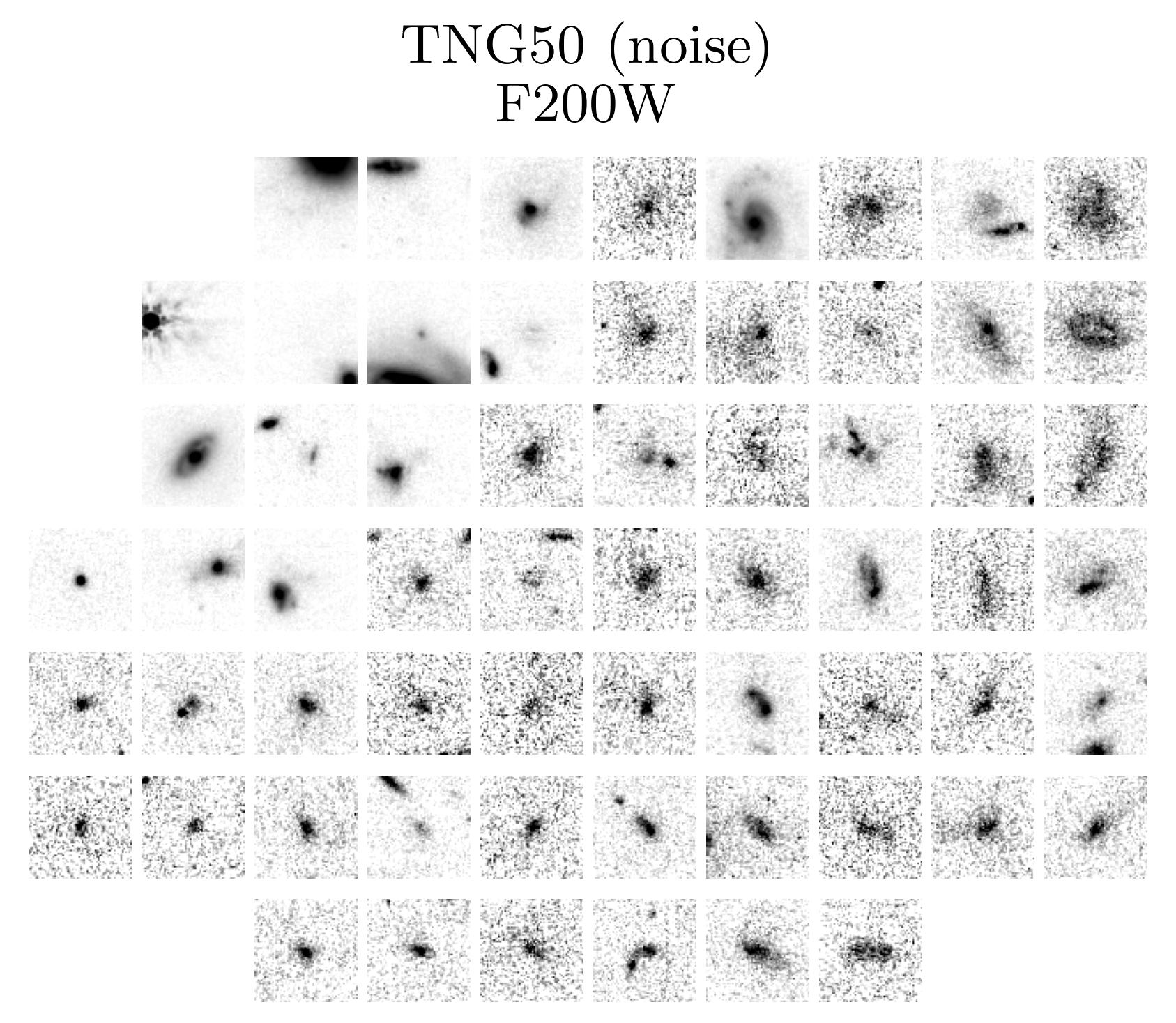}
	
    \caption{Randomly chosen images of the simulated TNG50 galaxies in the UMAP visualization. The UMAP space is binned and one galaxy image per bin is shown. The left-hand panel shows the noiseless versions of the training galaxies, while the reduced versions (TNG50 + random CEERS patch) are shown in the right-hand panel. Extended/disky galaxies lie in the upper-right, compact/rounder galaxies are located in the bottom-left, and bright companions (only in the right-hand panel) concentrate toward the upper-left section of the UMAP plane. Note that there is not a one-to-one correspondence between the galaxies shown in the two panels. Both panels correspond to galaxy images in the F200W filter.}
    \label{fig:umap_f356}
\end{figure*}

We can hence analyze the properties of the representation space learned by the SimCLR framework presented in the previous section. 

We start by visualizing how the TNG50 galaxy images, both with and without noise, are distributed in the representation space. It should be noted that, hereafter, the difference between the noiseless and the noise-added version of the TNG50 galaxy images is only the added patch of the CEERS paintings. Apart from the realistic noise added, no other augmentations are applied in this phase (only in the training phase described in the previous section). Since the representation space for each galaxy image is encoded into 1\,024 features, we perform a dimensionality reduction from the 1\,024 features to a 2D space to facilitate the visualization and interpretation of this representation. For that purpose, we use the Uniform Manifold Approximation and Projection \citep[UMAP;][]{McInnes2018} method with standard initial parameters (\textit{metric = euclidean}, \textit{n\_neighbors} $=15$ and \textit{min\_dist} $=0.1$). The UMAP algorithm seeks to learn the manifold structure of the input data and find a low-dimensional embedding that preserves the essential topological structure of that manifold. It is therefore a way to visualize in 2D the representations learned by the self-supervised network. Before applying the UMAP technique, we assume the same distance metric in the representation space as the one used to calculate the contrastive loss in the head projection space and, therefore, we normalize the representations with an $L^2$-norm such that the Euclidean and cosine distances between representations are equivalent. The two coordinate axes in the UMAP representation do not have any precise physical meaning and are a combination of the 1\,024 dimensions extracted by the contrastive learning setting. 

It is important to emphasize that the contrastive approach is not intended for dimensionality reduction but for obtaining a robust representation of galaxy morphologies in a different space than images, which explains the high dimensionality of the representation space. Several works have shown indeed that the performance of contrastive learning increases with representations of higher dimension (e.g.,\citealp{Chen2020a}).

In \autoref{fig:umap_f356}, we show random examples of galaxies in the F200W filter (both with and without noise) in the UMAP 2D space. Note that some stamps on the right-hand panel (along with the addition of observed noise) show one (or more) foreground/background sources in the field of view. The figure clearly shows that galaxies are not randomly organized in the plane, indicating that the network has learned some relevant morphological features. The distributions are also similar for galaxies with and without noise. Galaxies with extended light distributions and with clear signs of a disk component ---or interactions--- tend indeed to appear on the right and upper-right parts of the UMAP space, while more compact galaxies with smoother and concentrated light distributions tend to be placed on the left section of the plane. We can also see that galaxies showing more elongated shapes are found toward the bottom-right section of the representation space. Also interesting to notice is how several galaxies with bright companions (off-center sources) tend to be placed on the upper-left of the right-half panel (see \autoref{sec:noise} below for a more detailed discussion on this point).

\subsection{A morphological description of galaxies robust to noise and background/foreground contaminants}
\label{sec:noise}

\begin{figure*}[t!]
\centering
        \includegraphics[width=\columnwidth]{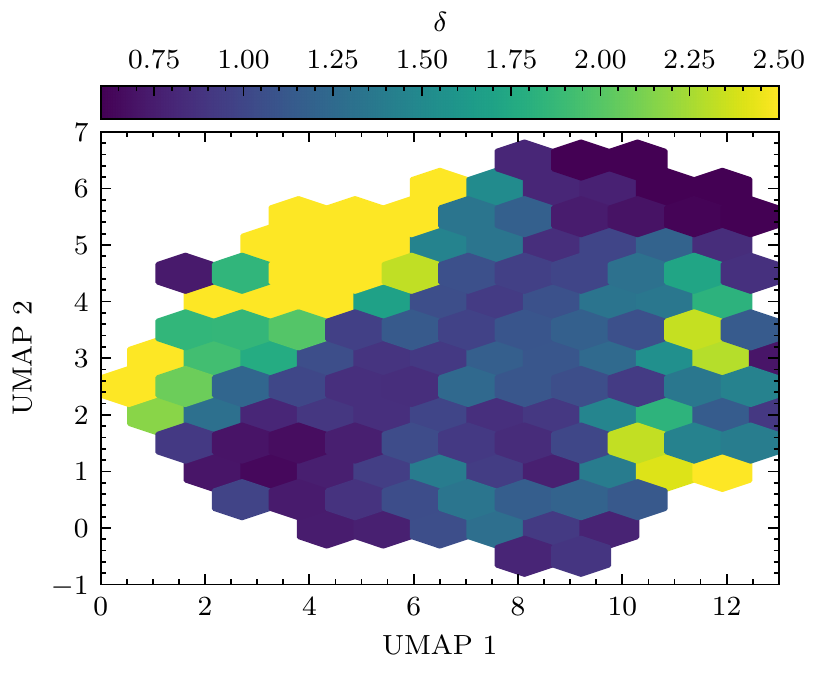}
	\includegraphics[width=\columnwidth]{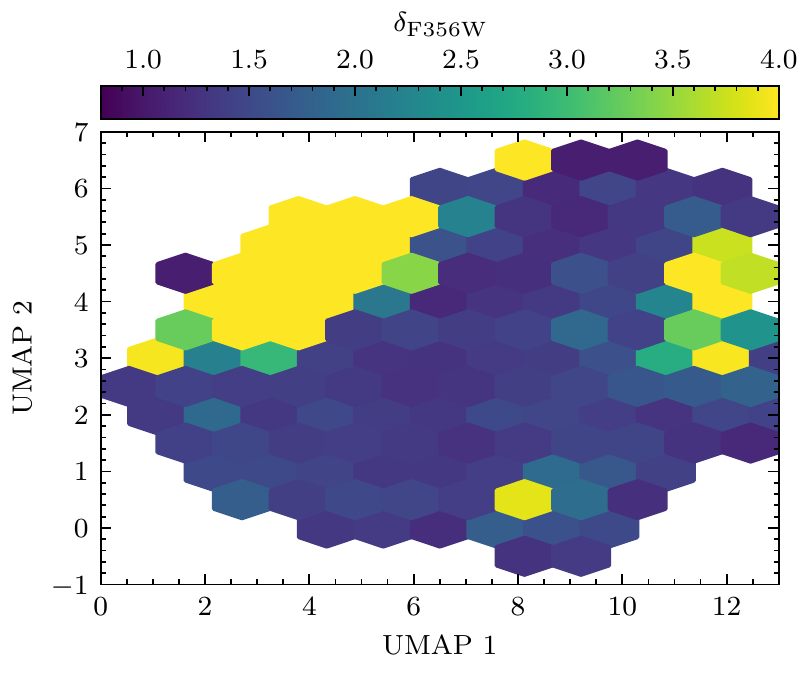}
    \caption{Left-hand panel: UMAP visualization for all the TNG50 galaxy images in our dataset color-coded by the mean value of the distance in the UMAP representation between the noiseless and the reduced images of each galaxy (denoted as $\delta$). Right-hand panel: UMAP visualization for all the TNG50 galaxy images in our dataset color-coded by the mean value of the ratio of the flux measured in the TNG50 stamps and the flux derived from the noiseless TNG50 stamps for the F356W filter (denoted as $\delta_\mathrm{F356W}$). Large values of $\delta_\mathrm{F356W}$ indicate the presence of a secondary (or even more) source. The larger $\delta_\mathrm{F356W}$ is, the brighter the companions are with respect to the central galaxy.}
    \label{fig:umap_CEERS_BACK356}
\end{figure*}

We now examine in more detail the differences between the representations of noise-added and noiseless TNG50 galaxy images, and how the different augmentations of the same galaxy are represented by our contrastive model. As described in~\autoref{sec:contrastive}, one of the reasons for using the SimCLR framework is to obtain a data-driven representation that is robust to noise and other observational effects such as foreground and background companions. 

On one hand, we quantify the effect of noise by computing the distance in the UMAP representation between the noiseless and the noise-added images of each galaxy, denoted as $\delta$. On the other hand, we check how our contrastive model behaves when more than one source (i.e., companions) is present in the stamp. To do so we measure the total flux in the noise-added galaxy images (TNG50 + random CEERS patch) and in the noiseless TNG50 stamps (therefore, the intrinsic flux of the central galaxy after the flux scaling is applied) for the F356W filter. Then we compute the ratio of the two, denoted as $\delta_\mathrm{F356W}$, as a proxy for the presence (or not) of companions and, if present, how bright they are with respect to the central galaxy.

In \autoref{fig:umap_CEERS_BACK356}, we show the UMAP plane for TNG50 galaxy images in our dataset color-coded by $\delta$ and $\delta_\mathrm{F356W}$. For a reference of the UMAP axis ranges, the horizontal axis (UMAP 1) spans within $(0.9,12.1)$, and the vertical axis (UMAP 2) spans within $(-1.7,5.5)$. The total area covered by the data points in the UMAP plane is approximately 60 (in the arbitrary UMAP units). It is interesting to note how the main yellow clump in the upper-left section of the UMAP plane where the stamps with bright companions (more than three times the flux than the flux of the central galaxy, i.e., $\delta_\mathrm{F356W} \gtrsim 4$) tend to concentrate, and also their correlation with large values $\delta \gtrsim 2.5$. Some of these cases can be seen in the right-hand panel in \autoref{fig:umap_f356}. For instance, there are several examples within these regions of $\delta_\mathrm{F356W} \gtrsim 4$ that correspond to TNG50 images for which the companion is so bright (such as a star) that the central galaxy cannot be even identified in the stamp. In these cases with bright companions around the central galaxy, the model detects the brightest component (thus, the companion) instead of the central galaxy and tends to represent it in a particular region of the UMAP plane. Additionally, in the right-hand panel, it is also clear a yellow clump with large values of $\delta_\mathrm{F356W}$ and moderate values of $\delta$. These cases correspond to nose-added images with a bright companion that, contrarily, are still close to their noiseless counterpart (i.e., $\delta \lesssim 2$). We inspect in detail several of these cases and find that their noiseless counterparts tend to be located on the bottom-right section of the UMAP plane. In any case, a galaxy image represented close to these regions of $\delta \gtrsim 2$ might be not well-represented and should be (at least) treated carefully or excluded from the subsequent analysis. It is also interesting to note the clump with large values of $\delta$ in the left edge of the representation space (i.e., $0.0 < \mathrm{UMAP~1} < 1.5$ and $2.0 < \mathrm{UMAP~2} < 3.5$) with no clear correlation with large values of $\delta_\mathrm{F356W}$. Although this clump does not include a large number of galaxies, we find that those objects have their noiseless counterparts displaced up in the UMAP plane. These are very compact galaxies that look even more compact in their noiseless versions.

\begin{figure}[t!]
\centering
	\includegraphics[width=\columnwidth]{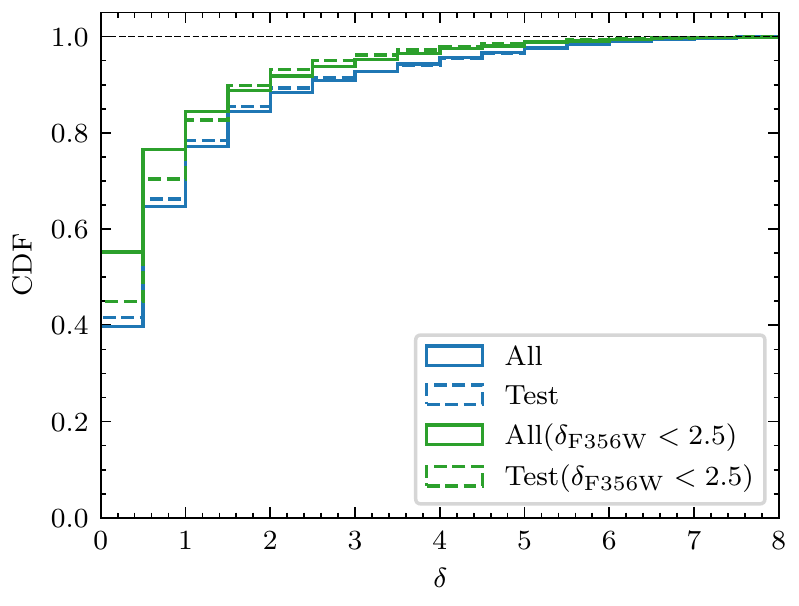}
    \caption{Cumulative distribution function of the distance in the UMAP plane between pairs of the same galaxy images, denoted as $\delta$. Blue histograms correspond to the distribution of $\delta$ for the training (solid) and the test (dashed) datasets. Green histograms correspond to the distribution of $\delta$ for the training (solid) and the test (dashed) datasets when only galaxy images with $\delta_\mathrm{F356W} < 2.5$ are considered.}
    \label{fig:delta_umap}
\end{figure}

By excluding objects falling within the yellow clump shown in  \autoref{fig:umap_CEERS_BACK356} it is possible to clean our dataset (or a JWST/CEERS sample of observed galaxies) of bright companions/contaminants that could bias our contrastive learning representations. As an alternative, we build up a dataset of simulated TGN50 galaxy images that include all the possible augmentations described before but keeping a value of $\delta_\mathrm{F356W} < 2.5$. By doing so we ensure our dataset does not include companions $1.5$ brighter (in flux) than the central galaxy. Nevertheless, these extremely bright contaminants are much rare in real CEERS observations than in our augmented set of galaxy images because, if the contaminant is so bright that it outshines the central galaxy, the galaxy would not be detected.

In \autoref{fig:delta_umap}, we show the cumulative distribution function of $\delta$ for the galaxy images in the training and the test sets, both including and excluding galaxy images with $\delta_\mathrm{F356W} > 2.5$. We find that $75\%$ and $90\%$ of the projections in the training dataset show values of $\delta \lesssim 1.5$ and $\delta \lesssim 3.0$ in the UMAP space, respectively. For the test set, $75\%$ and $90\%$ of the projections show values of $\delta \lesssim 1.4$ and $\delta \lesssim 2.7$ in the UMAP space, respectively. If we now exclude galaxy images with $\delta_\mathrm{F356W} > 2.5$: for the training dataset, $75\%$ and $90\%$ of the projections show values of $\delta \lesssim 1.2$ and $\delta \lesssim 2.4$ in the UMAP space, respectively; while for the test dataset, $75\%$ and $90\%$ of the projections show values of $\delta \lesssim 1.3$ and $\delta \lesssim 2.3$ in the UMAP space, respectively. A displacement of $\delta$ is analogous to saying that the noise and noiseless representations of the same galaxy pair are located within a circle of radius $\delta$. Converted into an area, this means $\approx 8\%$ displacement in the UMAP plane for $75\%$ of the galaxy images (i.e., $\delta \lesssim 1.2$) in the training set despite the level of noise, contamination, and augmentations applied to the input galaxy images, as can be seen in \autoref{fig:z3_ID3} and \autoref{fig:umap_f356}. It should be noted that excluding cases with $\delta_\mathrm{F356W} > 2.5$ does not remove all the cases with companions since there are still cases of companions with fluxes up to $1.5$ times the flux of the central galaxy. Even when these cases are included the model performs satisfactorily well for a large fraction of the galaxy images presented in this study. Besides, the distributions of $\delta$ for the training and test samples are very similar. Therefore, we emphasize the model is not suffering from over-fitting since none of the galaxy images in the test set has been shown previously to the network. 

\begin{figure}[t!]
\centering
	\includegraphics[width=0.47\textwidth]{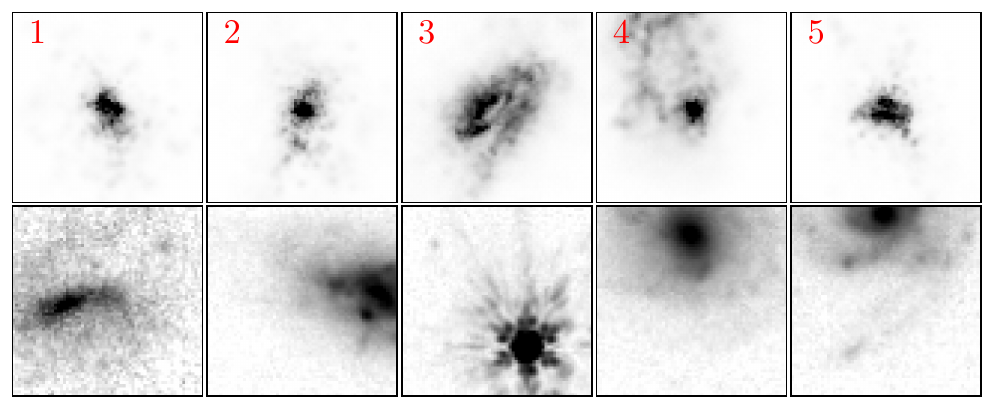}
	\includegraphics[width=0.47\textwidth]{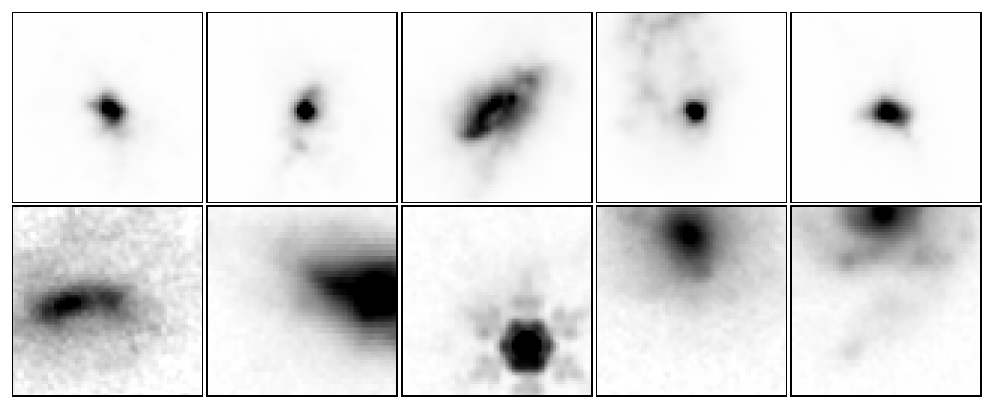}
    \caption{Randomly chosen examples of galaxy images with $\delta > 5$ and $\delta_\mathrm{F356W} > 10$ (see \autoref{fig:umap_CEERS_BACK356}). For each of the examples, we show noiseless and noise-added images in the F200W (top rows) and F356W (bottom rows) filters.}
    \label{fig:delta_umap_examples}
\end{figure}

To further illustrate the effect of companions and noise on the representation space, we show some examples of the most extreme cases ($\delta > 5$ and $\delta_\mathrm{F356W} > 10$) in \autoref{fig:delta_umap_examples}. The majority of images with the largest values of $\delta$ are mainly due to the presence of bright companions in the galaxy images (or artifacts). It is possible to identify compact bright companions (such as a star in case 3 and a compact galaxy in case 1), and more extended companions (extended galaxies in cases 2,4, and 5).

Therefore, training our contrastive model with a combination of noiseless and noise-added TNG50 images leads to a robust representation of TNG50 images even in the case of the presence of companions in the image (at least, for those cases in which the companion is not extremely bright compared to the central galaxy). For the cases in which the companion is much brighter than the central galaxy, their locations in the UMAP may certainly help to find them in observed images and to treat them carefully in subsequent analysis. %In fact, when we discuss clustering of the representation space in \autoref{sec:clustering}, it will become apparent how these cases may be easily identified. 

Hereafter, we show results for the representation of this `clean' dataset (i.e., only galaxies with $\delta_\mathrm{F356W} < 2.5$ are considered) for which the representations obtained are not affected by extremely bright contaminants in the galaxy images. 

\subsection{Dependence on physical and photometric parameters}
\label{sec:dependence}

An advantage of calibrating the neural network model with simulations is that we have access to a large number of physical properties of the galaxies. An additional test for our classification scheme is, therefore, to examine how the representation space is correlated with physical quantities as well as with other (more standard) morphological measurements. 

\subsubsection{Correlation with physical properties}

\begin{figure*}[t!]
\centering
	\includegraphics[width=\columnwidth]{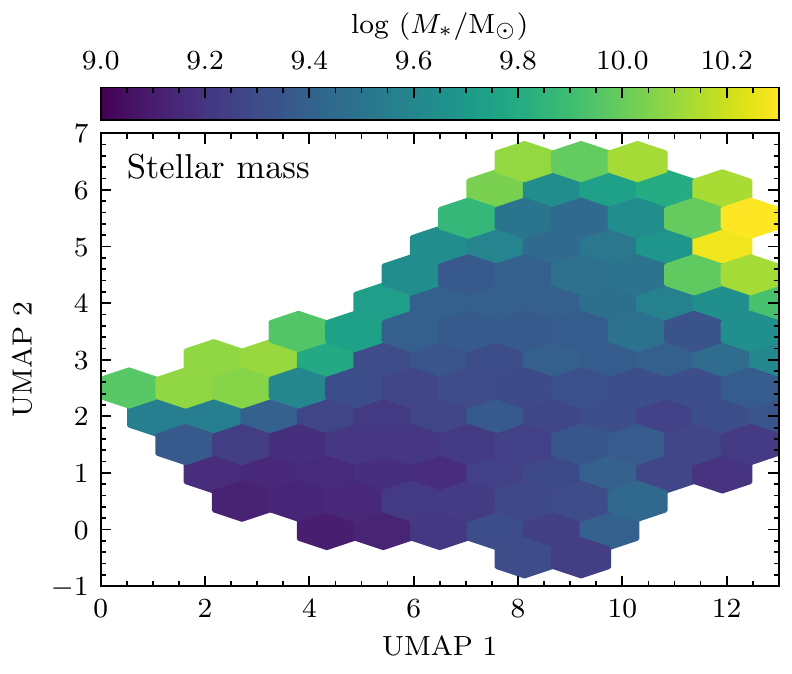}
	\includegraphics[width=\columnwidth]{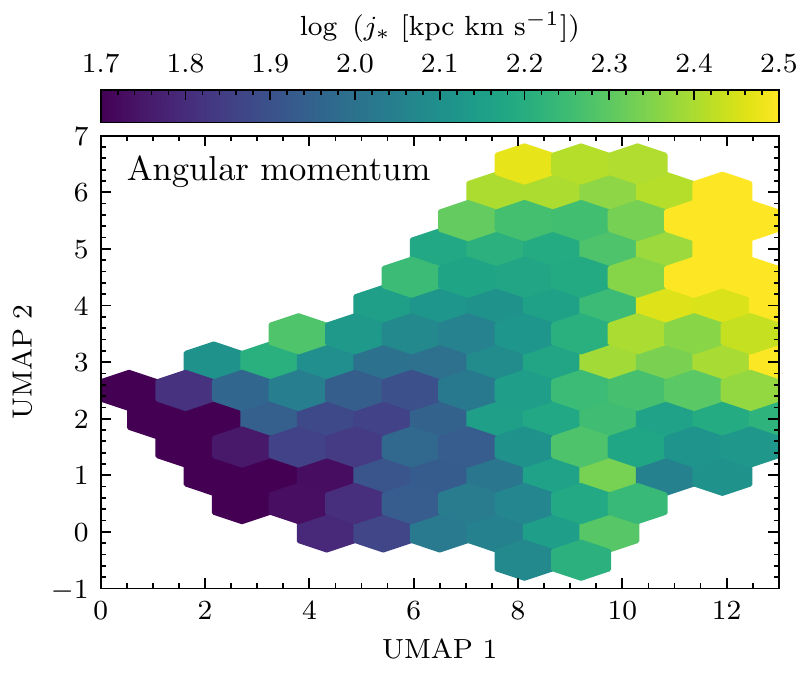}
        \includegraphics[width=\columnwidth]{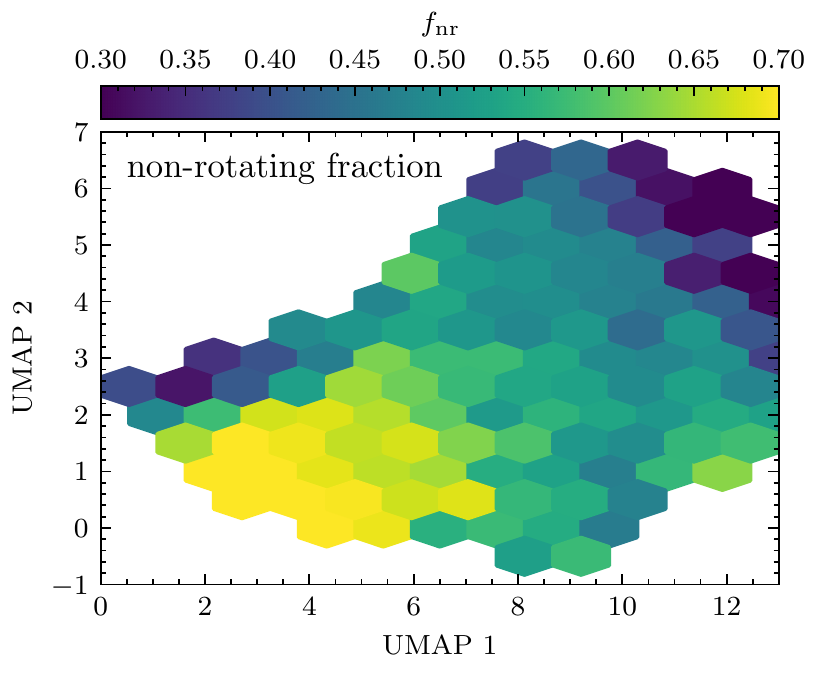}
	\includegraphics[width=\columnwidth]{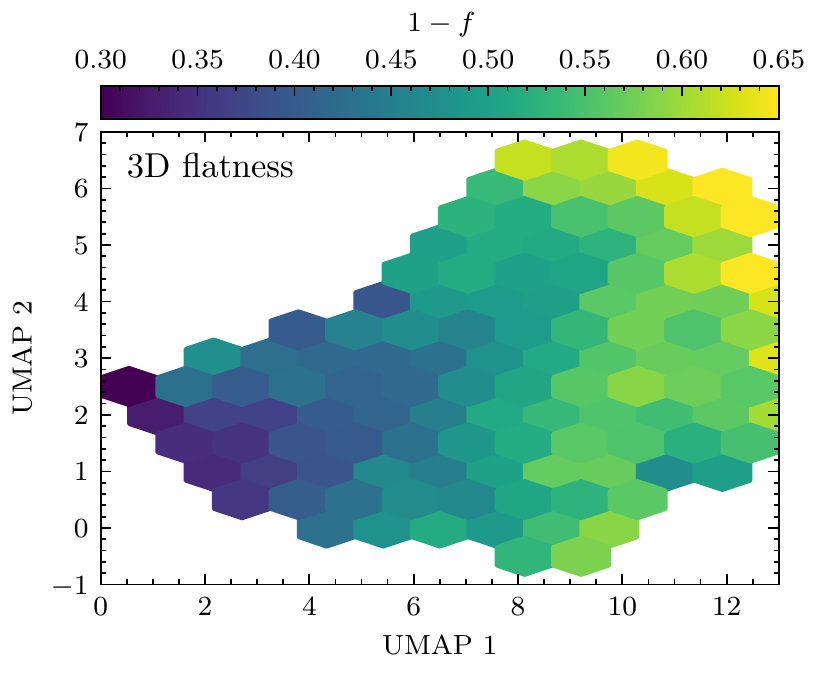}
	
    \caption{UMAP visualization for all the TNG50 galaxy images in our dataset color-coded by the distribution of several physical properties extracted from the TNG50 simulation. Color code corresponds to the median values in each hexagonal bin in the UMAP plane. From left to right and top to bottom, the different panels show: the logarithm of the total stellar mass ($\log M_* \mathrm{[M_\odot]}$), the logarithm of the specific angular momentum of the stars ($\log j_* \mathrm{[kpc~km~s^{-1}]}$), the mass fraction in non-rotating stars ($f_\mathrm{nr}$), and the galaxy flatness ($1-f$). The scatter maps of these parameters are presented in \autoref{app:dependence}.} 
    \label{fig:sims_hexbin_umap}
\end{figure*}

In this section, we discuss how some physical properties extracted from the TNG50 simulation correlate with the representation in the UMAP plane. In \autoref{fig:sims_hexbin_umap}, we show the dependence in the UMAP plane with the total stellar mass ($M_* \mathrm{[M_\odot]}$), the specific angular momentum of stars ($j_* \mathrm{[kpc~km~s^{-1}]}$), the mass fraction in non-rotating stars ($f_\mathrm{nr}$) and the flatness ($1-f$) of the galaxy. The mass fraction in stars that have no net angular momentum around the $z$-axis is defined using the circularity parameter $\epsilon=J_z/J(E)$, as in \cite{2014MNRAS.437.1750M}, for every star particle. It measures the maximum specific angular momentum possible at the specific binding energy E of the star. The mass fraction in non-rotating stars mass (denoted as $f_\mathrm{nr}$) is then defined as the fractional mass of stars with $\epsilon < 0$ multiplied by two. The flatness of the galaxy is computed as follows: $f = c / \sqrt{ba}$, where $c < b < a$ denote the principal axes obtained as the eigenvalues of the mass tensor of the stellar mass inside $2r_*$. The larger $1 - f$ is, the flatter the system is in 3D. Here and throughout the paper we refer to the definitions and measurements of \citet{Pillepich2019}. See also \autoref{sec:true_disks} for a more detailed discussion of the 3D shapes of the TNG50 galaxies.

\autoref{fig:sims_hexbin_umap} shows remarkable correlations between the position of galaxies in the UMAP and their average physical properties. Overall, galaxies with larger specific angular momentum and a flatter stellar distribution tend to populate the upper-right region of the UMAP. These galaxies are also the most massive ones although the correlation is less clear. Moreover, the galaxies with larger masses not only occupy the upper-right section of the UMAP plane but also extend along the upper edge towards the left corner of the UMAP plane. On the contrary, low-mass galaxies populate predominantly the bottom-left section of the UMAP representation. The left section of the UMAP plane is populated by rounder objects with lower specific angular momentum. It is also interesting to see that the transition between the variation of the physical properties is smooth, translating a continuum of galaxy morphology/structure. 

\autoref{fig:sims_hexbin_umap} only shows the median values of the physical properties in different regions of the UMAP. In order to quantify how constraining are these correlations, it is also important to measure the scatter of the different properties. This is shown in the appendix~\ref{app:dependence} (\autoref{fig:sigma_sims_hexbin_umap}). In most cases, the scatter represents less than $\sim20\%$ of the dynamical range, indicating that the distributions are overall relatively narrow and, therefore, the correlations with physical properties are informative.

We conclude that the representation space for images ---in addition to being robust to observational and instrumental effects--- carries information about the kinematics and intrinsic shapes of galaxies.

\subsubsection{Connection to standard morphological measurements}

\begin{figure*}[t!]
\centering
	\includegraphics[width=0.32\textwidth]{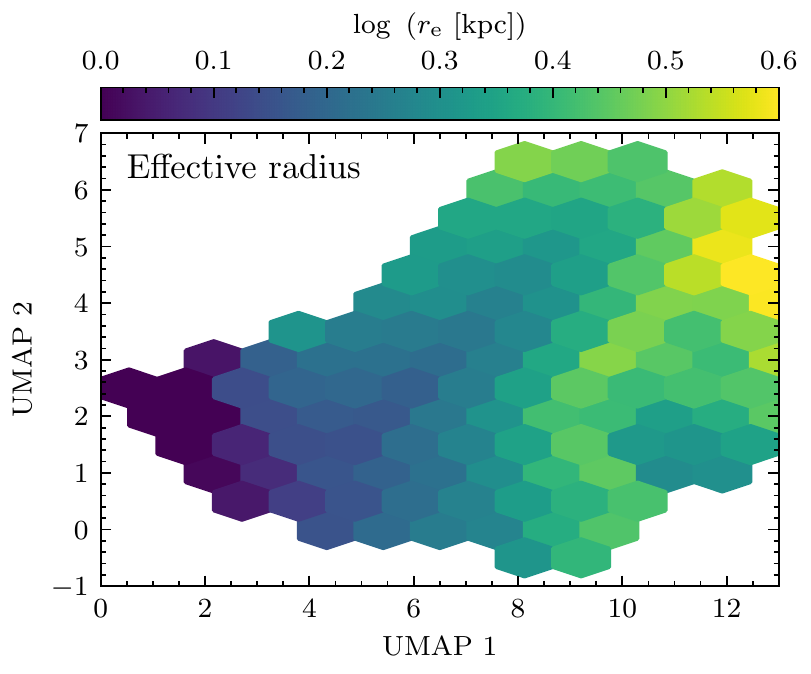}
	\includegraphics[width=0.32\textwidth]{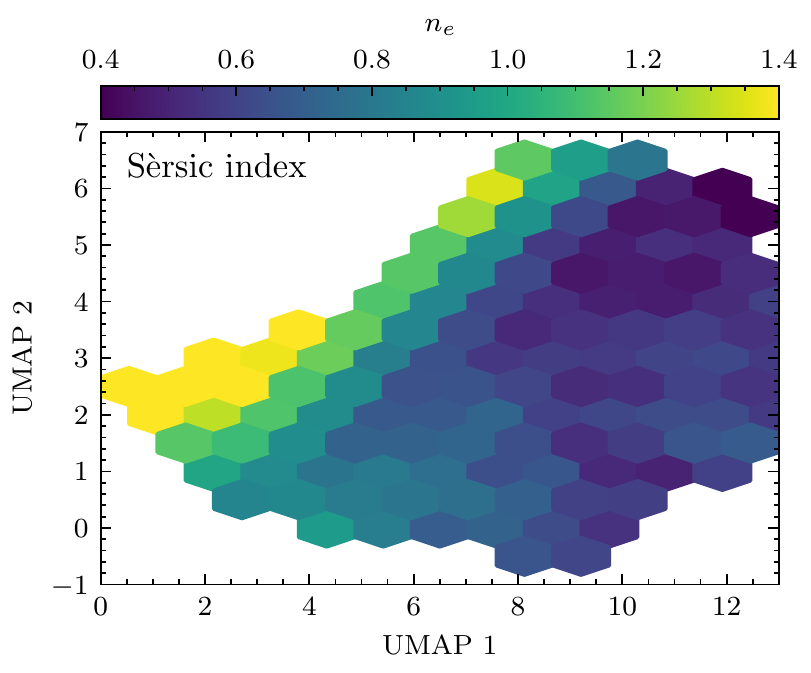}
	\includegraphics[width=0.32\textwidth]{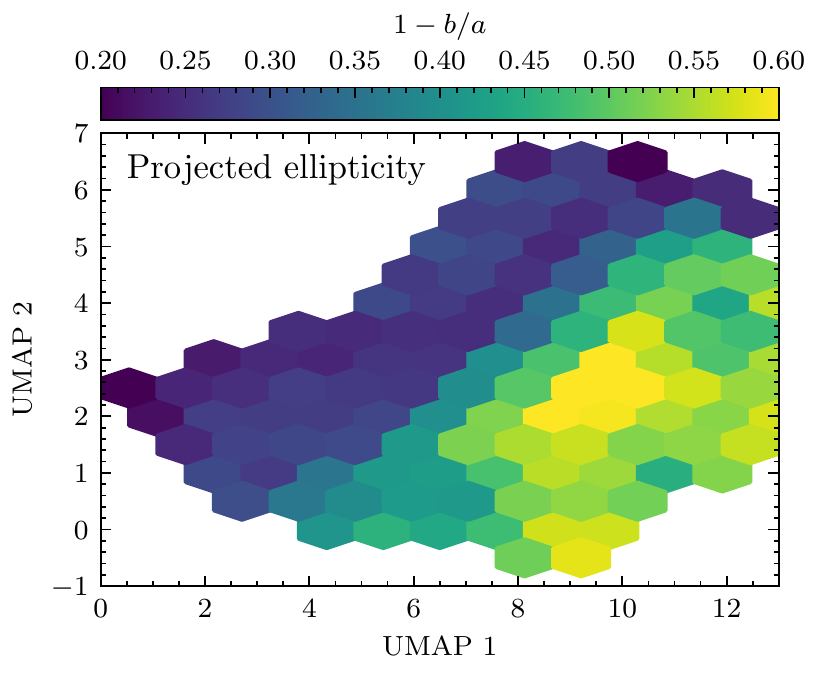}
	\includegraphics[width=0.32\textwidth]{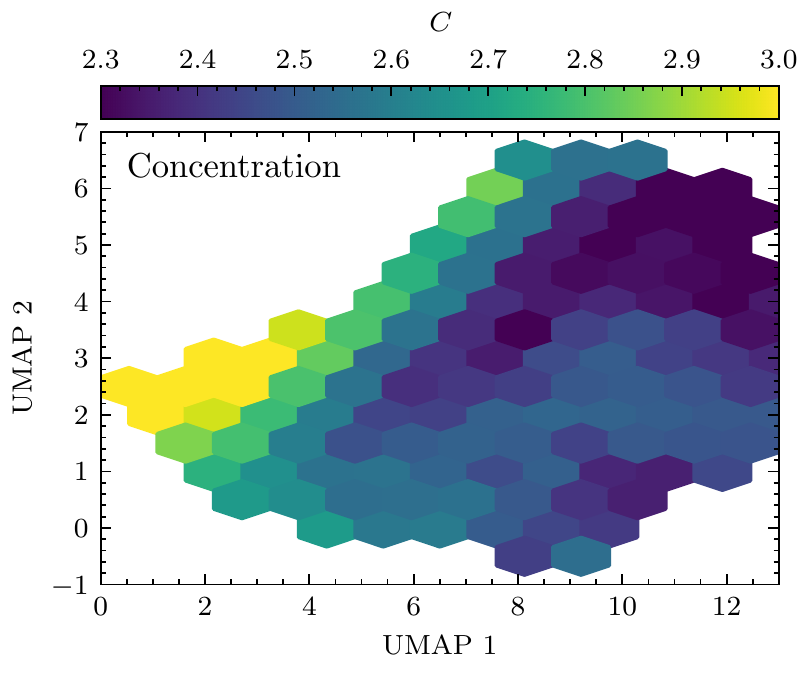}
	\includegraphics[width=0.32\textwidth]{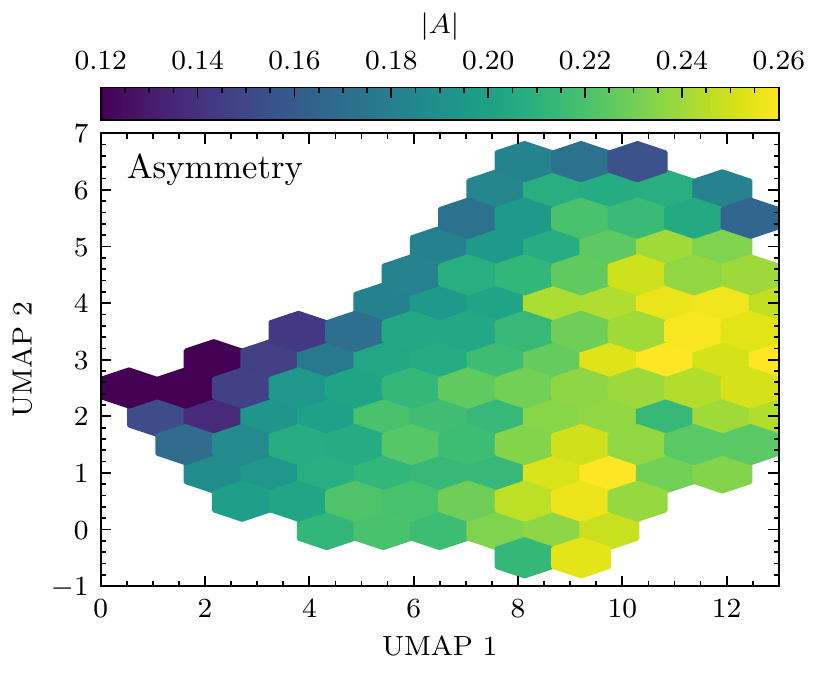}
	\includegraphics[width=0.32\textwidth]{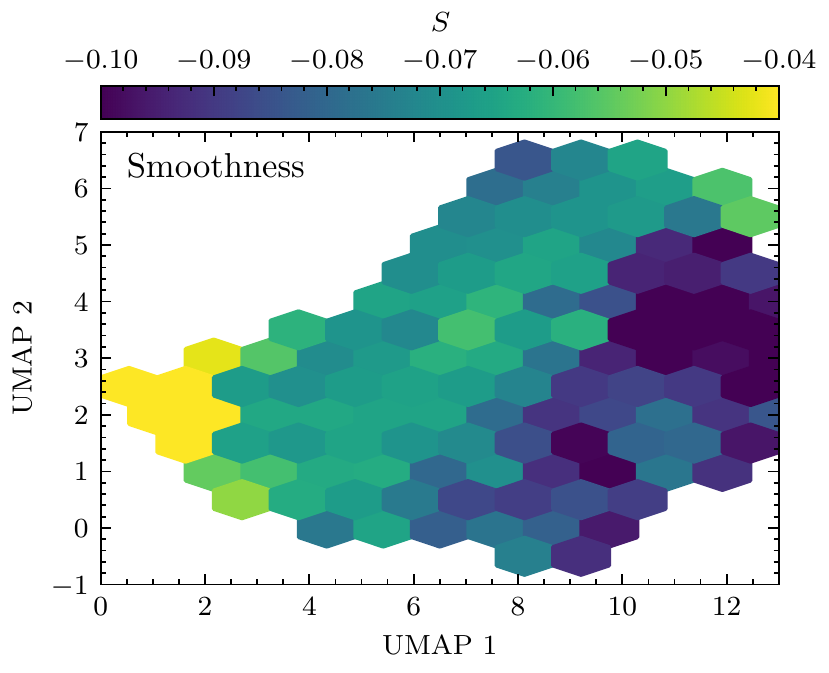}

    \caption{UMAP visualization for all the TNG50 galaxy images in our dataset color-coded by the distribution of several morphological and photometric parameters. Color code corresponds to the median values in each hexagonal bin in the UMAP plane. From left to right and top to bottom, the different panels show: the logarithm of the effective radius ($r_e \mathrm{[kpc]}$), in kpc), the S\`ersic index ($n_e$), the ellipticity based on S\`ersic fit ($1-b/a$), the concentration ($C$), the asymmetry ($A$) and the smoothness ($C$). The scatter maps of these parameters are presented in \autoref{app:dependence}.} 
    \label{fig:phot_hexbin_umap}
\end{figure*}

In \autoref{fig:phot_hexbin_umap}, we show the dependence on several photometric parameters estimated by \citet{Costantin2022a} in the F200W filter: the effective radius ($r_e$), the S\`ersic index ($n_e$), the ellipticity from the S\`ersic fit ($1-b/a$), the concentration parameter ($C$), the asymmetry ($|A|$) and the smoothness ($S$). There is a remarkable correlation between the position in the UMAP plane and $n_e$, $C$, $A$, and $S$. Gradually, $n_e$, $C$, and $S$ grow from right to left in the UMAP space, while the $A$ does it from left to right. Therefore, galaxy images with smoother, symmetric, and concentrated light distributions are found towards the left section of the UMAP plane. Also important is the correlation with the ellipticity ($1-b/a$), with more elongated galaxies lying on the right (bottom-right) section of the UMAP plane. To illustrate again the spread of these representations, we show in the appendix~\ref{app:dependence} (\autoref{fig:sigma_phot_hexbin_umap}) the scatter of the parameters shown in \autoref{fig:phot_hexbin_umap}.

It is important to notice the existing correlation with the physical effective radius, $r_e$, with the largest galaxies populating the right section of the UMAP plane. This correlation with the physical size reflects the known correlation between morphological appearance and physical size (e.g.,\citealp{2014ApJ...788...28V}).

Based on the previous maps calibrated with the TNG50 simulation, asymmetric, more extended, flatter, and rotationally-supported galaxies tend to populate the right and upper-right sections of the UMAP representation. In more detail, the more to the right in the UMAP plane a galaxy is, the more elongated it appears. Smoother, more compact, rounder, and non-rotating galaxies are located toward the left section of the UMAP representation. Also, less massive galaxies can be found predominantly towards the bottom and bottom-left sections of the UMAP plane. Although not shown, the results presented here are consistent (despite small variations) for the same morphological parameters measured in the F365W filter.

\section{Self-supervised learning representation of JWST galaxy images}
\label{sec:observations}

In this section, we apply the methodology described before to the two datasets of observed galaxies with JWST described in \autoref{sec:CEERS_data}.

\subsection{Representations of CEERS galaxy images}
\label{sec:CEERS}

\begin{figure*}[t!]
\centering
	\includegraphics[width=0.65\columnwidth]{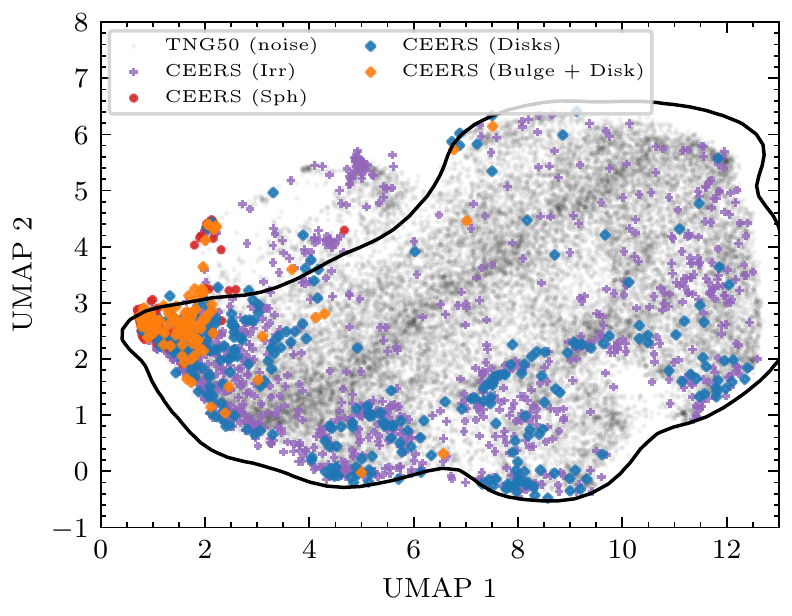}
	\includegraphics[width=0.65\columnwidth]{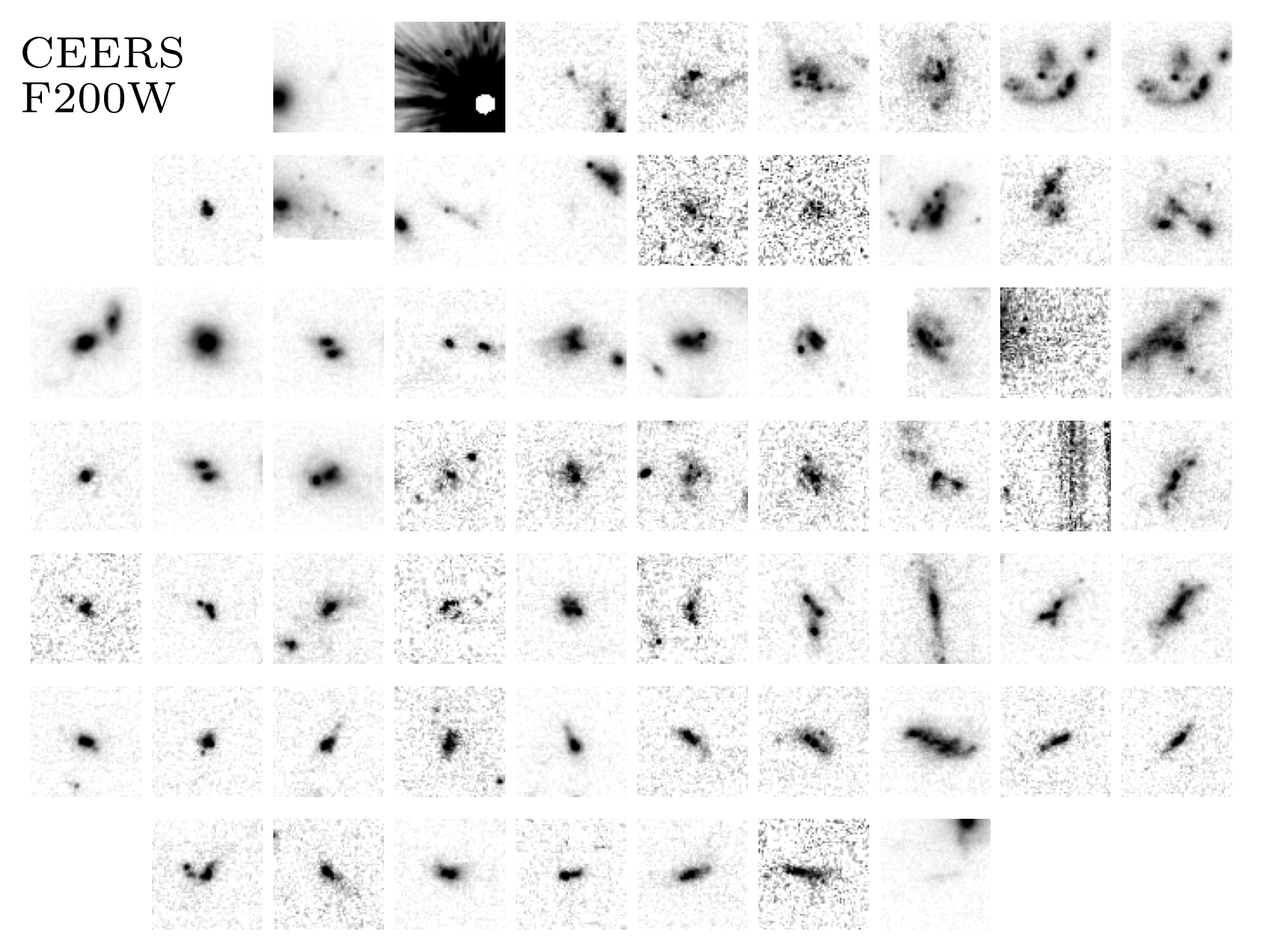}
	\includegraphics[width=0.65\columnwidth]{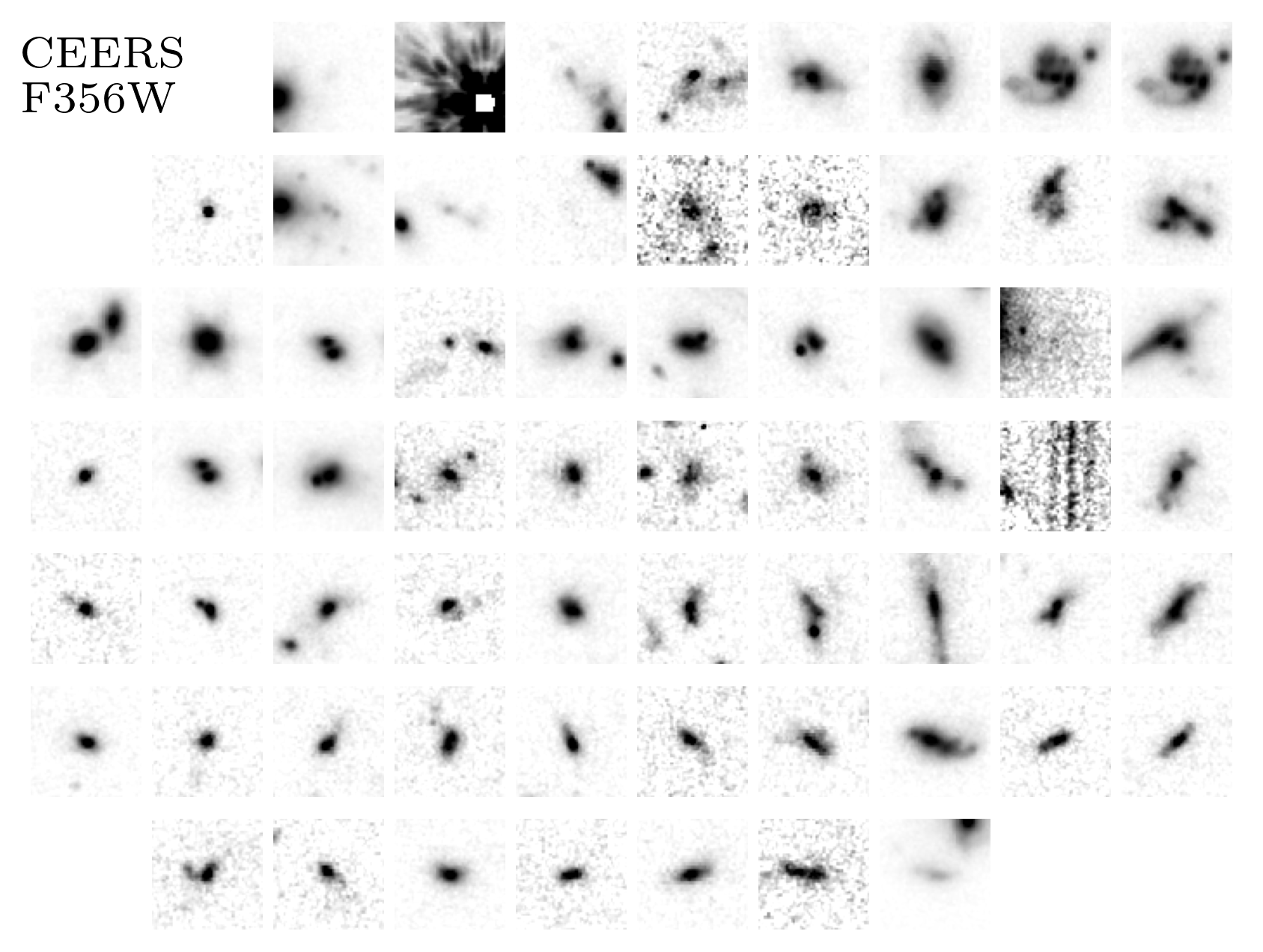}

  \includegraphics[width=0.65\columnwidth]{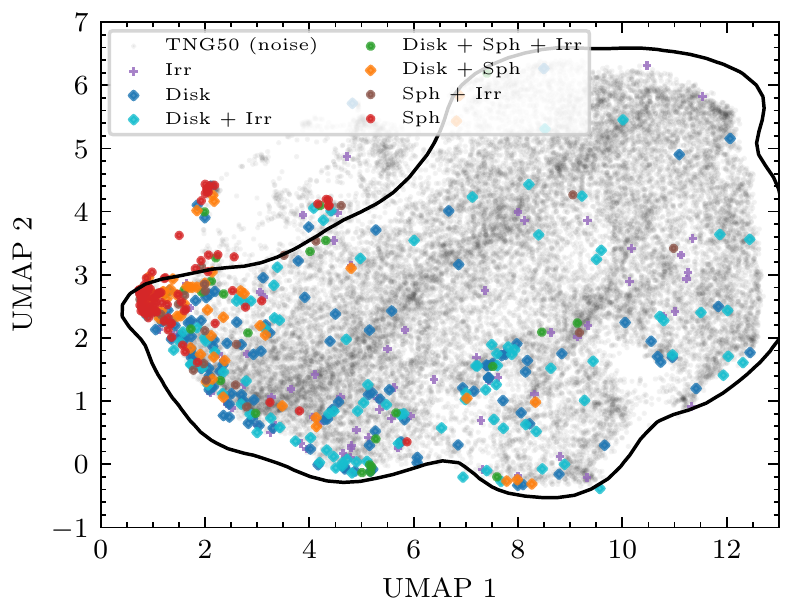}
    \includegraphics[width=0.65\columnwidth]{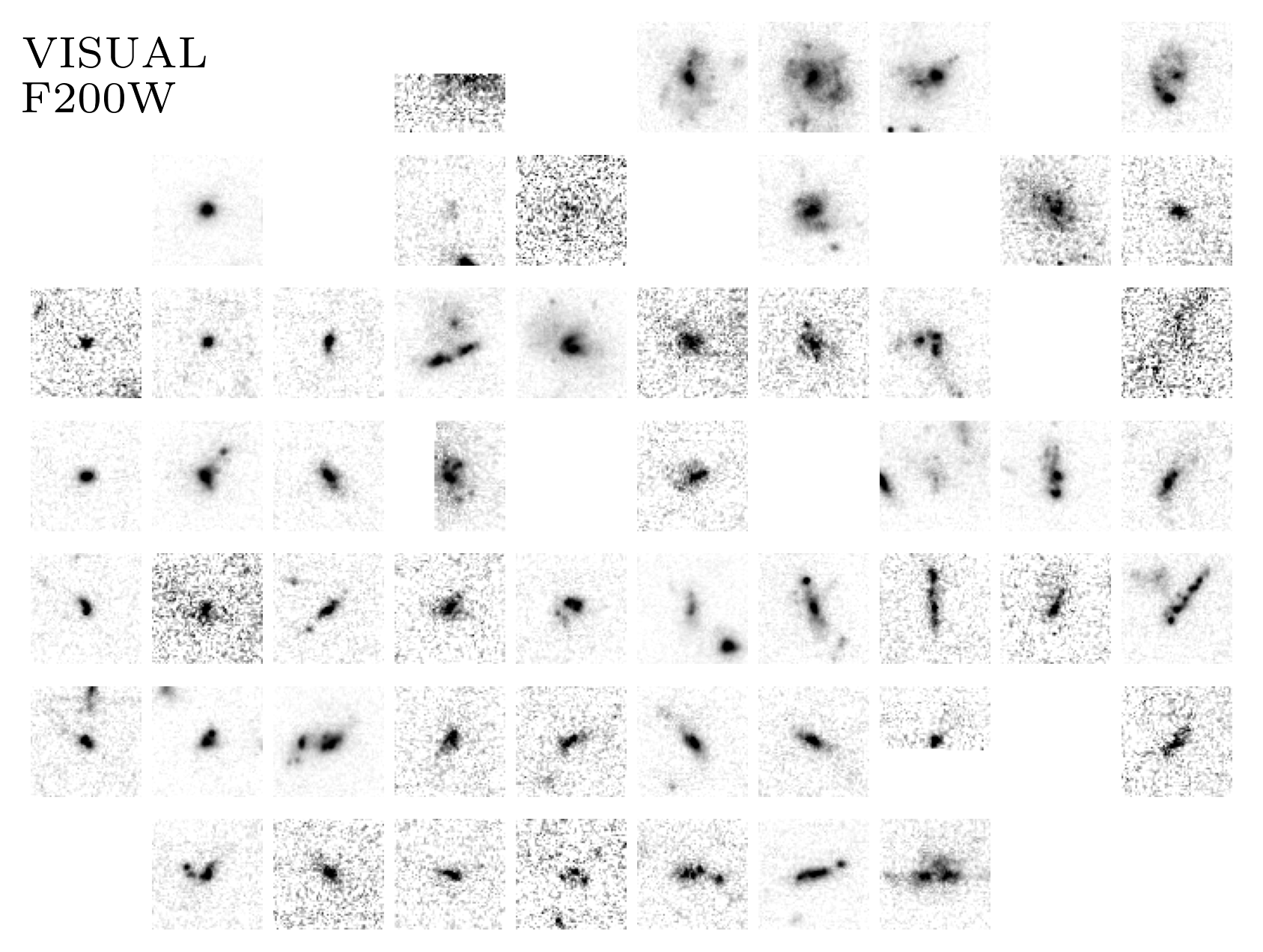}
    \includegraphics[width=0.65\columnwidth]{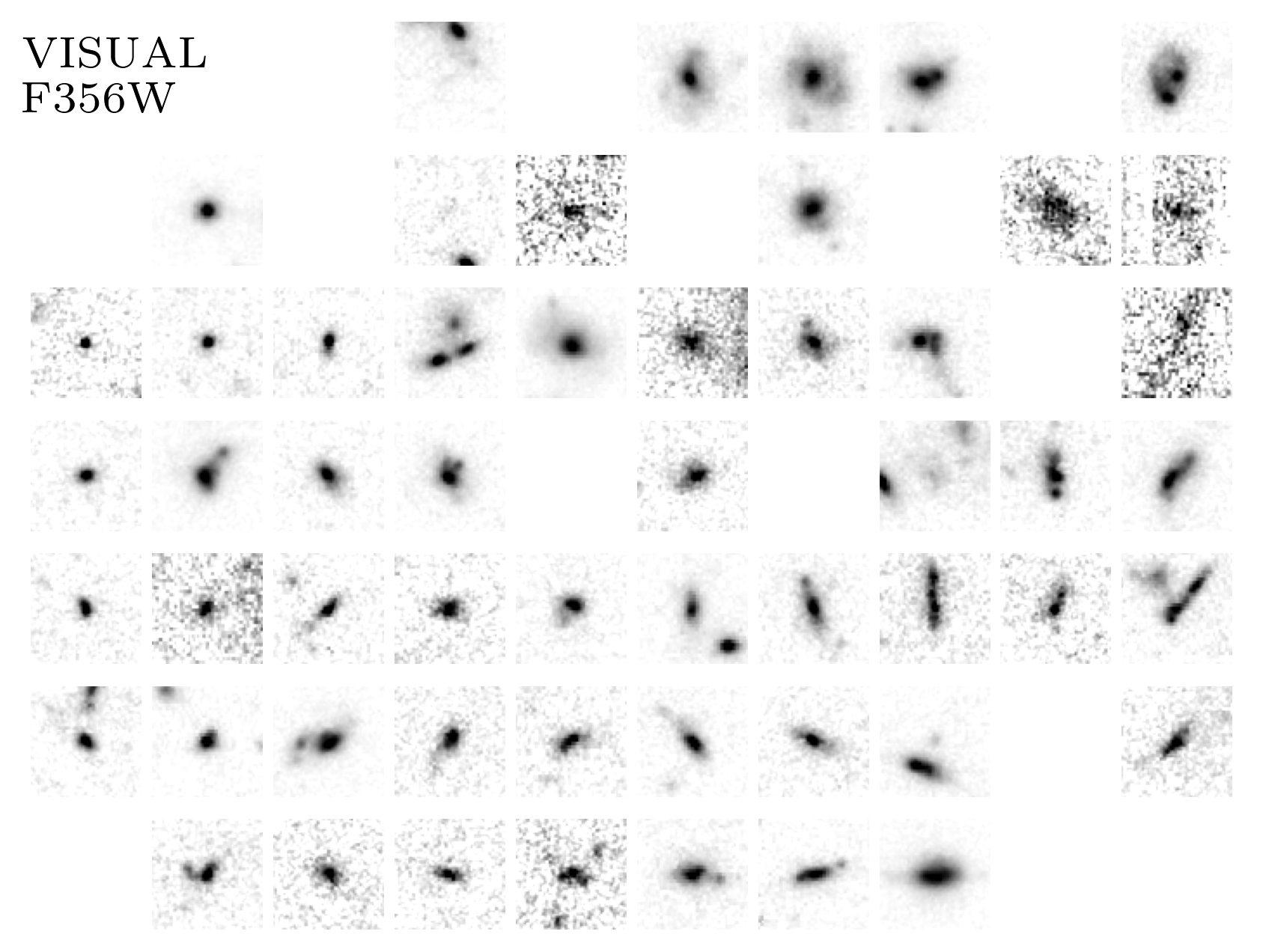}
	
    \caption{Comparison of distributions of observed and simulated galaxies in the representation space. The top row shows the CEERS mass-complete sample and the bottom row the VISUAL sample. Top left-hand panel: UMAP visualization for the observed CEERS galaxy images selected in mass and redshift (color-coded by the CNN-based morphological classes derived in \citealt{Huertas-Company2023} for the F356W filter) overlapped with the representation of noise-added TNG50 galaxy images. Black contour indicates the region that is not affected by extremely bright companions. Top middle panel: randomly chosen observed CEERS galaxy images in the UMAP visualization in the F200W filter. Top right-hand panel: randomly chosen observed CEERS galaxy images in the UMAP visualization in the F356W filter. Bottom left-hand panel: UMAP visualization for the observed VISUAL galaxy images selected in mass and redshift. Points are colored according to the visual classifications into several classes. Black contour indicates the region that is not affected by extremely bright companions. Bottom middle panel: randomly chosen observed VISUAL galaxy images in the UMAP visualization in the F200W filter. Bottom right-hand panel: randomly chosen observed VISUAL galaxy images in the UMAP visualization in the F356W filter.} 
    \label{fig:CEERS_umap}
\end{figure*}

We feed the 1\,664 observed CEERS galaxies to our contrastive model to retrieve their corresponding representations in the 1\,024 dimensions space. Then, we normalize the derived features and transform them into a 2D vector using the same UMAP embedding obtained for the features of the TNG50 galaxy images. 

In the top row of \autoref{fig:CEERS_umap}, we show the UMAP representation space for the observed CEERS dataset. Interestingly, observed galaxies tend to populate the complete UMAP plane which indicates that both samples share similar morphological diversity. The UMAP visualization is however a projection of a higher dimension space, which is not appropriate for outlier detection. Even if observed galaxies would not reside in the same manifold as simulated objects, the UMAP representation would tend to show them towards the edges of the plane but not outside. This is the behavior seen for observed CEERS galaxies which tend to be concentrated in the edges of the UMAP cloud (towards the bottom and bottom-left sections) independently of the source redshift. Given that the mass and flux distributions of both datasets are consistent ---even though we have not performed a careful one-to-one match between simulations and observations---  the differences in the distributions of points of both datasets are likely to originate in intrinsic differences in the morphological properties. Combining the distributions of points in \autoref{fig:CEERS_umap} with the information provided by \autoref{fig:sims_hexbin_umap} and \autoref{fig:phot_hexbin_umap}, we conclude that observed CEERS galaxies occupy more frequently than the simulated TNG50 galaxies the regions in the representation space where galaxies are more compact and with less specific angular momentum. We investigate these differences in more detail in \autoref{sec:comparison_representations}.

Also interesting is the presence of galaxy images with signs of interactions, multiple clumps, and gas accretion processes in the upper-right section of the UMAP in \autoref{fig:CEERS_umap} (more clear in the F200W filter because of its better spatial resolution compared to the F356W filter). Moreover, galaxy images with double nuclei (or even multiple clumps) closer in projection tend to appear in the bottom-right section of the UMAP plane. These systems are apparently more elongated and, therefore, lay into the region of the UMAP plane where the projected ellipticities are on average larger (as shown in \autoref{fig:phot_hexbin_umap}), but less flat than the systems located on the upper-right section of the UMAP plane (as shown in \autoref{fig:sims_hexbin_umap}).

Following \autoref{sec:noise}, hereafter, we only include galaxies located within the black contours shown in \autoref{fig:CEERS_umap}, for which it is unlikely to find bright companions or artifacts that could bias their representations. We find that $\sim90\%$ ($1\,481$) of the galaxies fulfill this criterion, while the remaining $\sim10\%$ are excluded from the subsequent analysis. In this dataset and according to the morphologies based on the F356W filter, 121 galaxies ($\sim 8\%$) are classified as \textit{Sph}, 297 galaxies ($\sim 20\%$) are classified as \textit{Disk}, 96 galaxies ($\sim 6\%$) are classified as \textit{Bulge + Disk} and 967 galaxies ($\sim 65\%$) are classified as \textit{Irr}. If we focus on the morphologies obtained from the F200W filter, the fraction of \textit{Irr} galaxies raises up to $\sim 82\%$ and the fraction of Disk galaxies decreases down to $\sim 6\%$.

\subsection{Representations of VISUAL galaxy images}
\label{sec:Zoo}

We also present a comparison of the representation obtained after applying our contrastive model to the VISUAL dataset for which visual morphological classifications are provided \citep{Kartaltepe2022}. After selecting those galaxies with $M_* \geq 10^9 \mathrm{M_{\odot}}$, $3 < z < 6$, and reliable visual classifications we end up with a dataset of 545 galaxies. To avoid including in the analysis galaxy images with contaminants or artifacts, hereafter, we only consider those galaxies located within the black contours shown in \autoref{fig:CEERS_umap}. In this case, we are confident about the representations obtained for $\sim 90\%$ (483) of the galaxy images, for which we find: 118 ($\sim24\%$) \textit{Disk} galaxies, 102 ($\sim21\%$) \textit{Disk+Irr} galaxies, 24 ($\sim5\%$) \textit{Disk+Sph+Irr} galaxies, 56 ($\sim11\%$) \textit{Disk+Sph} galaxies, 71 ($\sim14\%$) \textit{Sph} galaxies, 22 ($\sim4\%$) \textit{Sph+Irr} galaxies, 81 ($\sim16\%$) \textit{Irr} galaxies, and only 2 and 7 as \textit{point sources} and \textit{unclassifiable} galaxies, respectively.

In the bottom row of \autoref{fig:CEERS_umap}, we show the representation of the galaxy in the VISUAL dataset in the UMAP plane for the various morphological groups based on the provided visual classifications. Although the mass and redshift selection of the galaxies is based on different estimators (JWST photometry for the CEERS dataset and CANDELS photometry for the VISUAL dataset), the distribution of the representations in the UMAP plane for the VISUAL galaxies is similar to the CEERS representations (i.e., a significant fraction of galaxies occupy the bottom and bottom-left section of the UMAP plane). The figure reveals some expected correlations with the traditional visual morphology. It is reassuring that \textit{Disk+Sph} and \textit{Sph} groups from the VISUAL catalog populate the left corner of the UMAP plane, where compact, non-rotating galaxies with low angular momentum (according to TNG50 properties) are expected to be. However, we notice that galaxies classified as \textit{Disk}, \textit{Irr} and/or \textit{Disk+Irr} in VISUAL are distributed throughout the plane even towards the left section of the UMAP, very close to where spheroids lie. As shown in \autoref{fig:sims_hexbin_umap} and~\autoref{fig:phot_hexbin_umap}, the left region of the UMAP where, according to VISUAL, disk-like morphologies are located corresponds to galaxies in TNG50 with physical and photometric properties typically shared by spheroidal systems, such as low specific angular momentum, large mass fractions in a non-rotating component, low flatness, and larger S\`ersic indexes. This raises interesting questions about the true nature of these disks that we discuss in \autoref{sec:true_disks}.

%%%%%%%%%%%%%%%%%%%%%%%%%%%%%%%%%%%%%%%%%%%%%%%%%%%

\section{A comparison between simulated and observed self-supervised morphologies}
\label{sec:comparison_representations}

In this section, we examine in more detail the differences found in previous sections between the simulated TNG50 and the observed JWST galaxy images. 

\subsection{Distribution of self-supervised representations}
\label{sec:representations}

\begin{figure}[t!]
\centering
    \includegraphics[width=\columnwidth]{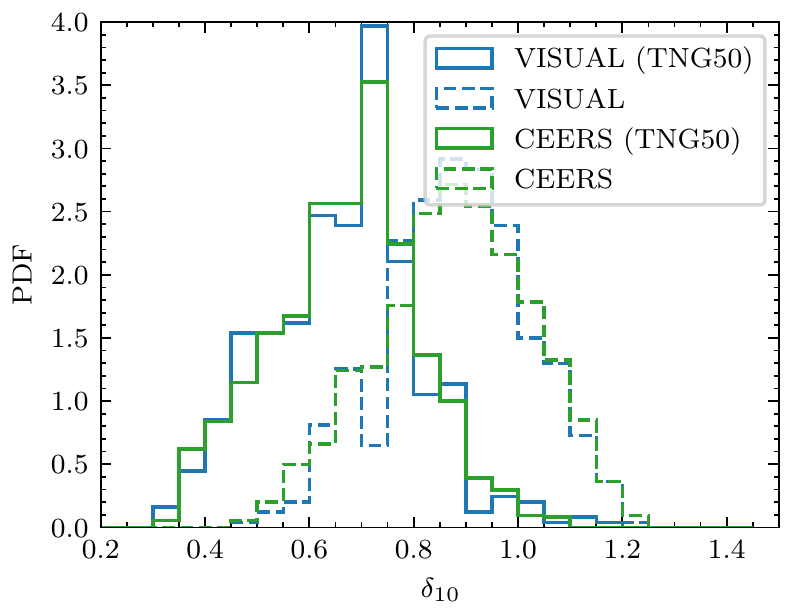}
		
    \caption{Probability density functions of the distances to the 10th closest neighbor in the 1\,024 dimensions of the representation space, denoted as $\delta_{10}$. Solid histograms correspond to the distance to the 10th closest neighbor in TNG50 of the closest TNG50 neighbor of each galaxy in the VISUAL (in blue) and the CEERS (in green) datasets. Dashed histograms correspond to the distance to the 10th closest neighbor in TNG50 of each galaxy in the VISUAL (in blue) and the CEERS (in green) datasets.} 
    \label{fig:dist_kn10}
\end{figure}

 \begin{figure*}[t!]
%\centering
    \includegraphics[width=0.24\textwidth]{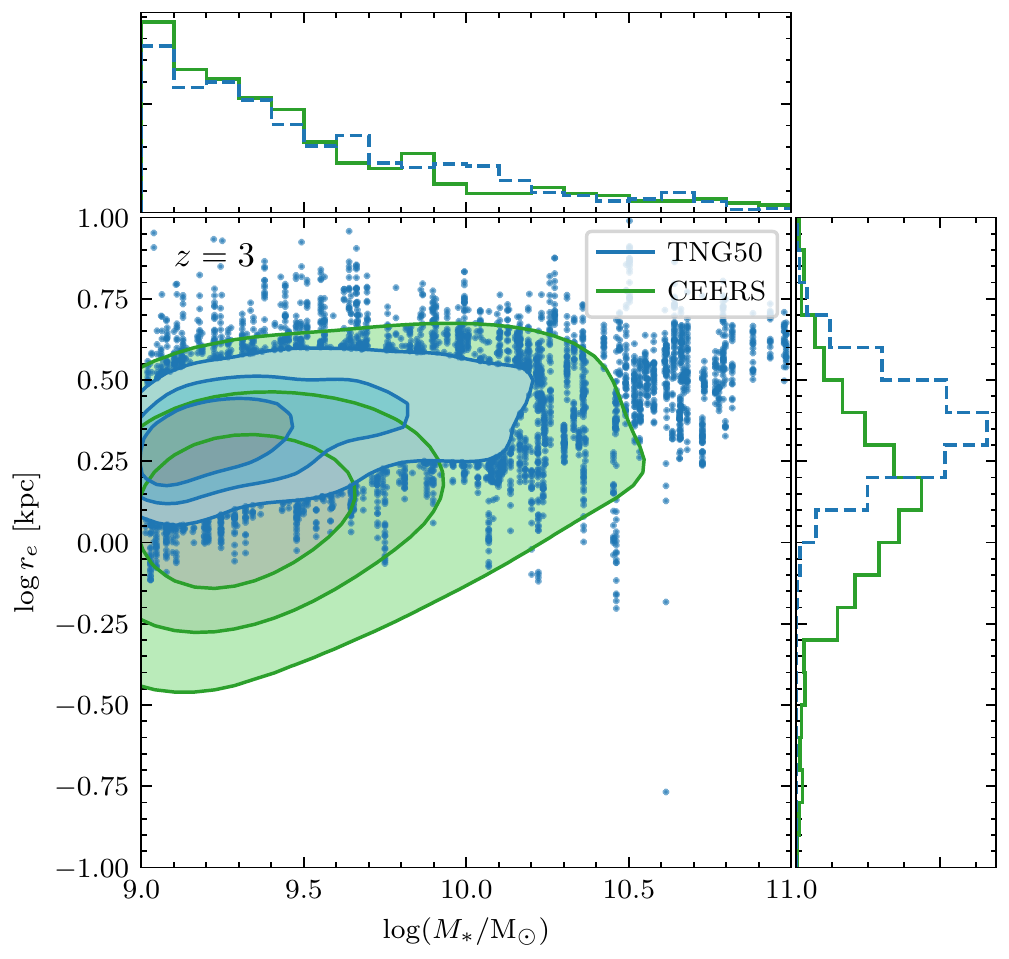}
    \includegraphics[width=0.24\textwidth]{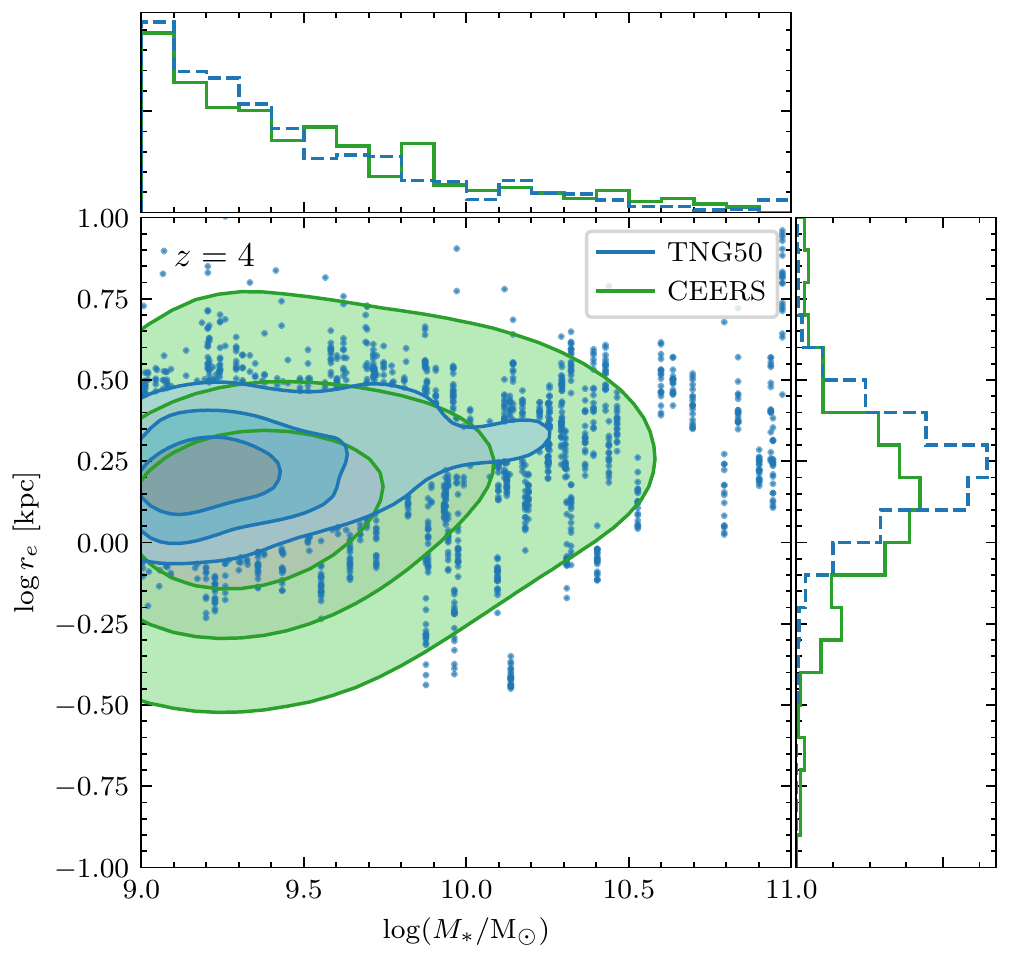}
    \includegraphics[width=0.24\textwidth]{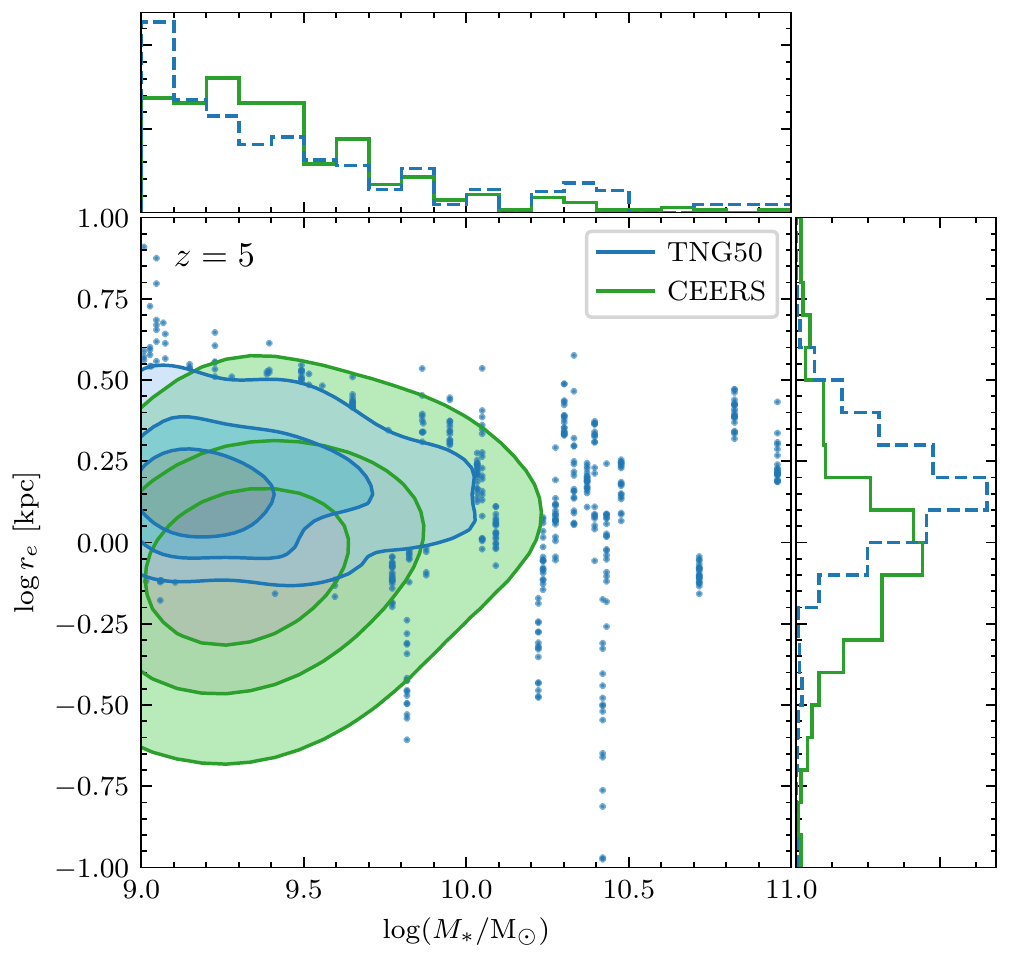}
    \includegraphics[width=0.24\textwidth]{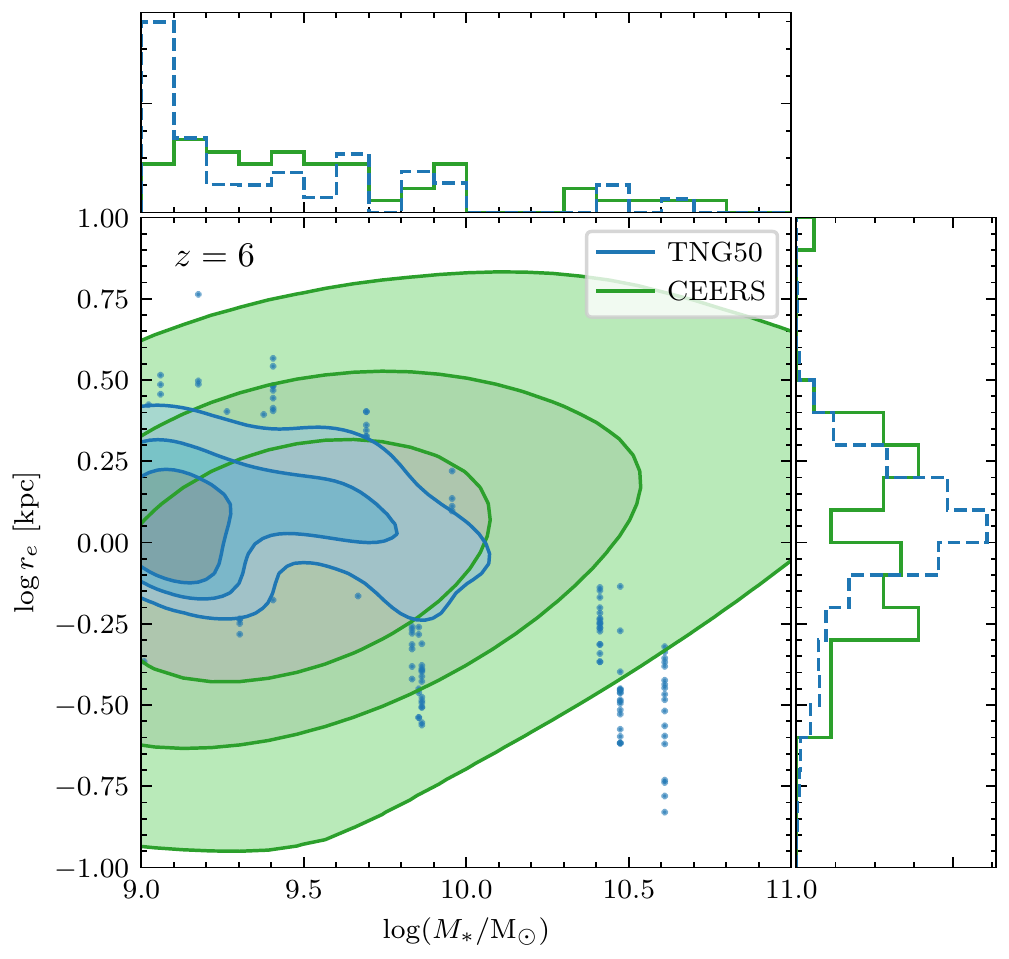}
    
    \includegraphics[width=0.24\textwidth]{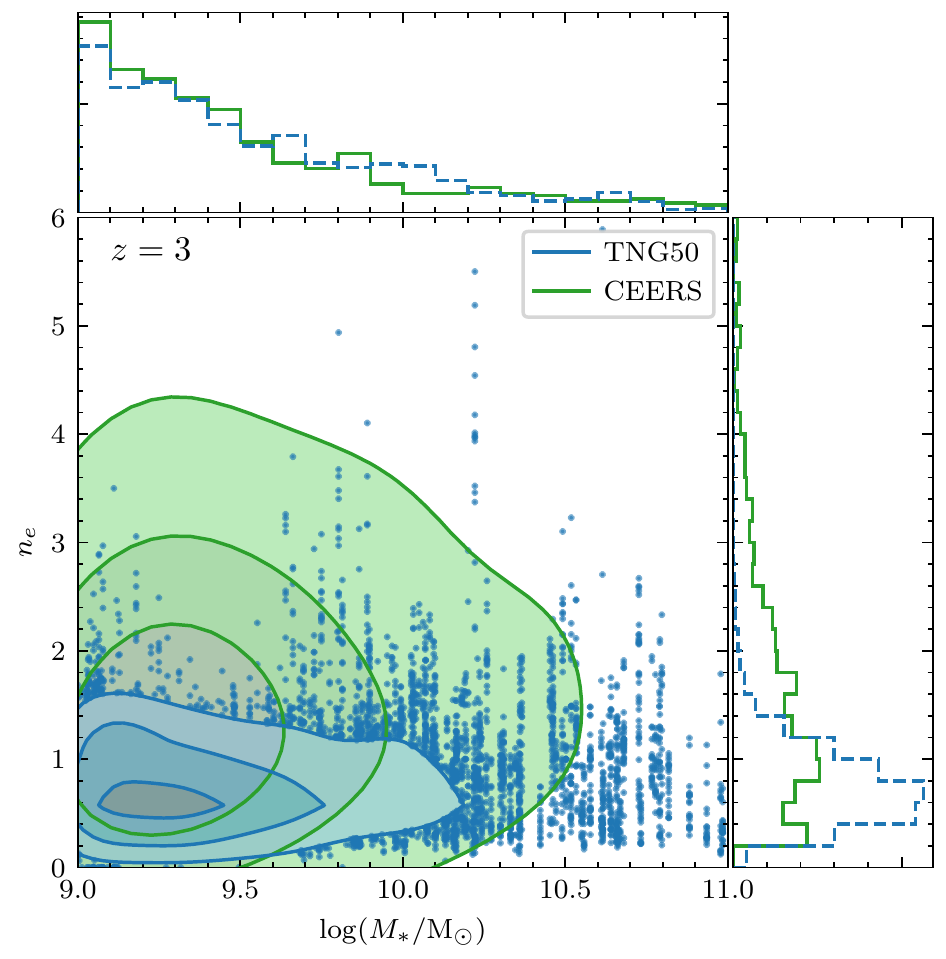}
    \includegraphics[width=0.24\textwidth]{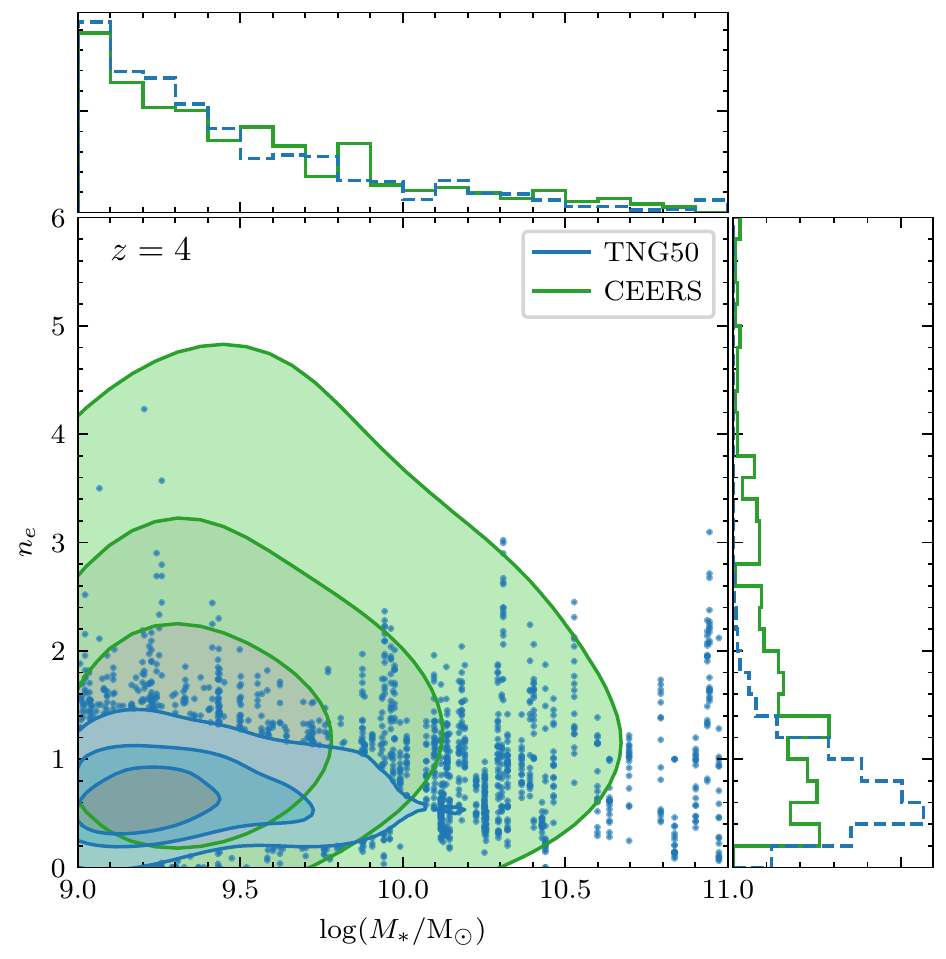}
    \includegraphics[width=0.24\textwidth]{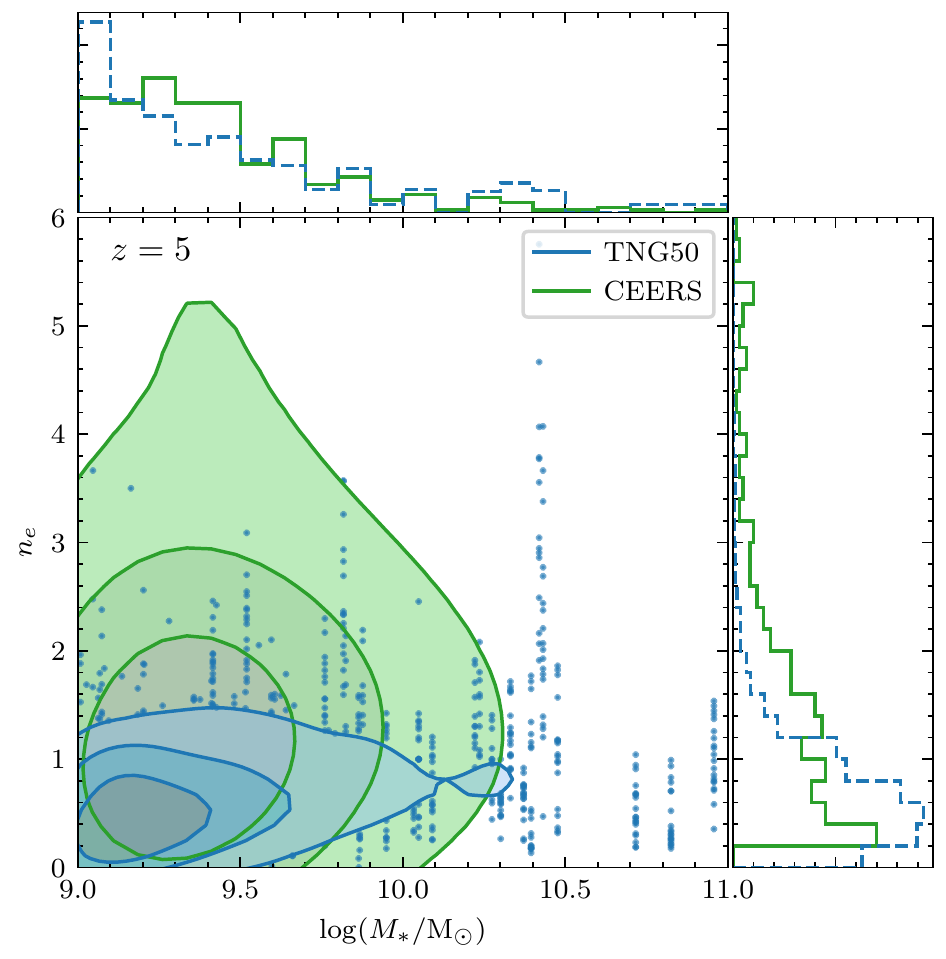}
    \includegraphics[width=0.24\textwidth]{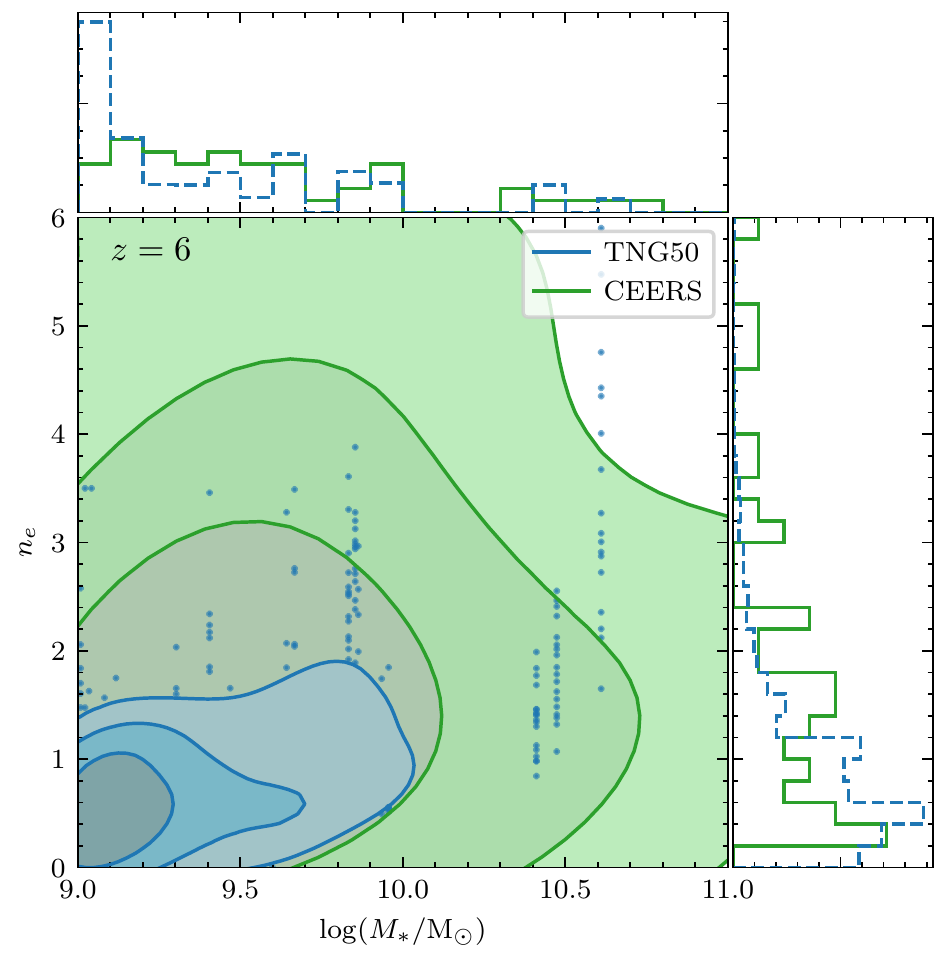}
    
    \includegraphics[width=0.24\textwidth]{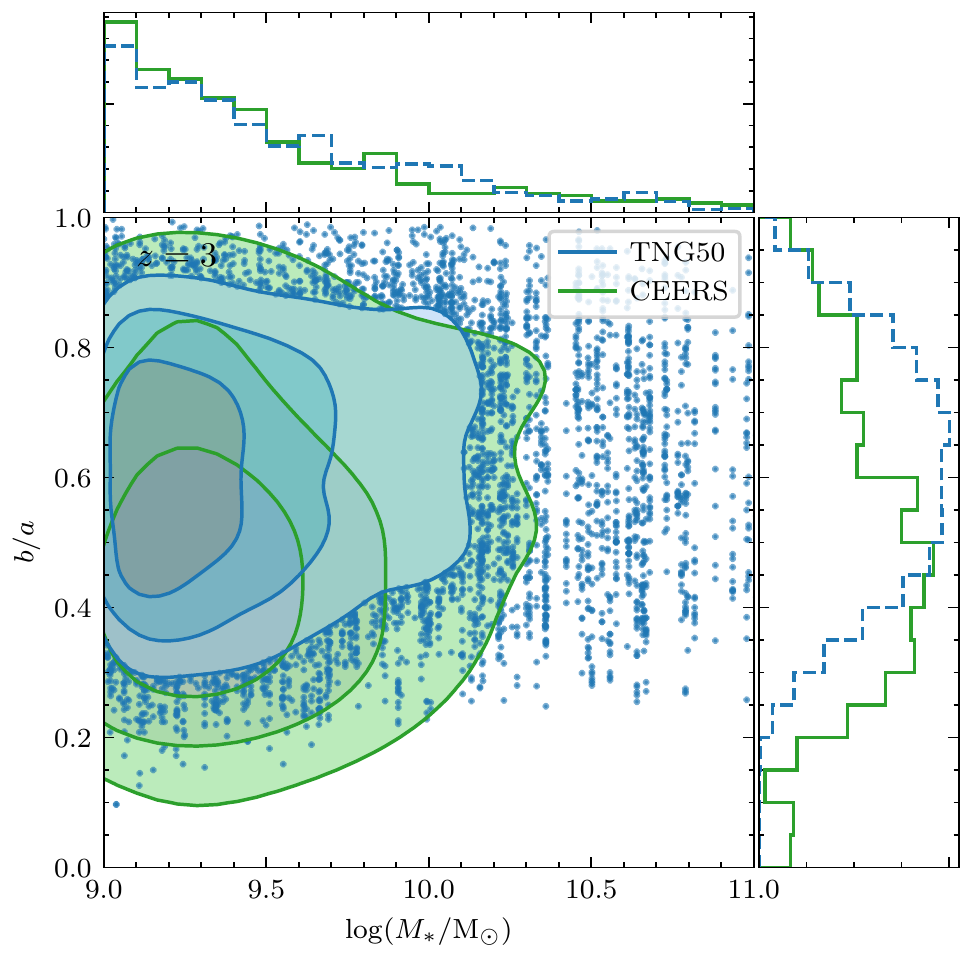}
    \includegraphics[width=0.24\textwidth]{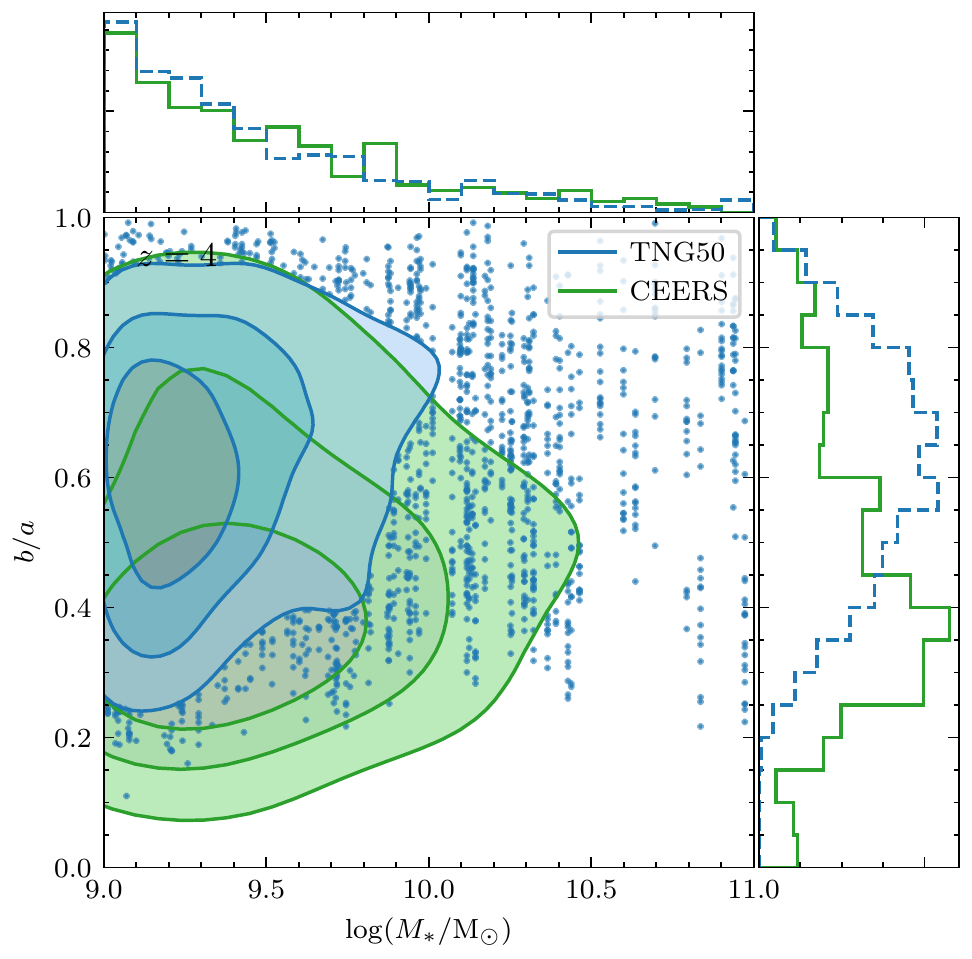}
    \includegraphics[width=0.24\textwidth]{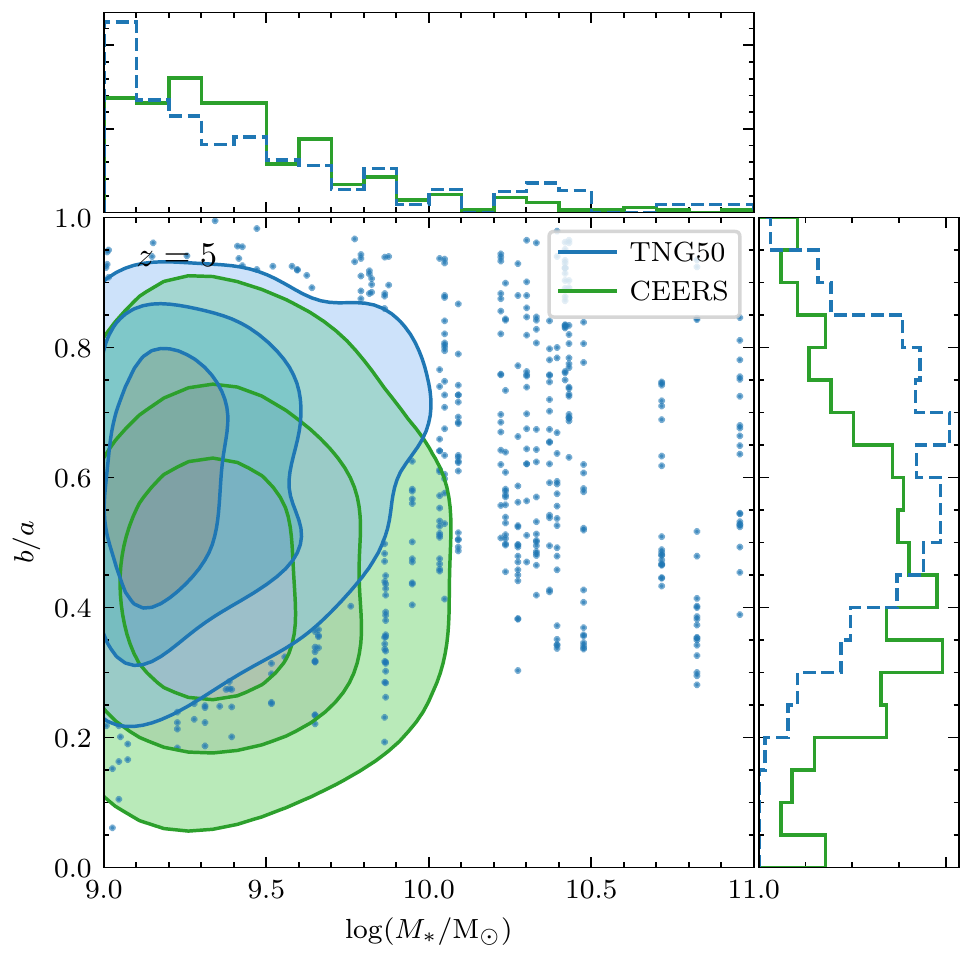}
    \includegraphics[width=0.24\textwidth]{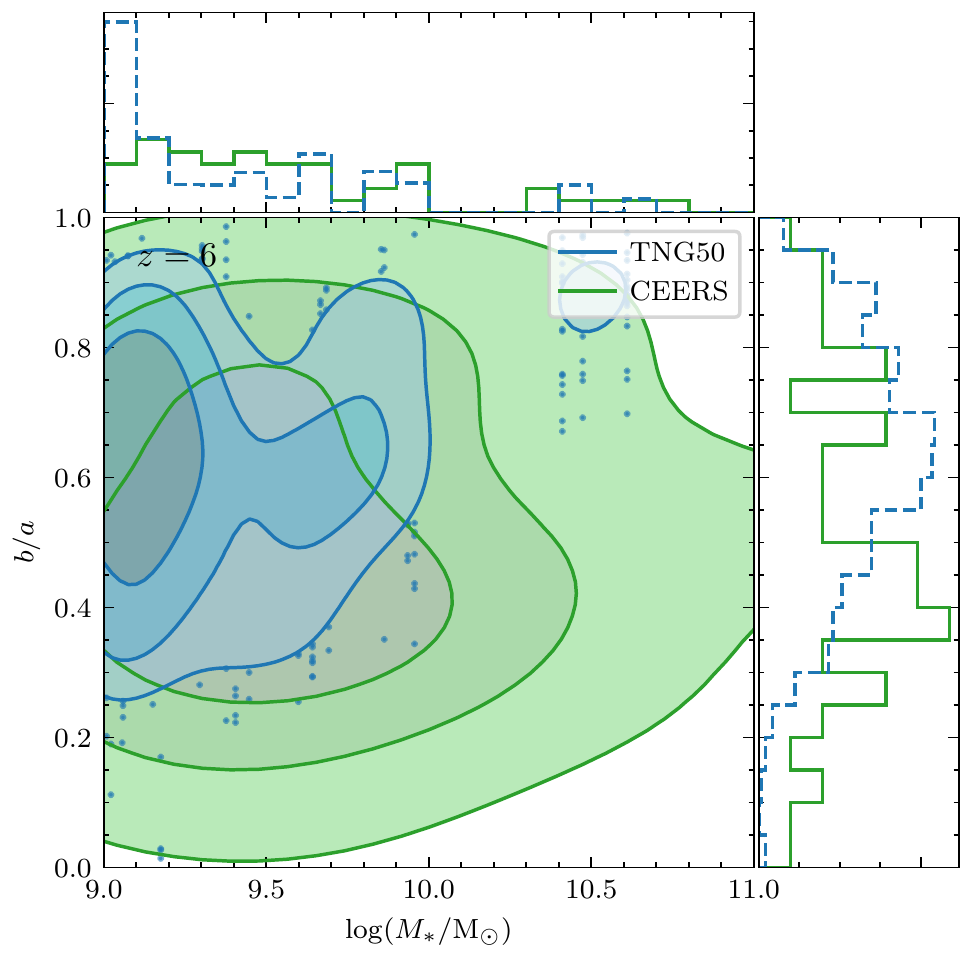}
%\epsscale{0.32}
%\plotone{figures/CEERS_vs_TNG50_radius_mass.pdf}
%\plotone{figures/CEERS_vs_TNG50_sersic_mass.pdf}
%\plotone{figures/CEERS_vs_TNG50_b_over_a_mass.pdf}

    \caption{Distribution of the logarithm of the effective radius ($\log r_e$ in kpc, top row), the S\`ersic index ($n_e$, middle row), and the axis ratio ($b/a$, bottom row) as a function of the logarithm of the stellar mass ($\log M_*$) for the TNG50 (in blue) and the CEERS (in green) datasets. From left to right, the panels show the distributions at $z = 3,4,5,6$ for the TNG50 dataset. For the CEERS dataset, galaxies are included in the closest redshift value. Contour levels enclose $25\%$, $50\%$, and $75\%$ of the data. Blue points correspond to galaxies outside the $75\%$ contour for the TNG50 dataset. Blue dashed and green solid histograms in the horizontal and vertical axes show the PDF of the TNG50 and CEERS datasets, respectively. The photometric parameters shown are measured in the F200W filter.} 
    \label{fig:TNG50_CEERS_morph_params}
\end{figure*}

The representations of the simulated TNG50 and the observed CEERS galaxy images inferred by our contrastive model seem to be distributed differently (\autoref{sec:CEERS} and \autoref{sec:Zoo}). Observed CEERS galaxies tend to concentrate in the left and bottom-left sections of the UMAP plane, while simulated TNG50 galaxies expand over the whole UMAP range with similar number densities. As previously mentioned, the UMAP representation is not well suited for the detection of outliers. Therefore, even if observed galaxies seem to (overall) lie in the same region as simulated ones, they can still live in different manifolds in the higher dimensionality representation space.

To further quantify this distribution shift, we first derive the distance ---in the 1\,024 dimensionality space--- to the 10th closest TNG50 neighbor for each galaxy in the VISUAL and CEERS datasets (denoted as $\delta_{10}$). In order to have a fair reference distribution, secondly, for each observed galaxy we find the closest simulated TNG50 neighbor in the representation space and compute the distance to its 10th closest TNG50 neighbor (also denoted as $\delta_{10}$). In other words, the former corresponds to the distance between the observed galaxy and its 10th closest TNG50 neighbor, while the latter corresponds to the distance to the 10th closest TNG50 neighbor of the closest TNG50 galaxy to each observed galaxy. If both datasets ---observed and simulated--- are distributed likewise in the same manifold the distribution of distances should be similar. If, on the contrary, observed galaxies occupy differently the parameter space, their representations should be disconnected and, therefore, we should measure larger values of $\delta_{10}$. The distributions of $\delta_{10}$ are shown in \autoref{fig:dist_kn10}. It can be clearly seen that the distributions for observed galaxies are shifted towards larger values compared to the reference distribution. This indicates that a significant fraction of observed galaxies are located in regions of the UMAP representation space with lower number densities (i.e., along the edges, not in the central regions) than the average of the TNG50 dataset. This separation could be interpreted as an additional indication that the representations obtained for the TNG50 and the JWST observations do not exactly live in the same manifold. Nevertheless, we find a small fraction of simulated TNG50 galaxies with values of $\delta_{10} \gtrsim 0.9$, meaning that the separation for the most extreme cases is also present for some galaxies in the TNG50 simulation. %As shown in the previous section, there is a trend for observed galaxies to be more compact which might be one explanation for the distribution shift. 

 To better understand these measured discrepancies, we quantify the differences between observed and simulated galaxies in terms of more standard morphological properties in \autoref{fig:TNG50_CEERS_morph_params}.  We show the distributions of observed and simulated galaxies in the $\log M_*-\log r_e$, $\log M_*-\log n_e$, and $\log M_*-b/a$ planes in four redshift bins. To divide in redshift, for the simulated dataset, we take all galaxies in a given snapshot, while, for the observations, we include all galaxies that are associated with the closest snapshot based on their photometric redshifts. It should be kept in mind that this figure (as the previous ones) does not include all TNG50 galaxies, but those for which the JWST mocks are available for a field-of-view larger than $64\times64$ pixels (see \autoref{sec:sims} and \autoref{fig:excluded_sample} for more details). Given the small number of galaxies removed, we do not expect the distributions to change significantly though.
 
It is manifest, firstly, that the TNG50 simulated galaxies overlap with CEERS observed ones in the parameter space of \autoref{fig:TNG50_CEERS_morph_params}. This is per se, again, a non-negligible confirmation of the zeroth-order good functioning of the underlying TNG50 model. However, it is also apparent, differently than what could be deduced from the representation space distributions, that the TNG50 galaxies studied here actually exhibit less galaxy-to-galaxy variation in sizes, S\`ersic indices, and shapes than CEERS observed galaxies, at fixed stellar mass and redshift. Furthermore, the TNG50 simulation predicts galaxies with larger sizes (at $z=3-4$, but not $z=5-6$), with smaller values of $n_e$ (at all $z=3-6$) and that are rounder in projection (more so the higher the redshift) than what is measured in the observed CEERS galaxies. These differences at least partly explain the different distributions in the representation space of contrastive learning and also go in the expected direction of observed galaxies mainly populating the left and bottom-left corner of the UMAP.
 
These reported differences could originate from a resolution-induced effect (see e.g., \citealp{2021MNRAS.501.4359Z}) or could be an indication of more fundamental physical differences. Resolution is certainly an important concern since galaxies at these redshifts are generally small. We recall that although the TNG50 has a softening length for stellar particles of $\sim 300~\mathrm{pc}$, it does not mean that galaxies' stellar disks cannot be thinner than that softening length. The interplay of the various numerical resolution choices (such as gravitational softening of the different matter components, hydrodynamical smoothing length of the gas out of which stars form, mass resolution, etc.) manifests itself in very complex manners in the final structures of the simulated galaxies (see e.g. Section 2.3 of \citealp{Pillepich2019} or Section 2.4 of \citealp{Pillepich2023}). As shown in \citet{Pillepich2019} (Figures 4 and B2), the half-light or half-mass heights of TNG50 galaxies can be smaller than $\sim 300~\mathrm{pc}$ depending on mass and redshift. Similarly, the stellar minor-to-major axis ratios of the stellar mass distributions of TNG50 galaxies can be smaller than $b/a \sim 0.3$, as shown in e.g. Figure 8 (top panels) of \citet{Pillepich2019}, again depending on mass and redshift. In fact, in \autoref{fig:TNG50_CEERS_morph_params}, it can be clearly seen how the axis ratios extend down to $b/a \sim 0.2$ (mainly at $z=3-4$). Moreover, as shown in Appendix B2 of \citet{Pillepich2019}, TNG50 disk heights can be considered converged to better than $20--40 \%$ across all studied masses and redshifts, when compared to the same galaxies simulated at worse numerical resolution.

We note as well that the stellar masses reported for the TNG50 simulation correspond to the 3D stellar mass, while those obtained for the CEERS dataset are based on the SED fitting to the JWST photometry. Also, the S\`ersic parameters for the TNG50 and the CEERS galaxies are derived using different methodologies: for the TNG50, the morphological parameters are obtained with \texttt{statmorph} (as described in \citealt{Costantin2022a}); while for the CEERS dataset, they are derived with \texttt{galfit}.

More in-depth comparisons of simulated and observed data ---likely beyond images--- are required to reach a final conclusion.

\subsection{Self-supervised clustering in TNG50 and CEERS}
\label{sec:morphological_classes}

\begin{figure*}[t!]
\centering
	\includegraphics[width=0.45\textwidth]{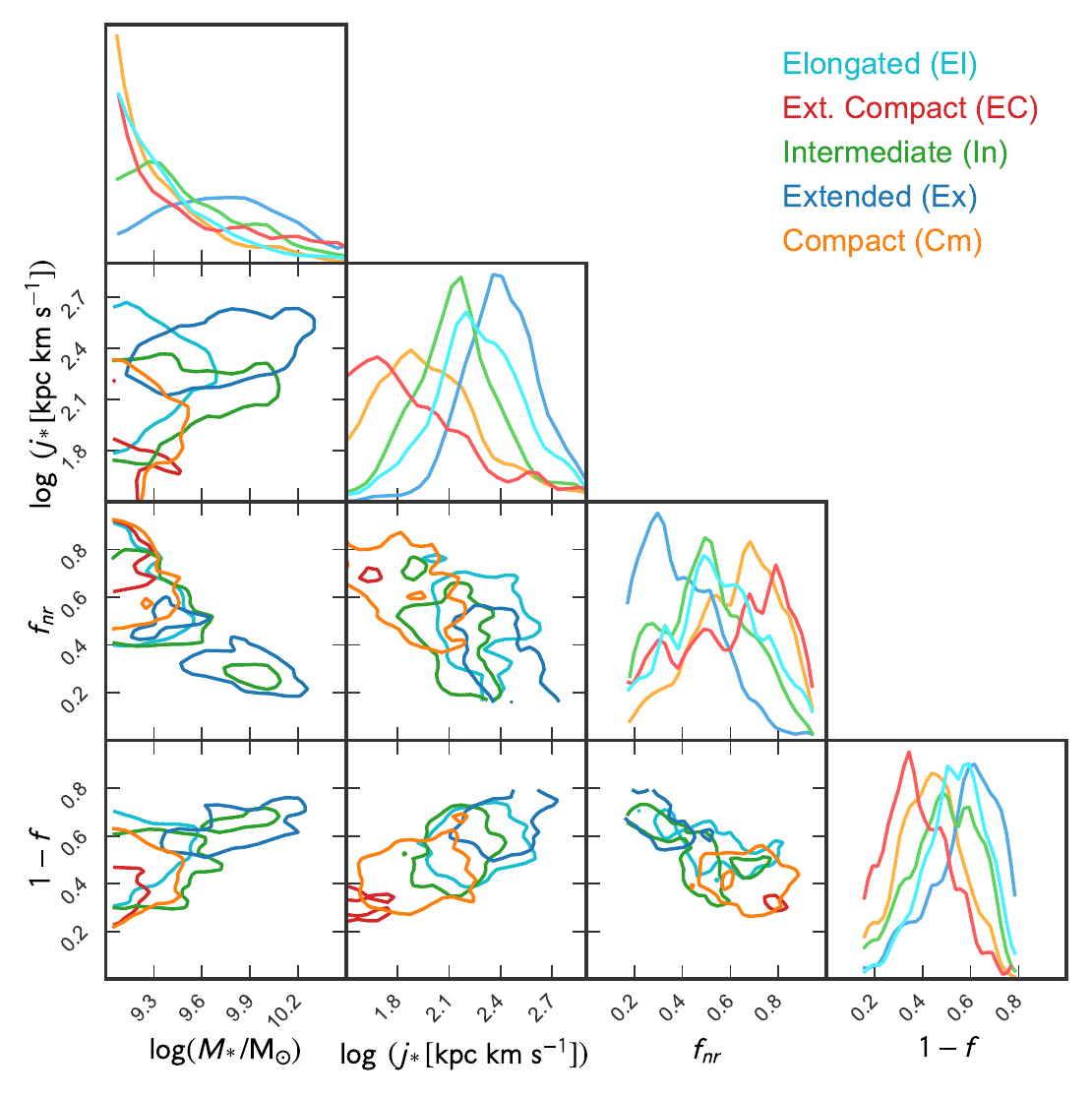}
	\includegraphics[width=0.45\textwidth]{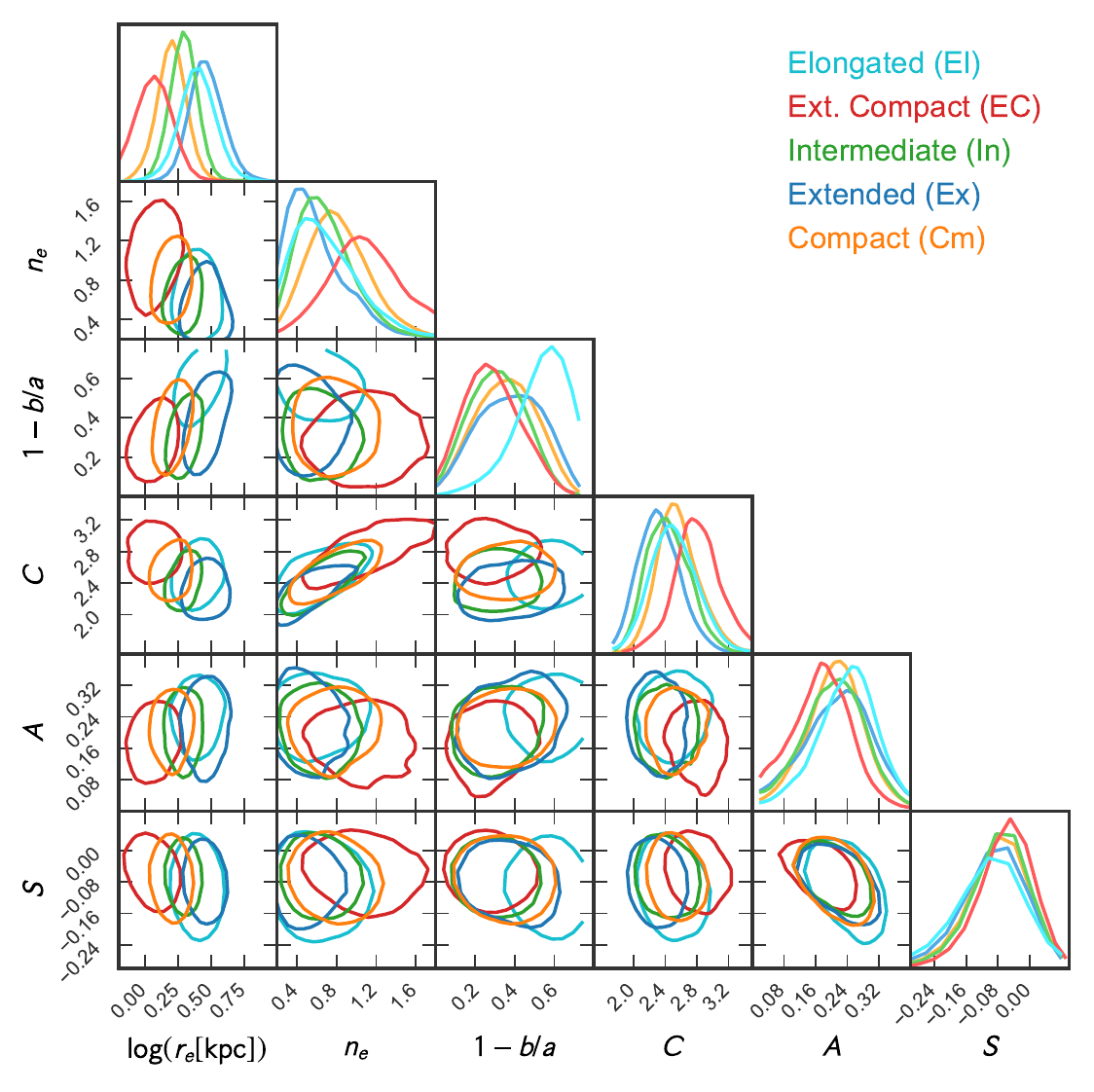}
	
    \caption{Triangle plot with physical properties extracted from the TNG50 simulation (left-hand panel) and morphological (right-hand panel) properties of the morphological clusters shown in the left-hand panel in \autoref{fig:frac_CEERS_2classes}. For clarity, only the $68\%$ contour levels are shown.} 
    \label{fig:morph_props}
\end{figure*}

As an additional way to quantify the differences between TNG50 galaxies and observed CEERS galaxies, we compare the abundances of TNG50 and CEERS galaxies retrieved from the separation into different classes using a clustering technique.

In particular, we apply the $k-$means algorithm to cluster data in the representation space by trying to separate samples in $k$ groups of equal variance, minimizing a criterion known as the inertia or within-cluster sum-of-squares (WCSS). We find $k = 5$ as the optimal number of clusters based on the Elbow curve. The elbow method is a graphical representation of finding the optimal $k$ in a $k-$means clustering. It works by finding WCSS (Within-Cluster Sum of Square), i.e. the sum of the square distance between points in a cluster and the cluster centroid. This result is also confirmed using an alternative method based on the Silhouette score. We implement the Elbow and Silhouette methods using the Yellowbrick package in \textsc{python} (see \citealt{bengfort_yellowbrick_2018}, for more details).

We label the different clusters according to the properties (photometric and morphological) of the galaxies belonging to each of them. In \autoref{fig:morph_props}, we show the properties (and correlations between them) of the different classes. Therefore, we define the following classes: \textit{Extremely Compact (EC)}, \textit{Compact (C)}, \textit{Intermediate (In)}, \textit{Elongated (El)} and \textit{Extended (EX)}. It is clear how \textit{EC} and \textit{C} galaxies show low mass, low angular momentum, large fractions of non-rotating stars, and low flatness. They are also smaller in size with larger S\`ersic index values, rounder in projection, and more compact than the rest of the classes. As going from the \textit{In} class to the \textit{El} and \textit{EX} classes, the masses and sizes of the galaxies progressively increase along with the angular momentum of the stars and the flatness. These classes also exhibit smaller S\`ersic indexes and are less concentrated and more asymmetric than the compact classes. Also interesting is the separation in the projected ellipticity of the \textit{El} class, showing extremely large values of the $1 - b/a$ compared to the rest of the classes.

\begin{figure*}[t!]
%\centering
    \includegraphics[width=0.32\textwidth]{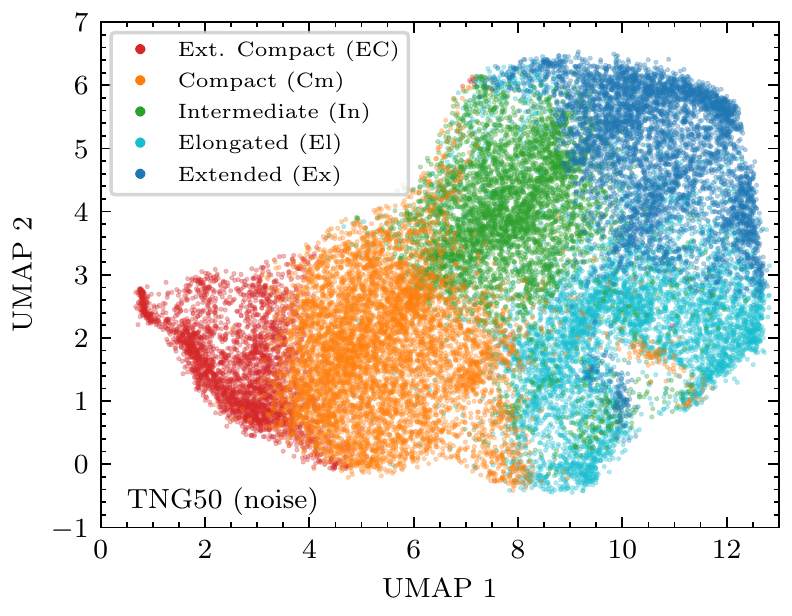}
    \includegraphics[width=0.32\textwidth]{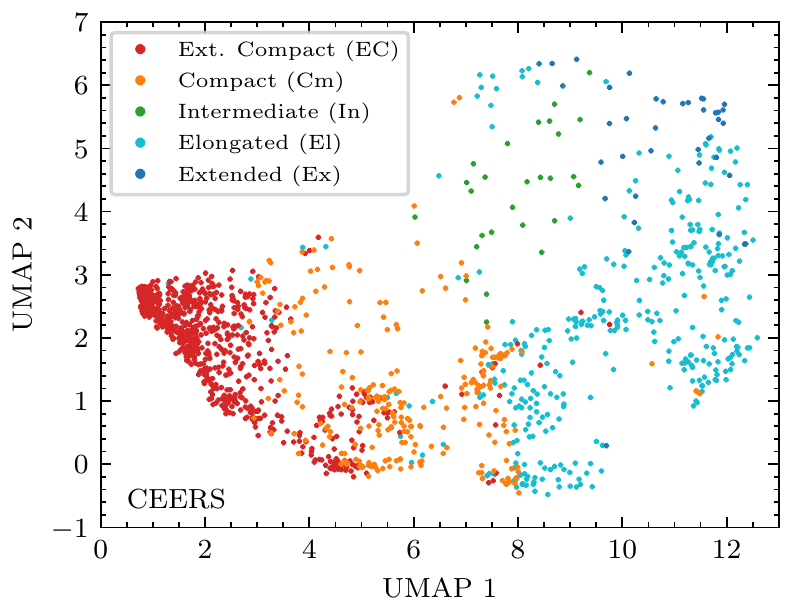}
    \includegraphics[width=0.32\textwidth]{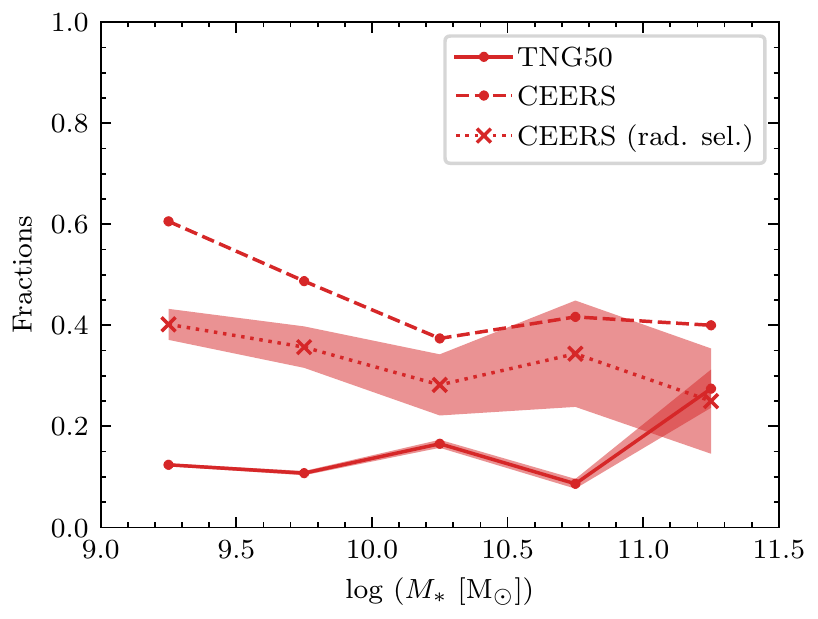}
		
    \caption{Left-hand panel: UMAP visualization of noise-added TNG50 galaxy images color-coded by classes according to the $k$-means method for five clusters. Morphological classes are labeled as: \textit{Extremely Compact} (\textit{EC}, in red), \textit{Compact} (\textit{C}, in orange), \textit{Intermediate} (\textit{In}, in green), \textit{Elongated} (\textit{El}, in cyan) and \textit{Extended} (\textit{EX}, in blue). Middle panel: same as the left-hand panel but for the CEERS galaxy images. Right-hand panel: fractions \textit{Extremely Compact (EC)} galaxies (i.e., those belonging to the EC cluster in red) in TNG50 (solid lines), CEERS (dashed lines) and CEERS with $\log r_e \mathrm{[kpc]} > 0$ (dotted lines) in 5 logarithmic mass bins of width 0.5 dex in the range $9 <= \log (M/\mathrm{M_\odot}) <= 11.5$. The shaded regions correspond to the fraction errors considering Poisson errors in the number of selected galaxies and the total number of galaxies in each mass bin. Note that for clarity we do not show the shaded region for the CEERS dataset without applying any cut in size.} 
    \label{fig:frac_CEERS_2classes}
\end{figure*}

For the simulated TNG50 galaxies, we find: $\sim 12\%$ of \textit{EC} galaxies, $\sim 26\%$ of \textit{Cm} galaxies, $\sim 19\%$ of \textit{In} galaxies, $\sim 22\%$ of \textit{El} galaxies and $\sim 20\%$ of \textit{EX} galaxies. While for the observed CEERS dataset, we find $\sim 55\%$ of \textit{EC} galaxies, $\sim 17\%$ of \textit{Cm} galaxies, $\sim 2\%$ of \textit{In} galaxies, $\sim 25\%$ of \textit{El} galaxies and $\sim 2\%$ of \textit{EX} galaxies. Therefore, and also clear from \autoref{fig:frac_CEERS_2classes}, there is a systematic lack of observed CEERS galaxies in the rest of the classes beyond the \textit{EC} class, with the exception of the \textit{El} class, for which the fractions of observed galaxies are slightly larger than for the simulated ones. In particular, the fractions of observed CEERS galaxies in the \textit{In} and \textit{EX} classes are significantly smaller than those for the simulated TNG50 ones.

In \autoref{fig:frac_CEERS_2classes}, we show the UMAP visualization color-coded by the five classes for the simulated TNG50 and the observed CEERS datasets. Note that galaxies with artifacts or bright companions are not included in the derivation of the different class fractions. Given the division and the correlations shown in \autoref{fig:sims_hexbin_umap} and \autoref{fig:phot_hexbin_umap}, we denote the galaxies located in the left section of the UMAP that belong to the cluster in red as \textit{Extremely Compact (EC)} galaxies, while the remaining galaxies (i.e., those not assigned to the red EC cluster) are considered as \textit{non-compact (NC)} galaxies. We find that on average $\sim 12\%$ of the TNG50 galaxies belong to the EC class. For the CEERS dataset, we find that $\sim 55\%$ of the CEERS galaxies belong to the EC class. 
 In order to mitigate the possible effects of resolution in the simulation, we additionally impose a minimum size threshold for CEERS galaxies and only include CEERS galaxies with $r_e > 1 \mathrm{kpc}$, measured in the F200W filter. This excludes the low-radius tail of $5\%$ of the TGN50 galaxies and $\sim 50\%$ of the CEERS dataset (mainly spheroids since they are typically smaller in size). We find that the fraction of galaxies in the \textit{EC} class for the CEERS dataset is reduced to $\sim 37\%$, but the discrepancy between observations and simulations is still significant.
 
 Our results tend to confirm that the TNG50 model systematically under-predicts the abundance of compact galaxies from a purely data-driven perspective, similar to the findings of e.g. \citet{Flores-Freitas2022}. The effect does not seem a pure consequence of resolution. 
 
\section{A comparison between self-supervised and supervised morphologies}
\label{sec:true_disks}

We now compare in more detail the CNN-based and visual classifications for the CEERS and VISUAL datasets with the self-supervised classifications ---which have been shown to correlate with physical properties--- and speculate about the abundance of disks at $z > 3$. As shown in section~\ref{sec:CEERS}, visually classified disks are spread all over the representation space which suggests that, according to the self-supervised representations, they represent a heterogeneous group of galaxies with different physical properties. 

\subsection{Self-supervised clustering vs. morphological classifications}
\label{sec:compact}

\begin{figure*}[t!]
\centering
\centering
    \includegraphics[width=0.45\textwidth]{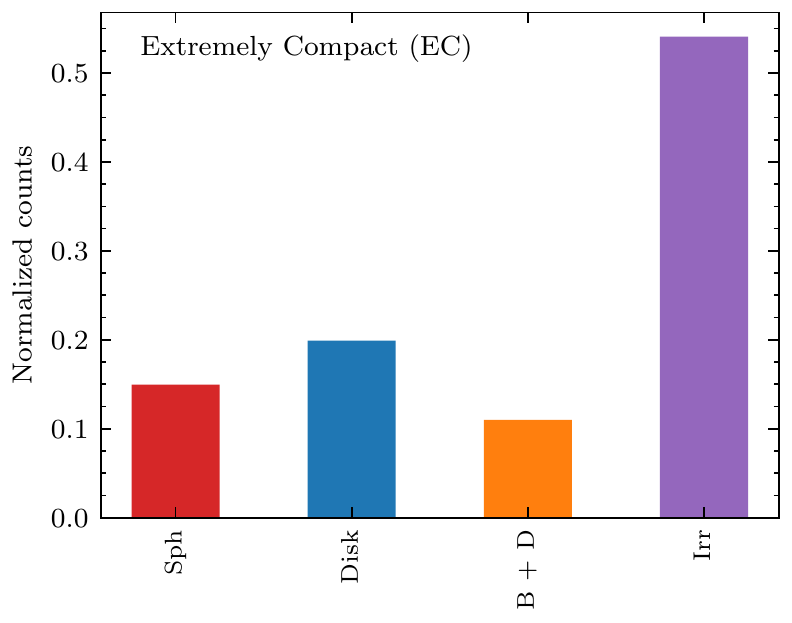}
    \includegraphics[width=0.45\textwidth]{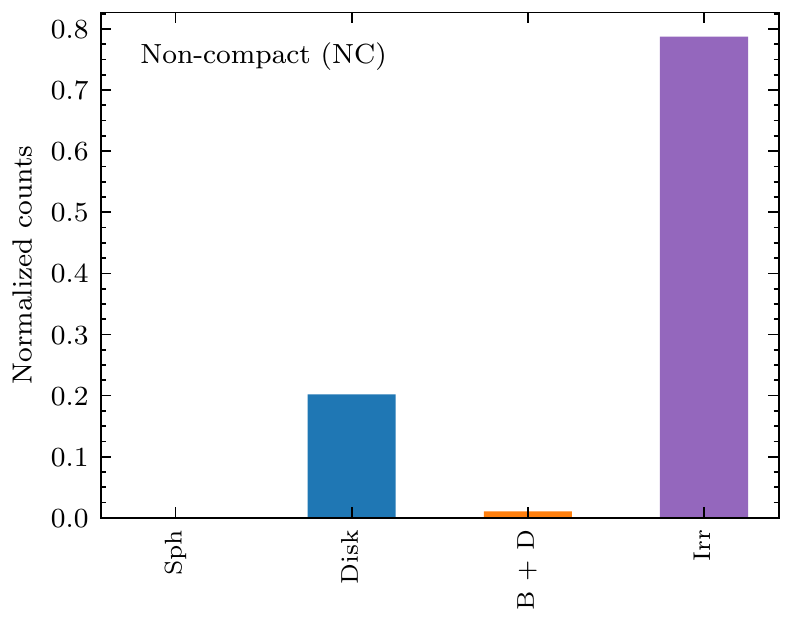}
    
    \caption{Comparison of the classification into \textit{Extremely Compact} (\textit{EC}, left-hand panel) and \textit{non-compact} (\textit{NC}, right-hand panel) galaxies in the CEERS dataset and the CNN-based classifications derived by \citet{Huertas-Company2023}. Histograms are color-coded according to the CNN-based morphological classes into: \textit{Sph} in red, \textit{Disk} in blue, \textit{Bulge + Disk (B+D)} in orange, and \textit{Disturbed (Irr)} in purple.}
    \label{fig:cluster_CEERS}
\end{figure*}

\begin{figure*}[t!]
\centering
\centering
    \includegraphics[width=0.45\textwidth]{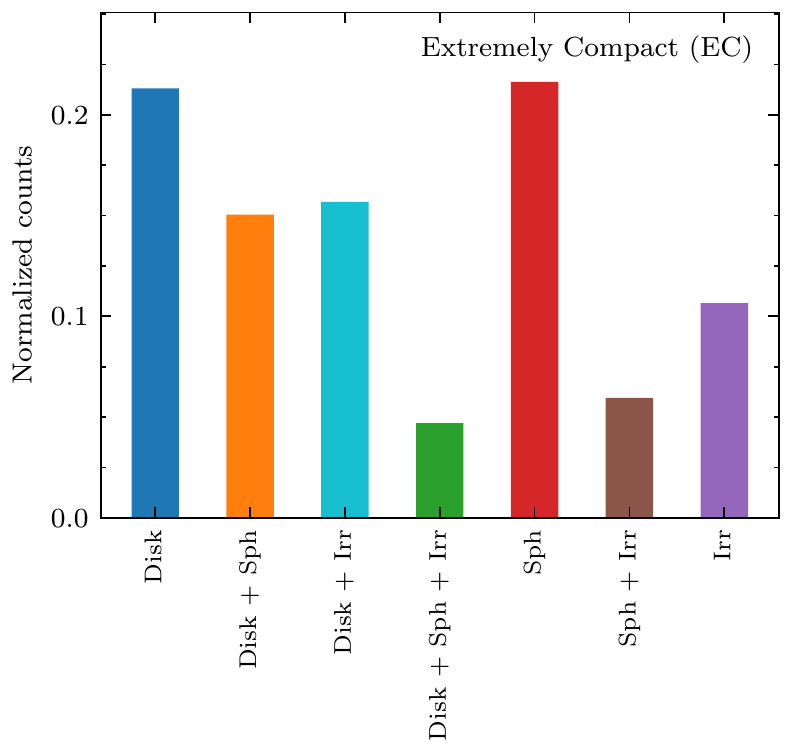}
    \includegraphics[width=0.45\textwidth]{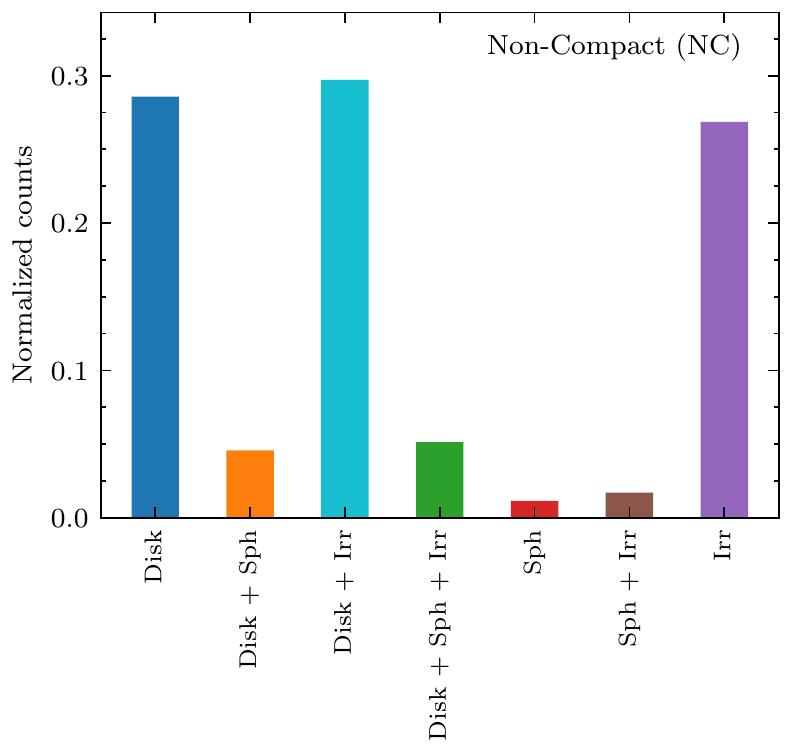}
    
    \caption{Comparison of the classification into \textit{Extremely Compact} (\textit{EC}, left-hand panel) and \textit{non-compact} (\textit{NC}, right-hand panel) galaxies in the VISUAL dataset and the visual classifications derived by \citet{Kartaltepe2022}. Histograms are color-coded according to the visual morphologies as in \autoref{fig:CEERS_umap}.}
    \label{fig:cluster_VISUAL}
\end{figure*}

In Figures ~\ref{fig:cluster_CEERS} and~\ref{fig:cluster_VISUAL}, we show a comparison between the CNN-based and visual classifications with the two broad contrastive learning clusters containing \textit{Extremely Compact (EC)} and \textit{Non Compact (NC)} galaxies, respectively.

The figures show that, for both the CNN and visual classifications, almost all galaxies classified as spheroids or with a bulge component belong to the EC cluster. This is expected as the EC cluster lies in a region of the representation space dominated by round and compact galaxies. However, the trend for disks and irregular galaxies is not so clear. The figures show that both the \textit{EC} and \textit{NC} clusters contain a large fraction of disks and irregular galaxies, which reflect the fact that disks and irregulars are distributed over all the representation space. In fact, we measure that each cluster contains roughly $\sim50\%$ of the \textit{Disk} and \textit{Irr} galaxies for both the CEERS and the VISUAL datasets. This discrepancy between the self-supervised and the traditional classes is somehow surprising and suggests that the population of visually classified disks and irregulars present a large spread of morphological properties that the contrastive learning representation is capturing. We quantify this further in the following. We emphasize that this discrepancy between supervised and self-supervised classifications is independent of the degree of agreement between simulations and observations discussed in~\autoref{sec:comparison_representations}.

\subsection{Two populations of visually classified disks?}
\label{sec:two_populations}

Based on the positions of the visual disks in the contrastive learning representation space, we identify two different populations of visually classified disks which we call \textit{EC Disks} and \textit{NC disks}, for \textit{Extremely Compact} and \textit{Non Compact} disks, respectively. 

\begin{figure*}[t!]
\centering
\centering
    \includegraphics[width=0.32\textwidth]{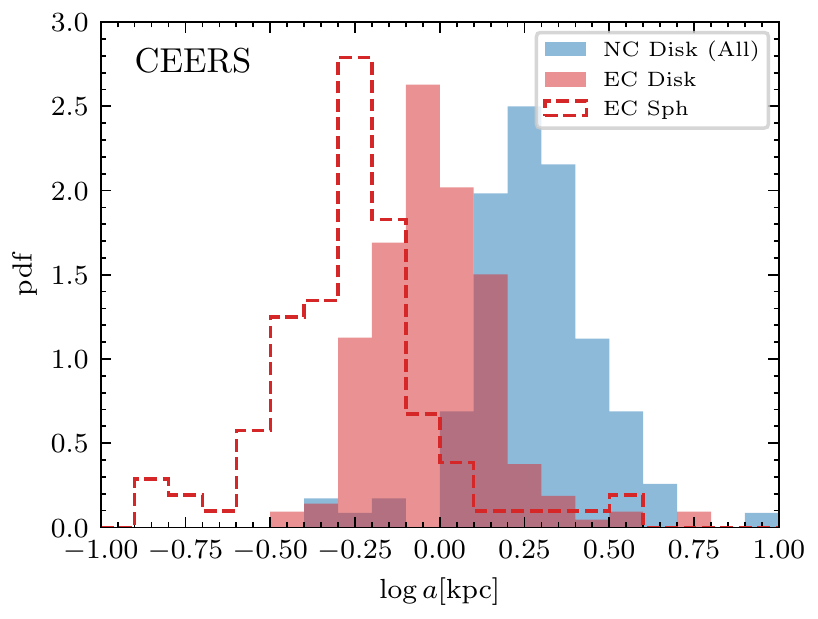}
    \includegraphics[width=0.32\textwidth]{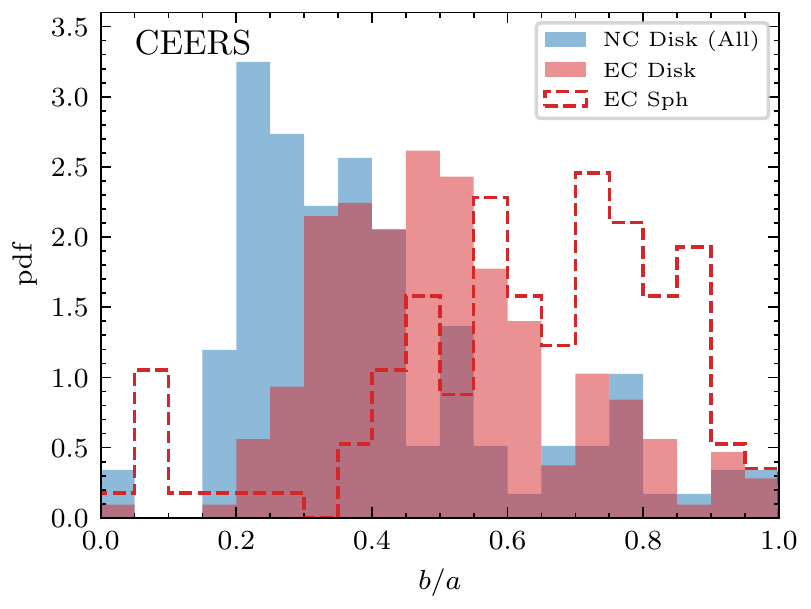}
     \includegraphics[width=0.315\textwidth]{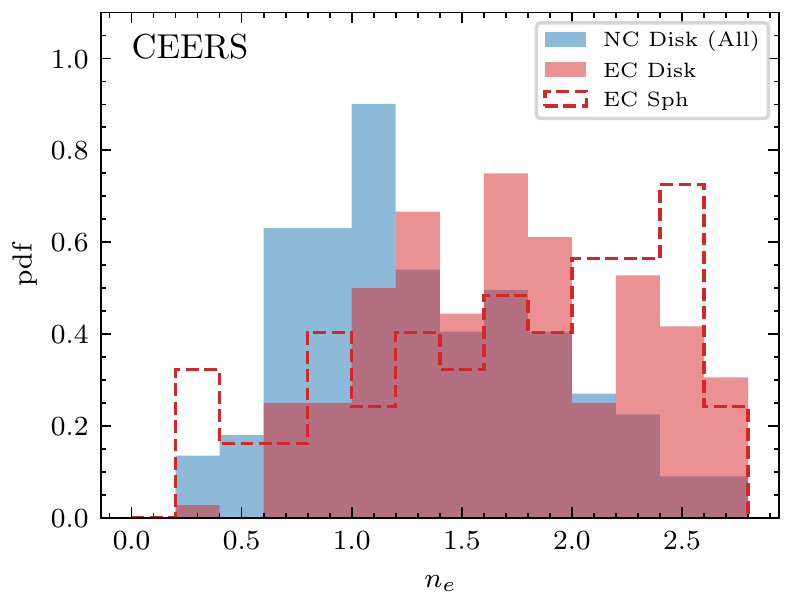}
    \includegraphics[width=0.32\textwidth]{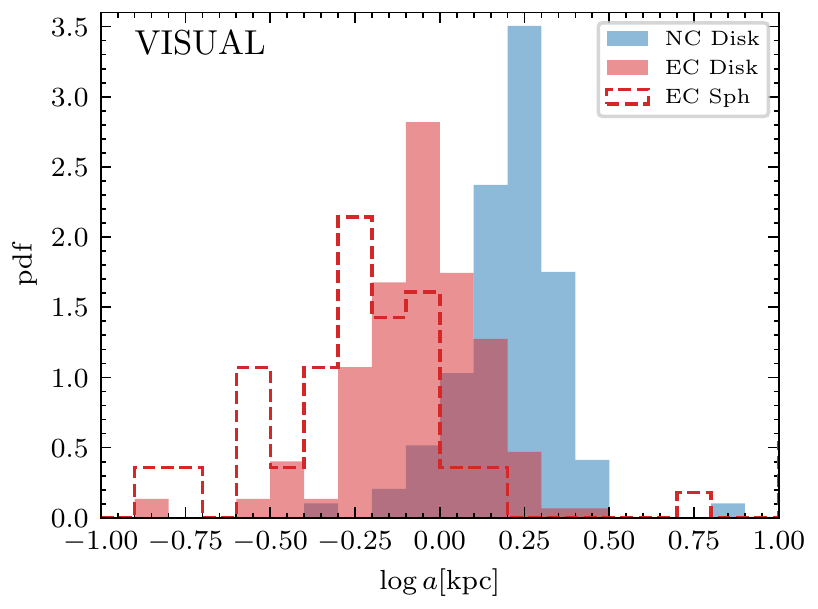}
    \includegraphics[width=0.32\textwidth]{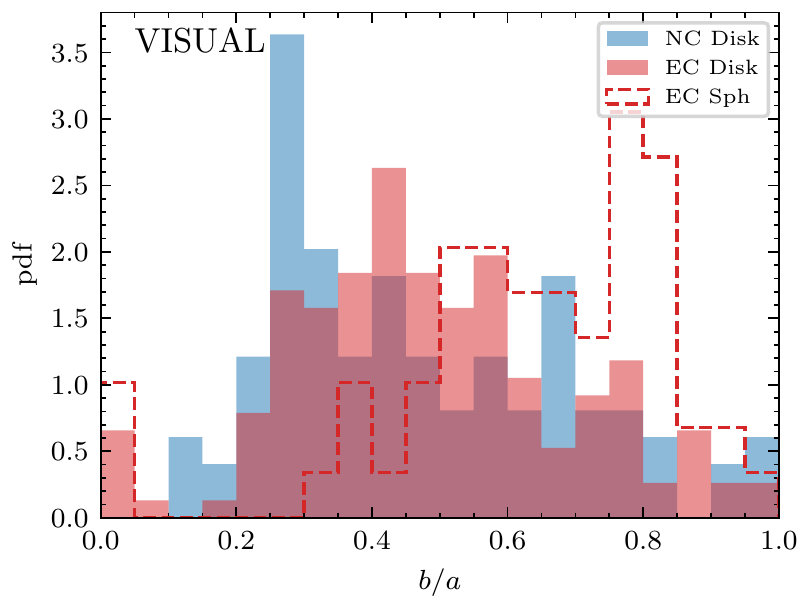}
    \includegraphics[width=0.315\textwidth]{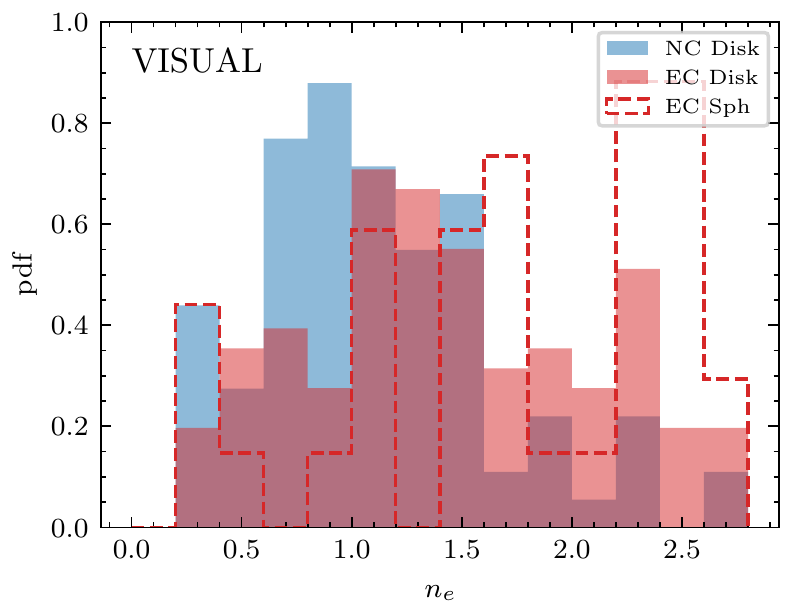}
    \caption{Morphological and photometric properties of the two populations of disk-like galaxies in the CEERS (top panels) and the VISUAL (bottom panels) datasets. From left to right, the different panels show the probability distribution function of the size of the galaxy ($\log a \mathrm{[kpc]}$), the projected ellipticity ($b/a$), and the S\`ersic index ($n_e$), respectively. The different histograms correspond to \textit{EC Disk} candidates (red shaded histograms), \textit{NC Disk} candidates (blue shaded histograms), and \textit{EC Sph} candidates (red dashed histograms). For the CEERS dataset, we consider as disk candidates those classified as \textit{Disk} and \textit{Bulge+Disk}, while for the VISUAL dataset, we consider as disk candidates the four classes of visually classified disks (\textit{Disk}, \textit{Disk + Sph}, \textit{Disk + Irr}, and \textit{Disk + Sph + Irr}). All quantities are derived from the S\`ersic fits in the F356W filter.}
    \label{fig:shape_CEERS}
\end{figure*}

In order to get more insights about why the self-supervised learning algorithm tends to locate them in different clusters, we then examine the properties of the \textit{EC Disks} and \textit{NC Disks} using parametric morphologies obtained via S\`ersic fitting in the F356W filter (see \autoref{sec:data} for more details). In \autoref{fig:shape_CEERS}, we show the distributions of axis-ratios ($b/a$), semi-major axes ($a$), and S\`ersic indices ($n_e$) for the two disk classes, and for the CEERS and VISUAL datasets. We also show, for reference, the same distributions for galaxies visually classified as spheroids.  

The figure clearly shows different distributions. As expected, \textit{EC Disks} have smaller effective radii. The \textit{NC Disks} exhibit a distribution of $\log a$ that peaks at $\log a \sim 0.25$, while for the  \textit{EC Disks} it peaks at smaller values of $\log a \sim -0.10$. The distribution of $n_e$ is also different and reflects that \textit{EC Disk} are more concentrated. For \textit{NC Disks}, it peaks at $n_e \lesssim 1.2$, characteristic of an exponential profile. However, for \textit{EC Disks}, the distribution is skewed towards larger values of the S\`ersic index, although smaller than for spheroids. Regarding the $b/a$ distribution, the \textit{NC Disks} tend to be more elongated, and, as expected, spheroids are rounder on average. The \textit{EC Disk} candidates are in between, with a peak at $b/a\sim0.5$. 

In view of these differences and the correlations between structural properties and the positions in the representation space highlighted in previous sections, it is expected that the \textit{NC} and \textit{EC Disks} fall in different regions of the parameter space, even if they share the same visual classification. It also makes sense that these \textit{EC Disks} are generally classified as disks by expert classifiers given the proposed classification scheme, in the sense that they do not show strong signs of irregularities and are more elongated and less concentrated than pure spheroids. Therefore, by default, they end up in the disk class. In \autoref{app:CEERS}, we show examples of several \textit{NC Disks}  and \textit{EC Disks} candidates (including some cases with $b/a < 0.3$) along with various \textit{EC Sph} candidates for comparison.

The difference between the completely supervised and self-supervised classifications illustrates how a purely data-driven approach can offer an alternative description of the information content of the data. It is also important to highlight that the difference between the two types of classifications is not only driven by differences in size. The distributions of $\log a$ plotted in \autoref{fig:shape_CEERS} overlap significantly. In addition, if we only include in the analysis galaxies with $r_e > 1 \mathrm{kpc}$, there is still a significant fraction ($\sim 40\%$) of visually classified disks in the \textit{EC} cluster.

An interesting question, therefore, is whether \textit{NC} and \textit{EC Disks} are all \emph{true} disks ---if by disk we understand a flat system predominantly supported by the rotation of the gas and stars--- as this has important implications on the frequency of disk formation at these very early cosmic epochs. It is certainly impossible to unambiguously answer this question with the data at hand. However, we can speculate based on the information inferred from the TNG50 simulation. As shown in the previous subsections, there is a strong correlation between the contrastive learning representations and the stellar specific angular momentum. Interestingly, the location of \textit{EC Disks} is predominately populated by low angular momentum galaxies, which would suggest that these systems are preferentially velocity dispersion supported. 

As an additional attempt to shed some light on the physical properties of \textit{EC} and \textit{NC} Disks, we explore in more detail how the contrastive learning representation space distributes galaxies with different 3D stellar structures. We characterize the shape of galaxies in the TNG50 sample as done and studied in \citet{Pillepich2019}, i.e., with an ellipsoid with three semi-axes $c < b < a$ and use the axial ratios $q = b/a$ (intermediate-to-major) and $s = c/a$ (minor-to major) to define three main 3D shape classes: oblate, spheroid and prolate following the definitions of \cite{vanderWel2014} and \cite{Zhang2019}. The axial ratios are derived after diagonalizing the stellar mass tensor in an iterative way while keeping the major axis length fixed to $2r_*$. We consider oblate or disk galaxies those with $a\sim b>c$, elongated or prolate objects those with $a>b\sim c$, and spheroidal systems those with similar values for the three semi-axes. Note that, by definition, the 20 projections of the TNG50 galaxies have the same 3D shape. At these redshifts, we find that $\sim 67\%$ of the galaxies in the simulation have a spheroidal shape according to this definition. Only $18\%$ and $15\%$ of the galaxies have oblate and prolate shapes, respectively. The fact that at high redshift and low stellar masses, galaxies tend to present a more prolate structure has been found both in observations~\citep{vanderWel2014,Zhang2019} and simulations~\citep{2016MNRAS.458.4477T,Pillepich2019}. 

\begin{figure*}[t!]
\centering
    \includegraphics[width=0.32\textwidth]{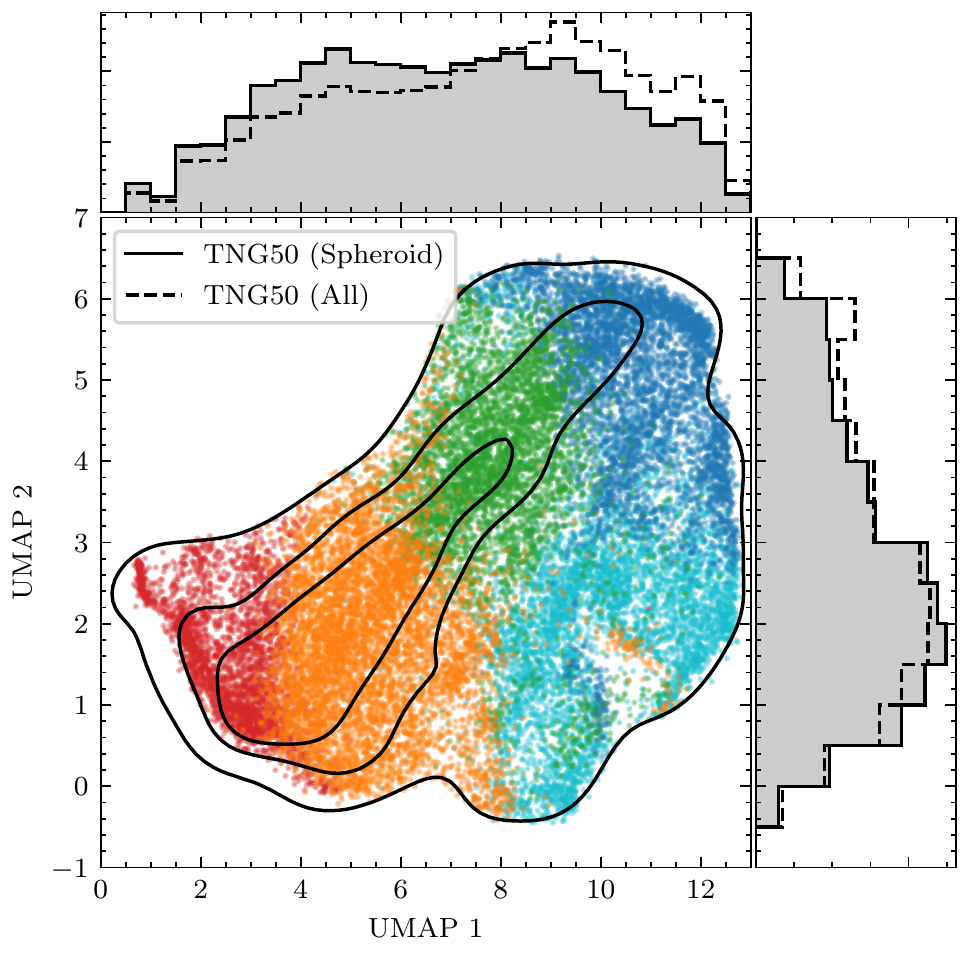}    
    \includegraphics[width=0.32\textwidth]{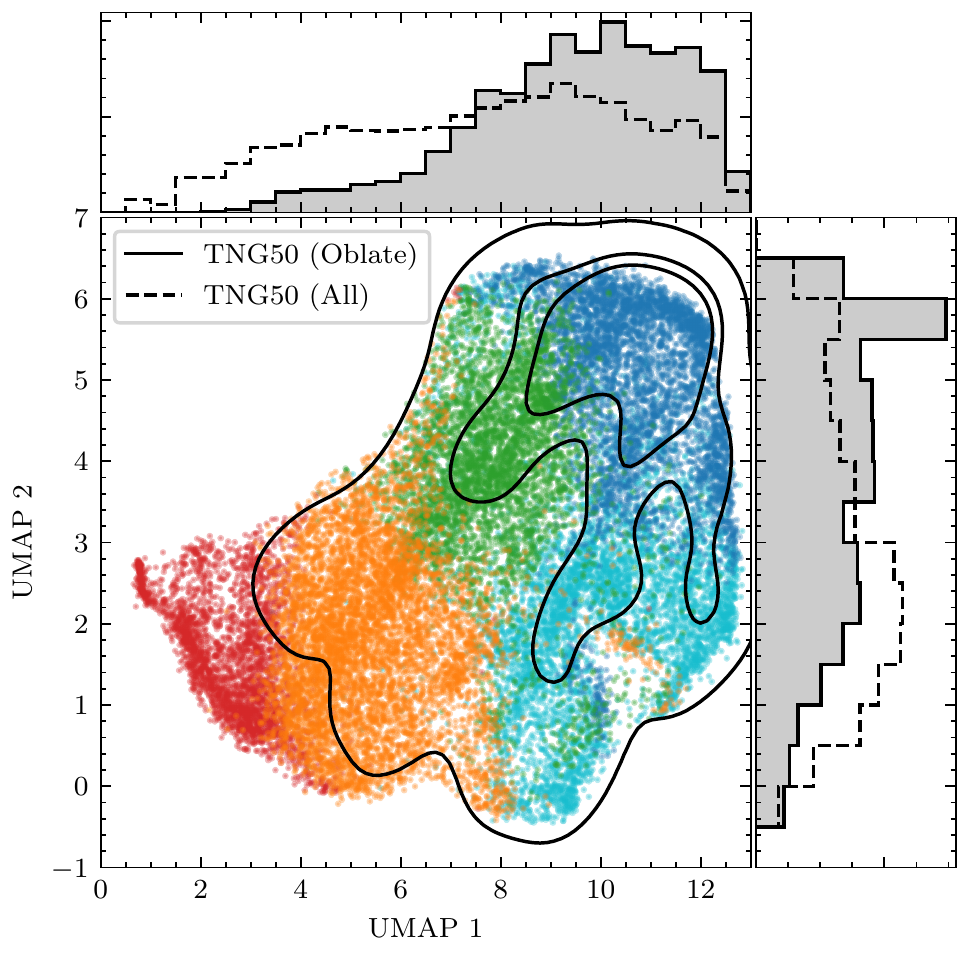}    
    \includegraphics[width=0.32\textwidth]{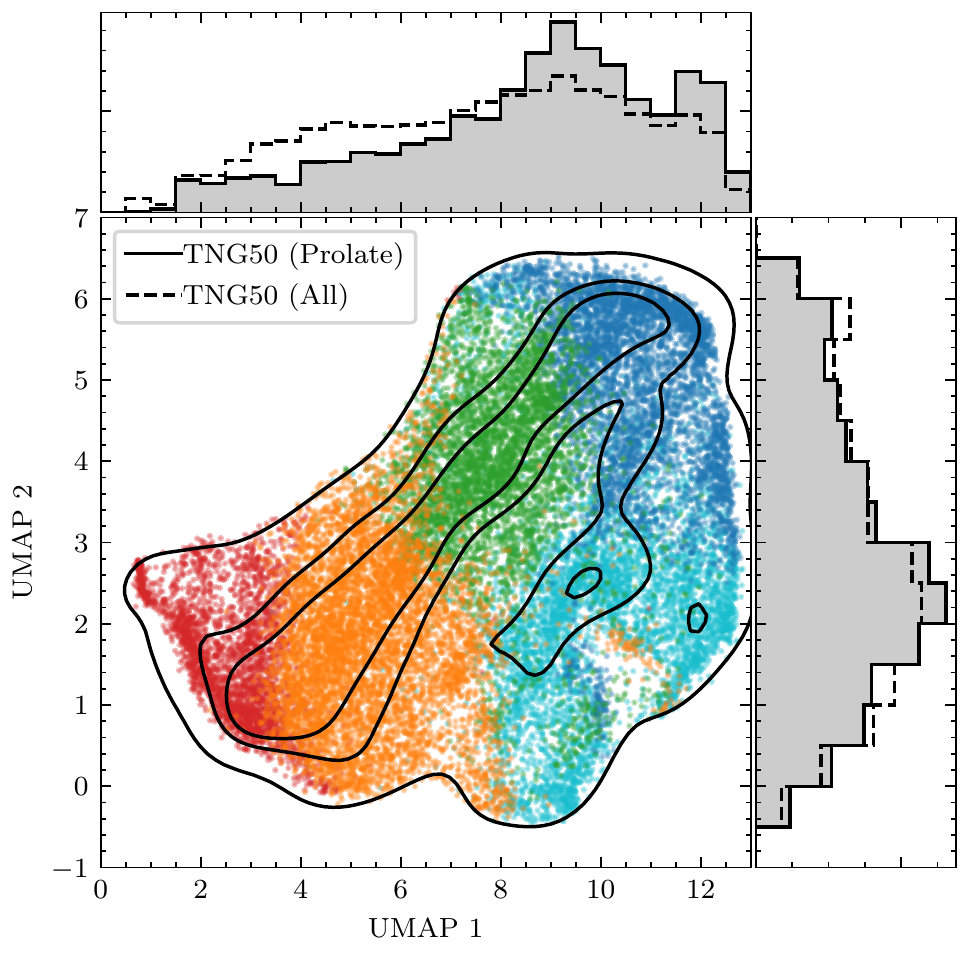}    
		
    \caption{Location in the UMAP plane of TNG50 galaxies according to the 3D shape inferred from the stellar particles: spheroid (left-hand panel), oblate (middle panel), and prolate (right-hand panel). Points are color-coded according to the morphological classes described in \autoref{sec:morphological_classes}. The contour levels indicate the $25\%$, $50\%$, and $95\%$ probabilities. Unfilled black dashed and filled solid histograms in the horizontal and vertical axes show the PDF of the whole dataset and the shape-selected galaxies of the TNG50 dataset, respectively. Morphologically disk (i.e., oblate) galaxies tend to populate the right and upper-right regions of the UMAP plane, while it is unlikely to find them in the left section of the UMAP plane.} 
    \label{fig:umap_3D_shape}
\end{figure*}

\begin{figure}[t!]
\centering
    \includegraphics[width=\columnwidth]{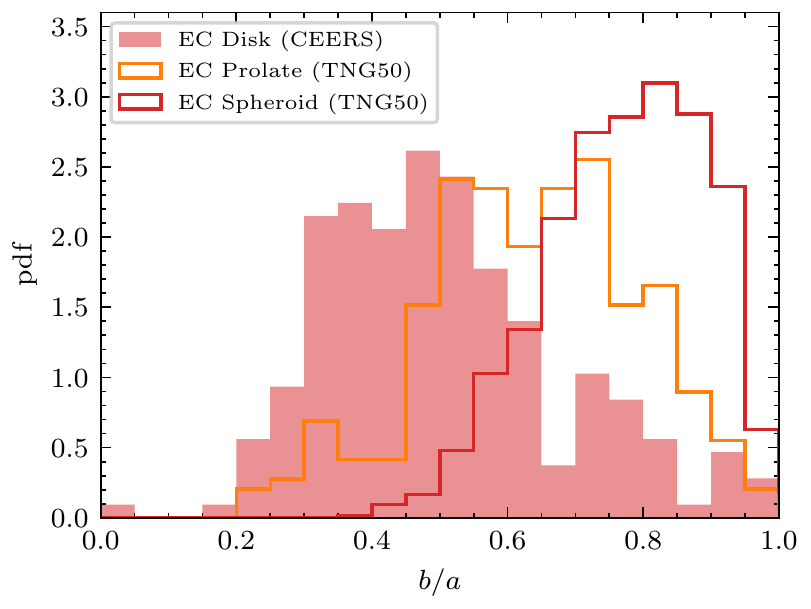}    
		
    \caption{Distributions of projected ellipticity ($b/a$) for \textit{EC Disk} candidates in CEERS (red shaded histogram), and \textit{EC} prolate and spheroid TNG50 candidates (orange and red empty histograms, respectively). Elongated galaxies in projection (i.e., $b/a \lesssim 0.4$) are compatible with being prolate in shape according to the TNG50 \textit{EC} prolate candidates.} 
    \label{fig:TNG50_compact_prolate}
\end{figure}

We show in~\autoref{fig:umap_3D_shape} the distribution of the mock images of oblate galaxies in the UMAP projection of the representation space obtained with contrastive learning. For clarity, we do not show the distributions of prolate and spheroidal galaxies. Interestingly, the bottom right corner, where the \textit{EC Disks} lie, does not contain almost any oblate system. The result would suggest that \textit{EC Disks} are very unlikely to be true disks (i.e. flat rotating systems) even if they appear as elongated in the images. Despite the small number of TNG50 galaxies with $b/a \lesssim 0.4$, we find that \textit{EC} TNG50 galaxies with $b/a \lesssim 0.4$ are more likely to be prolate than spheroid in shape, as shown in \autoref{fig:TNG50_compact_prolate}.

However, a big caveat is that the previous statements rely of course on the calibration with the TNG50 simulation, which as shown in the previous sections, does not properly reproduce the observed morphological diversity. Therefore, no firm conclusion can be established without more observations and comparisons with other state-of-the-art simulations. As a matter of fact, an alternative explanation for the different representations of \textit{EC Disks} and \textit{NC Disks} could be that the simulation cannot generate compact disks because of resolution issues, as already acknowledged. This would imply that the location of the \textit{EC Disks} in the representation space cannot be interpreted in terms of physical properties, as these systems simply do not exist in the simulation. We expect the fine-tuning of the model presented in \autoref{sec:contrastive} should mitigate this effect, but cannot be guaranteed. Nevertheless, our data-driven analysis suggests that robustly establishing the fraction of disk galaxies at $z>3$ remains an open issue. 

%%%%%%%%%%%%%%%%%%%%%%%%%%%%%%%%%%%%%%%%%%%%%%%%%%%
\section{Summary and Conclusions}
\label{sec:clonlusions}

This work presents a novel data-driven method based on contrastive learning to infer the morphological properties of galaxies observed with JWST. The method is calibrated on mock JWST galaxy images extracted from the TNG50 cosmological simulation that, thanks to its large number of qualitatively realistic galaxies, allows us to produce a morphological description ---without any assumption on galaxy classes--- robust to background noise, signal-to-noise, color, and orientation. In addition, we show that the obtained representations of galaxies based on their images correlate well with some other physical properties inferred from the simulation (such as the specific angular momentum of stars, $j_*$, and the intrinsic 3D shape) along with some measured photometric and structural properties (such as S\`ersic index and the projected ellipticity). 

We have applied the method to JWST images from the CEERS survey in the F200W and F356W bands of: 1) a mass-complete sample ($M_* \ge 10^9 \mathrm{M_{\odot}}$) of galaxies at $3<z<6$ in the CEERS survey for which CNN-based classifications are available; and 2) a mass- and a redshift-selected sample of CEERS galaxies at $3<z<6$ with $M_* \ge 10^9 \mathrm{M_{\odot}}$ for which visual morphological classifications are available. 

Our main results are:

\begin{itemize}

\item Simulated galaxies from the TNG50 cosmological simulation seem to cover well the observed morphological diversity at $z>3$. However, the morphological distributions of CEERS and simulated galaxies are measured to be different. When compared at the pixel level, simulated and observed galaxies seem to populate in different proportions in the different regions of the TNG50-trained manifolds. We show that these differences can be at least partly explained because observed galaxies can be more compact and more elongated than simulated ones. In fact, the galaxy-to-galaxy variation in sizes, S\`ersic indices, and shapes at fixed stellar mass and redshift are larger in the observed CEERS population than in TNG50 simulated ones. These differences might be partly explained by the limited resolution of the TNG simulation, but not completely since the discrepancies are not erased when only large galaxies are considered. 

\item Our morphological description also suggests that CNN-based and visually classified disks comprise two different populations: one made of \textit{extremely compact (EC) Disks} and another of \textit{non-compact (NC) Disks}. Half of the galaxies that are classified as disks are indeed located in the representation space very close to compact spheroids and, therefore, are more consistent with not being pure disk-like galaxies (i.e., having a prolate or spheroidal stellar structure). Although some of these conclusions might be affected by the calibration with the TNG50 model, our study suggests that some \textit{EC Disk} candidates can be misclassified as disks, as they appear (on average) more elongated in the images than \textit{EC Sph} candidates. The coexistence of prolate and oblate systems at high redshift is in qualitative agreement with the predictions of other models (e.g. zoom-in simulations) which also found that low-mass galaxies at high-$z$ tend to present a prolate shape \citep{2015MNRAS.453..408C, 2016MNRAS.458.4477T}. More in-depth follow-up of these two populations of galaxies, possibly with spectroscopy and additional comparisons with other simulations, is required to further constrain their true nature.

\end{itemize}

%%%%%%%%%%%%%%%%%%%%%%%%%%%%%%%%%%%%%%%%%%%%%%%%%%%
\section*{Acknowledgments}

The authors acknowledge the referee for their comprehensive and constructive comments that played a crucial role in enhancing the accuracy, clarity, and exhaustiveness of the paper. MHC thanks Shy Genel, David Koo, and Sandy Faber for insightful discussions. JVF, MHC, RS, and JHK acknowledge financial support from the State Research Agency (AEI\-MCINN) of the Spanish Ministry of Science and Innovation under the grant ``Galaxy Evolution with Artificial Intelligence" with reference PGC2018-100852-A-I00 and under the grant ``The structure and evolution of galaxies and their central regions" with reference PID2019-105602GB-I00/10.13039/501100011033, from the ACIISI, Consejer\'{i}a de Econom\'{i}a, Conocimiento y Empleo del Gobierno de Canarias and the European Regional Development Fund (ERDF) under grants with reference PROID2020010057 and PROID2021010044, and from IAC projects P/300724 and P/301802, financed by the Ministry of Science and Innovation, through the State Budget and by the Canary Islands Department of Economy, Knowledge and Employment, through the Regional Budget of the Autonomous Community. JVF and FB also acknowledge support from the grant ``Galactic Edges and Euclid in the Low Surface Brightness Era (GEELSBE)" with reference PID2020-116188GA-I00 financed by the Spanish Ministry of Science and Innovation. LC acknowledges financial support from Comunidad de Madrid under Atracci\'on de Talento grant 2018-T2/TIC-11612 and Spanish Ministerio de Ciencia e Innovaci\'on MCIN/AEI/10.13039/501100011033 through grant PGC2018-093499-B-I00. This work was supported by the AWS Cloud Credits for Research Program.

%%%%%%%%%%%%%%%%%%%%% REFERENCES %%%%%%%%%%%%%%%%%%


\bibliographystyle{aasjournal}
\bibliography{bibliography} % if your bibtex file is called example.bib

%\begin{thebibliography}{99}
%\end{thebibliography}

%%%%%%%%%%%%%%%%%%%%%%%%%%%%%%%%%%%%%%%%%%%%%%%%%%%
%
%%%%%%%%%%%%%%%%%% APPENDICES %%%%%%%%%%%%%%%%%%%%%
%
\restartappendixnumbering

\appendix

\section{Scatter of physical and photometric parameters in the UMAP visualization}
\label{app:dependence}

\begin{figure*}[t!]
\centering
	\includegraphics[width=0.45\textwidth]{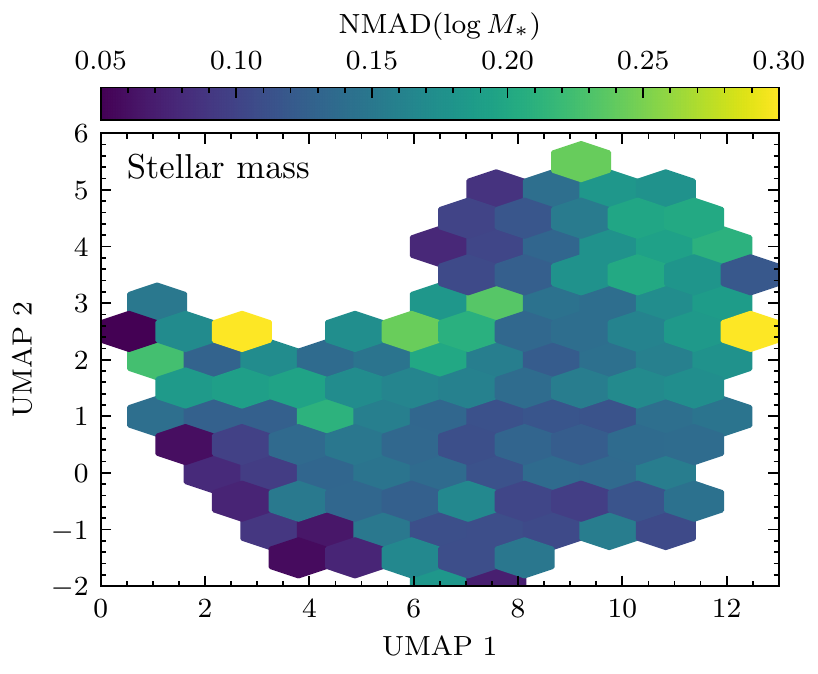}
	\includegraphics[width=0.45\textwidth]{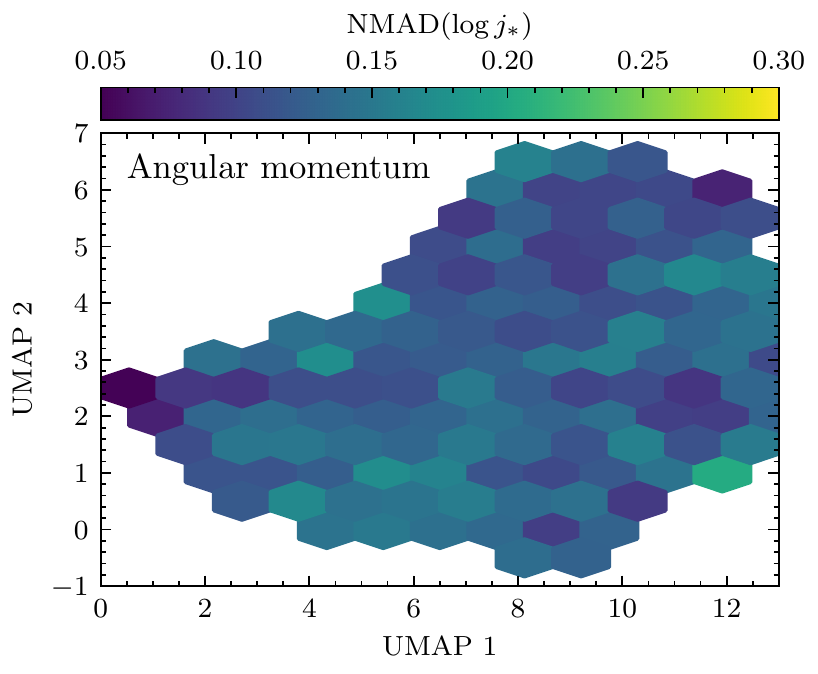}
        \includegraphics[width=0.45\textwidth]{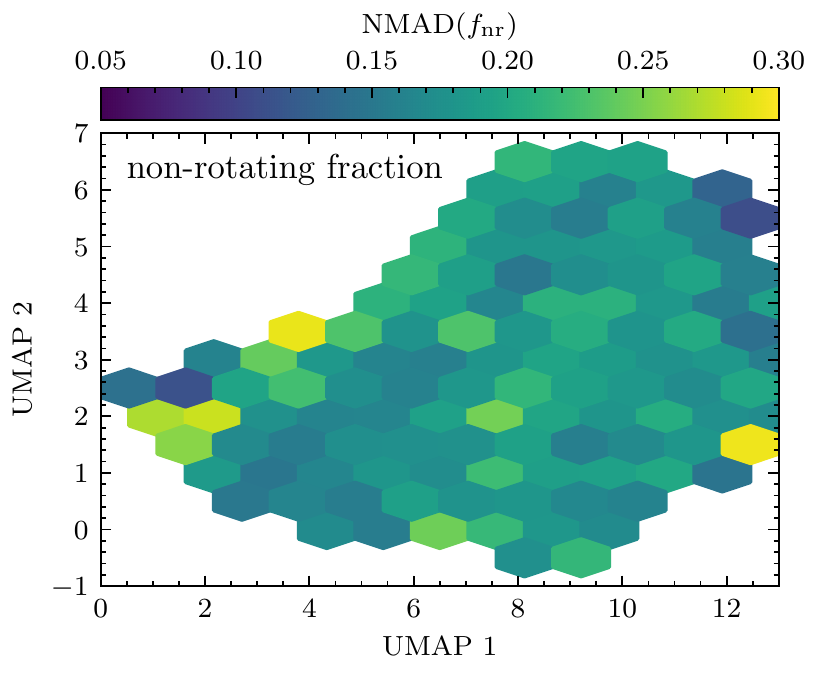}
	\includegraphics[width=0.45\textwidth]{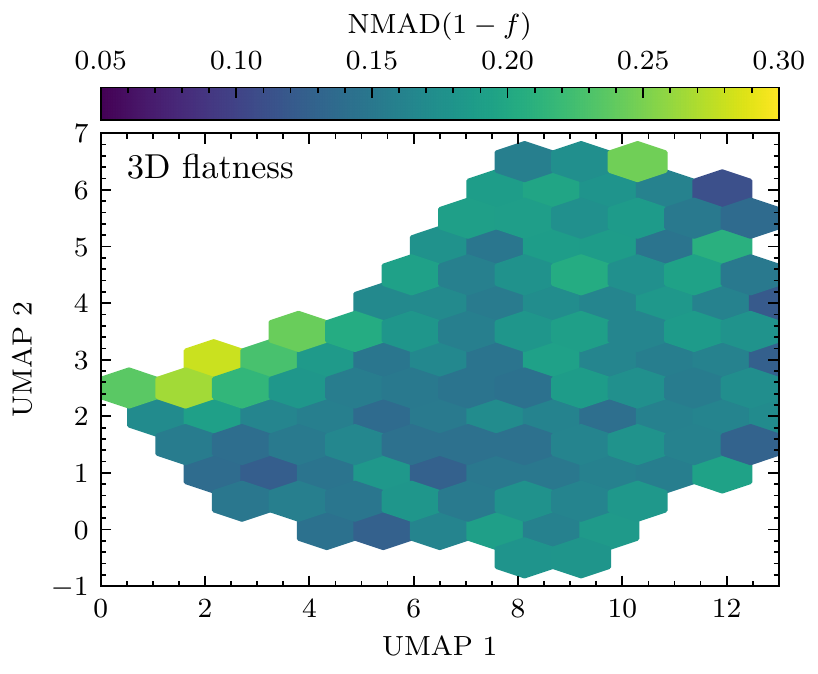}
	
    \caption{UMAP visualization for all the TNG50 galaxy images in our dataset color-coded by the distribution of several physical properties extracted from the TNG50 simulation. From left to right and top to bottom, the different panels show the UMAP plane color-coded by the NMAD of: the logarithm of the total stellar mass ($\log M_* \mathrm{[M_\odot]}$), the logarithm of the specific angular momentum of the stars ($\log j_* \mathrm{[kpc~km~s^{-1}]}$), the mass fraction in non-rotating stars ($f_\mathrm{nr}$) and the galaxy flatness ($1-f$).} 
    \label{fig:sigma_sims_hexbin_umap}
\end{figure*}

\begin{figure*}[t!]
\centering
	\includegraphics[width=0.32\textwidth]{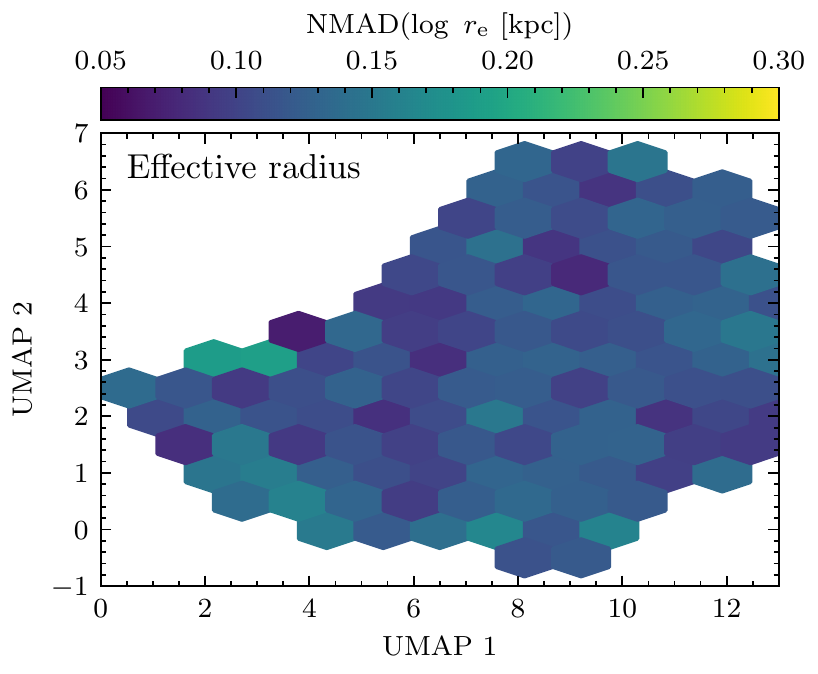}
	\includegraphics[width=0.32\textwidth]{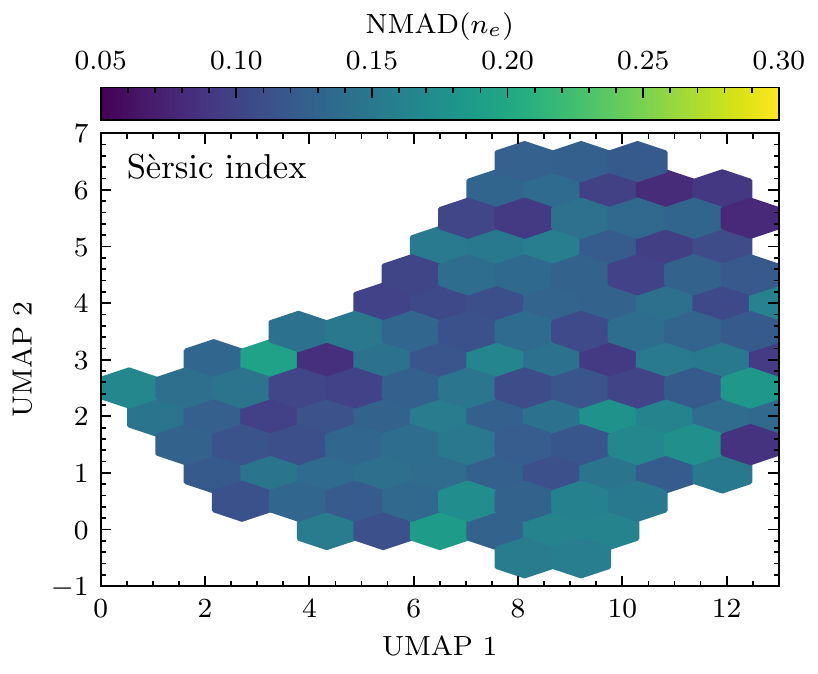}
	\includegraphics[width=0.32\textwidth]{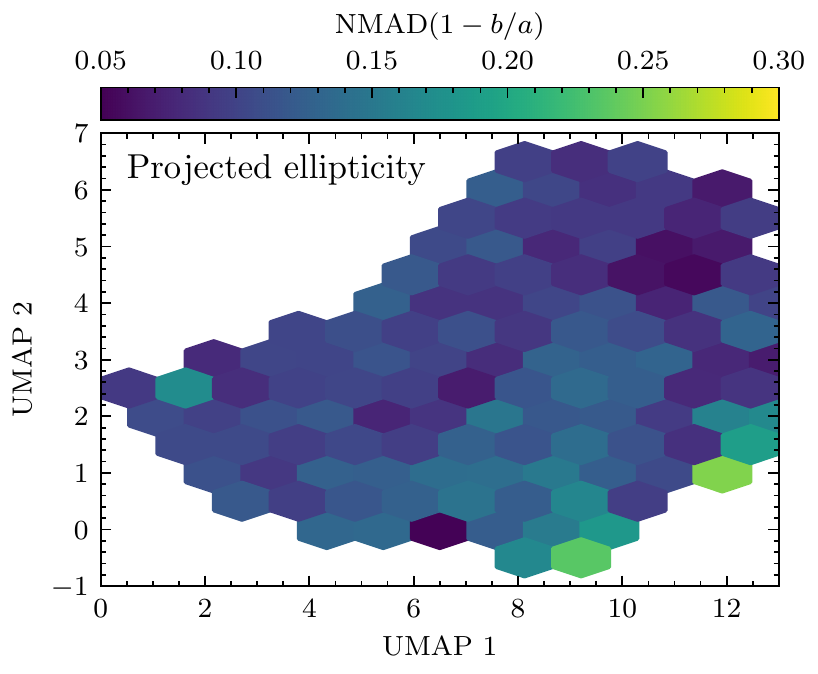}
	\includegraphics[width=0.32\textwidth]{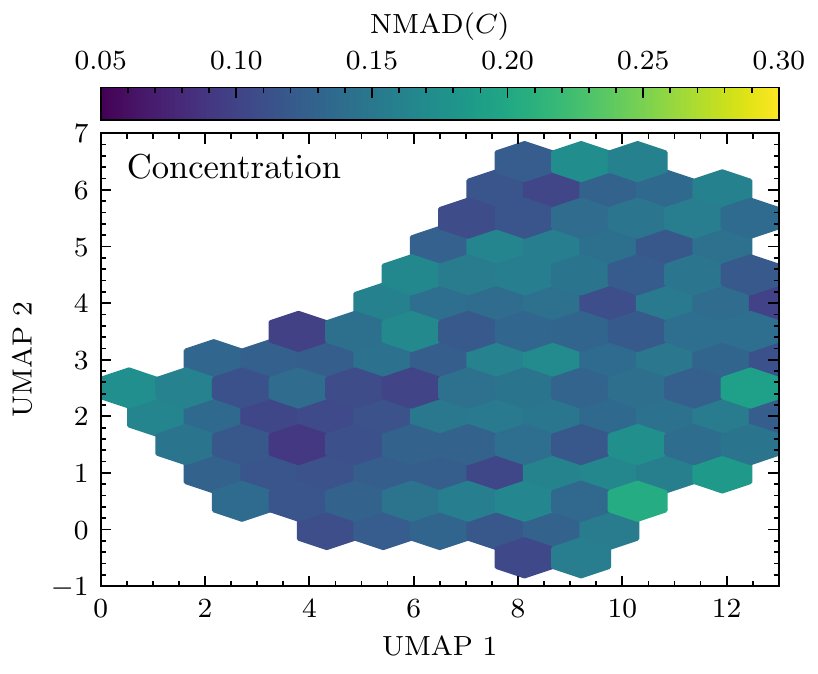}
	\includegraphics[width=0.32\textwidth]{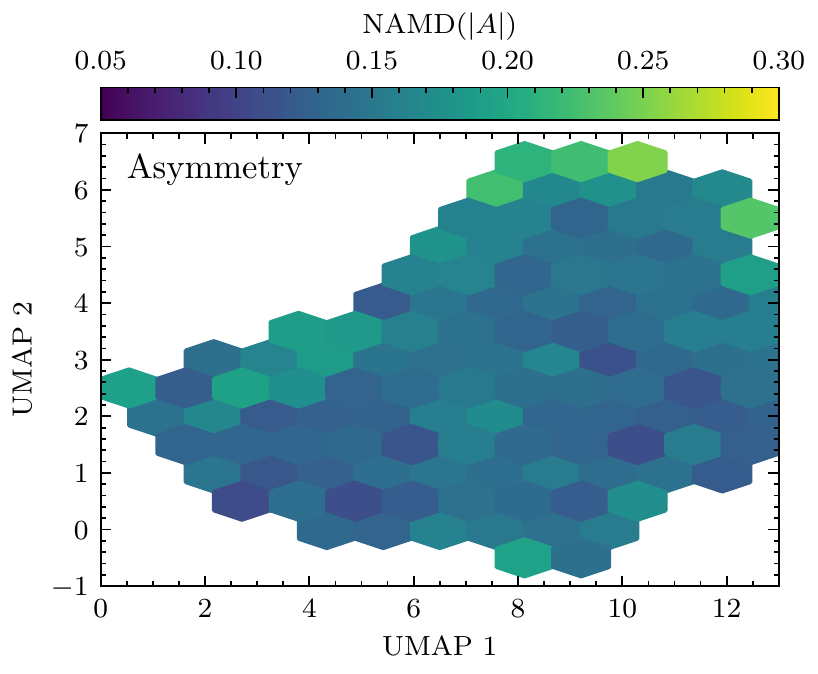}
	\includegraphics[width=0.32\textwidth]{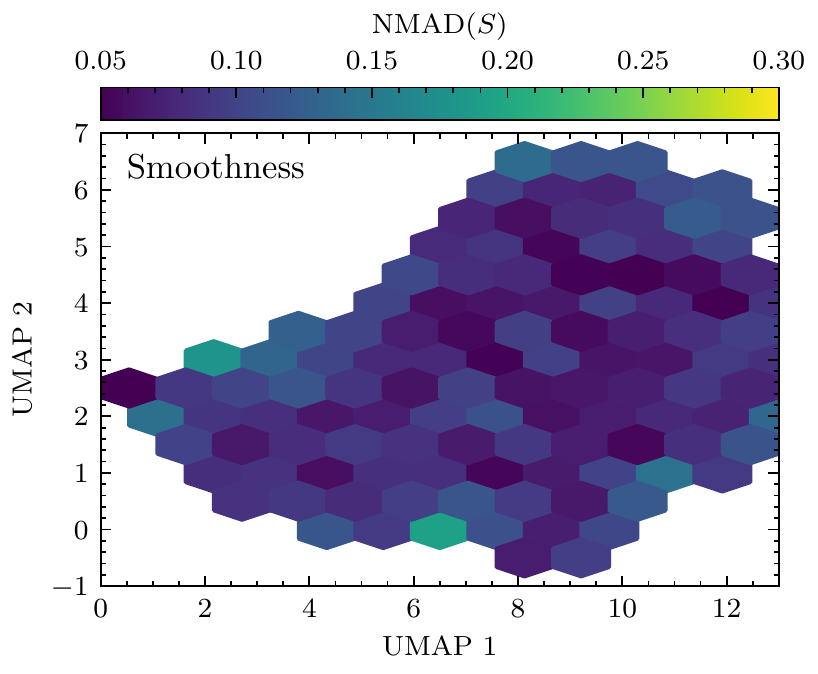}

    \caption{UMAP visualization for all the TNG50 galaxy images in our dataset color-coded by the distribution of several morphological and photometric parameters. From left to right and top to bottom, the different panels show the UMAP plane color-coded by the NMAD of: the logarithm of the effective radius ($r_e \mathrm{[kpc]}$), the S\`ersic index ($n_e$), the ellipticity based on S\`ersic fit ($1-b/a$), the concentration ($C$), the asymmetry ($A$) and the smoothness ($S$).} 
    \label{fig:sigma_phot_hexbin_umap}
\end{figure*}

In \autoref{fig:sigma_sims_hexbin_umap} and \autoref{fig:sigma_phot_hexbin_umap}, we show the variability in the physical and photometric parameters, respectively, in the UMAP visualization shown in \autoref{fig:phot_hexbin_umap} in \autoref{sec:dependence}. The scatter is quantified as a normalized median absolute deviation, denoted here as NMAD. The median absolute deviation (MAD), defined as $\mathrm{MAD (y)} = median(|y - median(y)|)$, is a robust measure of the variability of a univariate sample of quantitative data. The MAD is less affected by outliers and non-gaussianity than the typical variance and standard deviation. To facilitate the comparison between different variables, we normalized the MAD by the dynamical range of the data, defined as the percentile range containing $98\%$ of the data. The resulting normalized MAD, denoted as NMAD, is an indicator of the variability of the data that, in this case, shows how informative the correlation with the different parameters shown in \autoref{fig:sims_hexbin_umap} and \autoref{fig:phot_hexbin_umap} are. Values of NMAD $\lesssim 0.2$ are indicative of a low variability of the data.

%%%%%%%%%%%%%%%%%%%%%%%%%%%%%%%%%%%%%%%%%%%%%%%%%%%

\section{Examples of observed JWST galaxy images}
\label{app:CEERS}

\begin{figure*}[t!]
\centering
    \includegraphics[width=0.75\textwidth]{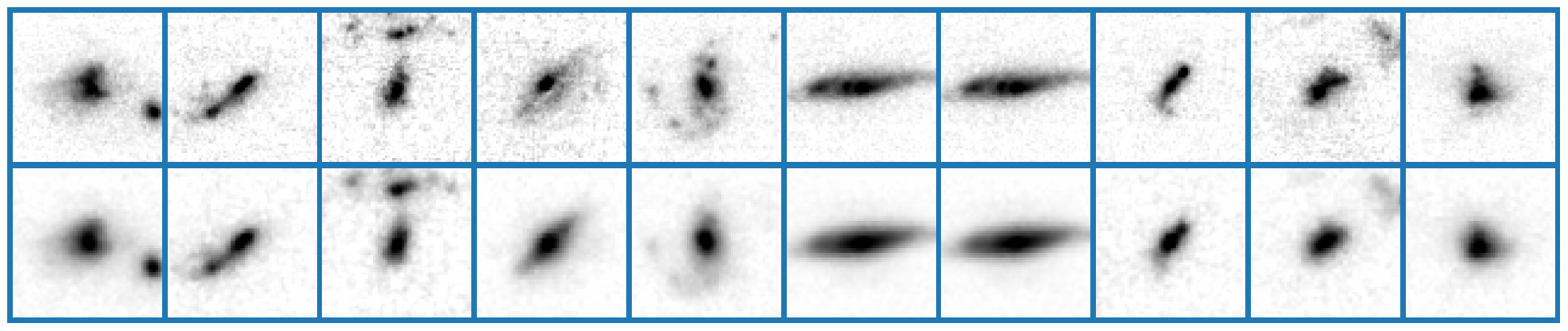}
    \includegraphics[width=0.75\textwidth]{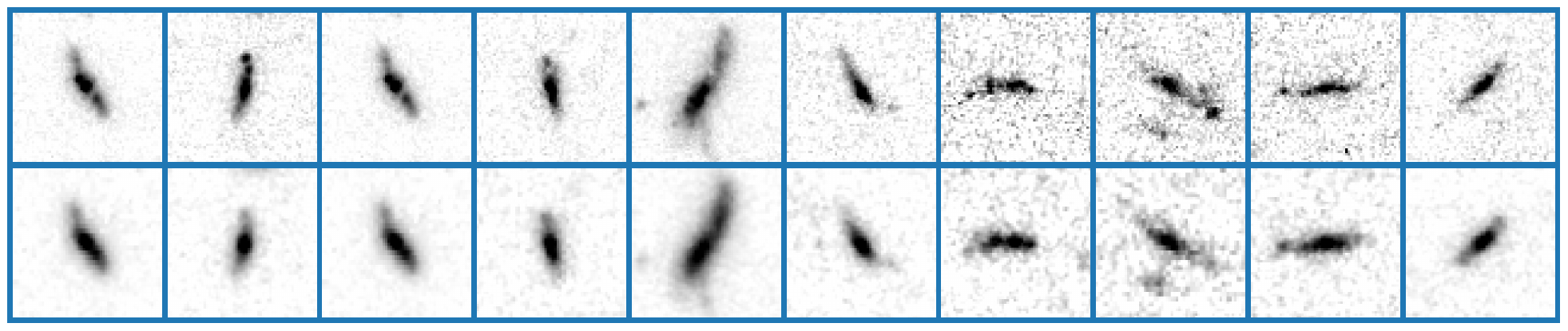}
    \includegraphics[width=0.75\textwidth]{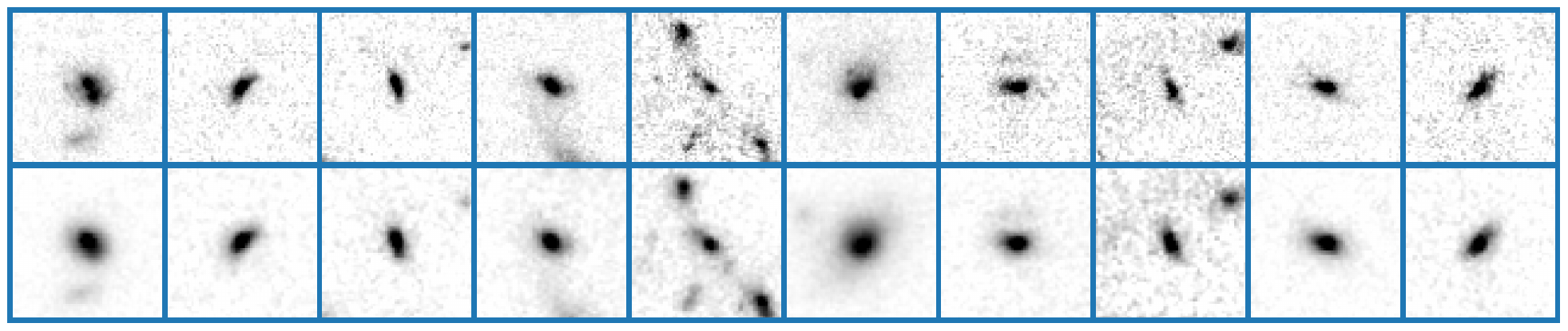}
    \includegraphics[width=0.75\textwidth]{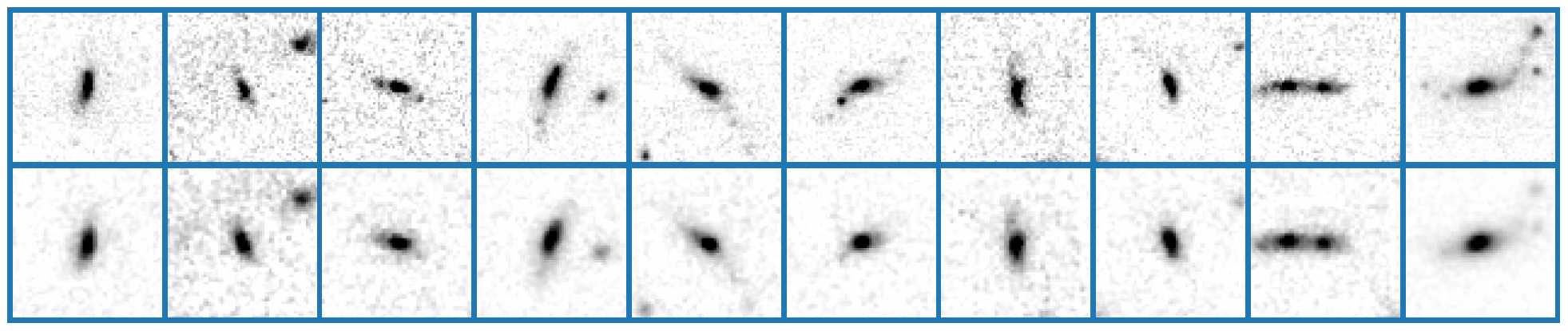}
    \includegraphics[width=0.75\textwidth]{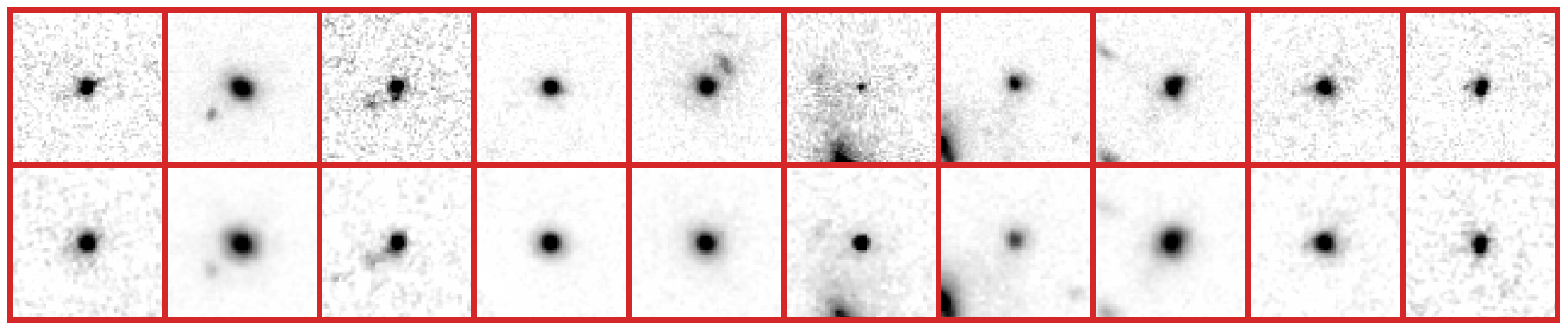}
    \caption{Examples of galaxy images in the CEERS dataset considered as: \textit{NC Disks} (first row), \textit{NC Disks} with $b/a < 0.3$ (second row), \textit{EC Disks} (third row), \textit{EC Disks} with $b/a < 0.3$ (fourth row), and \textit{EC Sph} (fifth row) candidates. Images are framed in color according to the visual classification: blue for \textit{Disk} and red for \textit{Sph} candidates. Each galaxy image is shown in the two F200W (top row) and the F356W (bottom row) filters.} 
    \label{fig:CEERS_compact_examples}
\end{figure*}

In \autoref{fig:CEERS_compact_examples}, we show examples of galaxies in the CEERS dataset that are classified into \textit{EC} and \textit{NC}, according to the criterion described in \autoref{sec:compact}. For comparison, we show examples of \textit{Disk} and \textit{Sph} galaxies following the CNN-based morphological classification presented in \citep{Huertas-Company2023}. It is clear the differences between the \textit{NC Disks} and the \textit{EC Disks}, with more extended and elongated (in projection) light distributions for the \textit{NC Disks} candidates than for the \textit{EC Disks} candidates. We also include examples of \textit{NC Disks} and \textit{EC Disks} candidates that contribute to the low end of the $b/a$ distribution in \autoref{fig:shape_CEERS}. In fact, all these examples have values of $b/a < 0.3$. For the \textit{NC Disks} candidates with $b/a < 0.3$, they show in almost all the examples signs of multiple clumps and/or in interaction with the central galaxy. For several cases of the \textit{EC Disks} candidates with $b/a < 0.3$, there are also signs of multiple clumps that could lead to an underestimation of the true $b/a$ values.

Moreover, it is difficult to distinguish between some \textit{EC Disks} and \textit{EC Sph} candidates, as they exhibit compact and round light distributions, although the \textit{EC Disk} galaxies appear slightly more elongated (on average) than the \textit{EC Sph} ones.

%%%%%%%%%%%%%%%%%%%%%%%%%%%%%%%%%%%%%%%%%%%%%%%%%%

\end{document}